\documentclass[a4paper,11pt]{article}

\usepackage{jcappub} 

\usepackage[T1]{fontenc} 

\usepackage{graphicx}
\usepackage{amsmath,bm,amssymb,amsfonts,dsfont,xspace}
\usepackage{textgreek}
\usepackage{derivative}
\usepackage[usenames,dvipsnames]{xcolor}
\usepackage[normalem]{ulem}
\usepackage{url}
\usepackage{footmisc}
\usepackage{multirow}
\usepackage{hyperref}
\usepackage{scrextend}
\usepackage{booktabs}
\usepackage{adjustbox}
\usepackage{array}
\usepackage{comment}

\AtBeginDocument{%
  \heavyrulewidth=.08em
  \lightrulewidth=.05em
  \cmidrulewidth=.03em
  \belowrulesep=.65ex
  \belowbottomsep=0pt
  \aboverulesep=.4ex
  \abovetopsep=0pt
  \cmidrulesep=\doublerulesep
  \cmidrulekern=.5em
  \defaultaddspace=.5em
}

\newcommand{\lsim}{\mathrel{\mathop{\kern 0pt \rlap
  {\raise.2ex\hbox{$<$}}}
  \lower.9ex\hbox{\kern-.190em $\approx$}}}
\newcommand{\gsim}{\mathrel{\mathop{\kern 0pt \rlap
  {\raise.2ex\hbox{$>$}}}
  \lower.9ex\hbox{\kern-.190em $\approx$}}}

\newcommand{\ie}{{\it i.e.}}
\newcommand{\eg}{{\it e.g.}}
\definecolor{darkblue}{RGB}{50,0,150}
\newcommand{\dbl}[1]{{\color{darkblue}#1}}	

\newcommand{\maddm}{{\sc MadDM}\xspace}
\newcommand{\mg}{{\sc MadGraph\_aMC@NLO}\xspace}

\newcommand{\Pythia}{{\sc Pythia}\xspace}

\newcommand{\vincia}{{\sc Vincia}\xspace}
\newcommand{\pppc}{{PPPC}\xspace}
\newcommand{\qcdunc}{{\sc QCDUnc}\xspace}
\newcommand{\hdms}{{\sc HDMS}\xspace}
  
\newcounter{daggerfootnote}

\title{\dbl{CosmiXs}: \dbl{Cos}mic \dbl{m}essenger spectra for \dbl{i}ndirect \dbl{dark matter} \dbl{s}earches}

\author[a]{Chiara Arina}
\author[b]{\!\!, Mattia Di Mauro{$^\dagger$}}
\author[c,b]{\!\!, Nicolao~Fornengo}
\author[d,e]{\!\!, Jan~Heisig}
\author[f]{\!\!, Adil~Jueid{$^\dagger$}}
\author[g]{\! and Roberto Ruiz de Austri$^\dagger$}

\affiliation[a]{Centre for Cosmology, Particle Physics and Phenomenology (CP3), Universit\'e catholique de Louvain, Chemin du Cyclotron 2, 1348 Louvain-la-Neuve, Belgium}
\affiliation[b]{Istituto Nazionale di Fisica Nucleare, Sezione di Torino, Via P. Giuria 1, 10125 Torino, Italy}
\affiliation[c]{Department of Physics, University of Torino, Via P. Giuria 1, 10125 Torino, Italy}
\affiliation[d]{Institute for Theoretical Particle Physics and Cosmology, RWTH Aachen University, D-52056 Aachen, Germany}
\affiliation[e]{Department of Physics, University of Virginia, Charlottesville, Virginia 22904-4714, USA}
\affiliation[f]{Particle Theory and Cosmology Group, Center for Theoretical Physics of the Universe, Institute for Basic Science (IBS), Daejeon, 34126, Republic of Korea}
\affiliation[g]{Instituto de F\'{\i}sica Corpuscular, CSIC-Universitat de Val\`encia, E-46980 Paterna, Valencia, Spain}

\emailAdd{chiara.arina@uclouvain.be}
\emailAdd{dimauro.mattia@gmail.com}
\emailAdd{nicolao.fornengo@unito.it}
\emailAdd{heisig@physik.rwth-aachen.de}
\emailAdd{adiljueid@ibs.re.kr}
\emailAdd{rruiz@ific.uv.es}

\abstract{
The energy spectra of particles produced from dark matter (DM) annihilation or decay are one of the fundamental ingredients to calculate the predicted fluxes of cosmic rays and radiation searched for in indirect DM detection. We revisit the calculation of the source spectra for annihilating and decaying DM using the \vincia shower algorithm in \Pythia to include QED and QCD final state radiation and diagrams for the EW corrections with massive bosons, not present in the default \Pythia shower model. We take into account the spin information of the particles during the entire EW shower and the off-shell contributions from massive gauge bosons. Furthermore, we perform a dedicated tuning of the \vincia and \Pythia parameters to LEP data on the production of pions, photons, and hyperons at the $Z$ resonance and discuss the underlying uncertainties. To enable the use of our results in DM studies, we provide the tabulated  source spectra for the most relevant cosmic messenger particles, namely antiprotons, positrons, $\gamma$ rays and the three neutrino flavors, for all the fermionic and bosonic channels and DM masses between 5 GeV and 100 TeV, on \href{https://github.com/ajueid/CosmiXs.git}{github}.}

\keywords{Dark Matter, Indirect Detection experiments, Monte Carlo event generators.}

\begin{document}
{\let\thefootnote\relax
\footnotetext{$^\dagger$\,Contact authors} }

\maketitle

\flushbottom

\section{Introduction}
\label{sec:intro}

The particle origin of dark matter (DM) remains one of the most puzzling mysteries in Physics. Different strategies are pursued to search for a particle physics signal originated by DM  interactions beyond the gravitational one (see, \eg,~\cite{Freese:2008cz} for a review).
Indirect detection searches seek to find excesses in the fluxes of cosmic messenger particles, like positrons ($e^+$), antiprotons ($\bar{p}$), $\gamma$ rays, neutrinos ($\nu$) and antinuclei~\cite{Gaskins:2016cha}, that stem from DM annihilation into SM particles and their subsequent decays in locally overdense regions like the Galactic center. They constitute an important pillar for testing the mechanism behind DM genesis in the early Universe because they directly probe the annihilating nature of DM required by thermal freeze-out.

A theoretically well-motivated DM candidate that allows for thermal freeze-out is a Weakly Interacting Massive Particle (WIMP). WIMPs emerge as new particles in several Beyond the Standard Model (BSM) theories, for instance, in Supersymmetry. A signal of cosmic messengers generated by WIMPs in the GeV-TeV energy range can be detected by different operating experiments such as \eg~AMS-02, {\it Fermi}-LAT, Imaging and Water Cherenkov detectors, SuperK and IceCube (see \eg~\cite{Leane:2020liq}).

The theoretical calculation of the flux of messenger particles produced by DM annihilation or decay is based on the {\it energy spectra} at source\footnote{The energy spectra at source are the one produced directly from DM particles annihilating or decaying at the astrophysical target of interest. Therefore, they are not the spectra of particles that reach Earth, which instead need to account for their propagation from the source, thus involving additional processes, like oscillation in the case of neutrinos, absorption in the case of $\gamma$ rays or energy redistribution in the case of charged particles.}
(hereafter simply called spectra) of these particles, which are typically calculated using codes for the generation of high-energy physics collision events also known as multi-purpose Monte Carlo event generators.
One of the most widely used reference for the source DM spectra is Ref.~\cite{Cirelli:2010xx} (hereafter \pppc).
The latest version of \pppc\ has been produced by using the \Pythia event generator version 8.135 to calculate DM spectra for different annihilation channels and masses from 5 GeV to 100 TeV\@. \pppc employed a process where DM creates a resonance, with a center-of-mass energy equal to twice the DM mass, which subsequently decays into a couple of SM particles. These results can be applied to a wide range of models and have established themselves as a standard tool used by the community for indirect DM searches.

For the most widely considered DM candidates, like the WIMP, messenger particles are mainly produced by three processes: hadronization, leading order electroweak (EW) processes and EW corrections.
The hadronization is initiated with the production of quarks and/or gluons, either produced directly or through the decays of heavy resonances, which subsequently generate gluons and other quarks.  After a timescale of the order of a Fermi ($10^{-15}$ m), these particles hadronize and produce mesons and baryons before decaying into stable particles at cosmological scales (except, $\bar{p}$ which is stable). Hadronization constitutes one of the main processes for producing $\bar{p}$, $\gamma$ rays, through the production and subsequent decay of $\pi^0$, and $\nu$  and $e^{\pm}$ coming from the decay of $\pi^{\pm}$, kaons and hyperons. Leading order EW processes take into account all EW decays, such as the one of $\tau$ and $\mu$ leptons or EW gauge bosons. These processes are the main production mechanisms for electrons and positrons.

EW corrections include initial state radiation (ISR), final state radiations (FSR) and internal bremsstrahlung (IB).  When DM is a singlet under the SM gauge symmetries, it does not directly emit radiation and thus ISR is not considered. However, when DM particles couple to the SM gauge bosons, EW ISR can provide a sizeable contribution to the final spectra, see \eg~\cite{Cirelli:2005uq,Ciafaloni:2011gv,Bringmann:2013oja,Baumgart:2018yed,Beneke:2019vhz,Beneke:2022pij,Baumgart:2023pwn}. In general, ISR exhibits a degree of model dependence. Since we focus on a model-independent approach, we will not discuss ISR further in this work. Instead, we will take into account the radiative emissions from FSR processes, that can produce final state particles in terms of photons, gluons and EW bosons. The \Pythia standard shower algorithm by default includes FSR of photons from fermions and gluons from quarks. It includes also the emission of $W^{\pm}$ and $Z$ bosons, named as EW showers. This latter contribution was not included in the \Pythia version employed in \pppc.
Therefore, the authors have included EW showers semi-analytically to first order, following the method of Ref.~\cite{Ciafaloni:2010ti}.

\begin{figure*}
    \centering
    \includegraphics[width=0.65\linewidth]{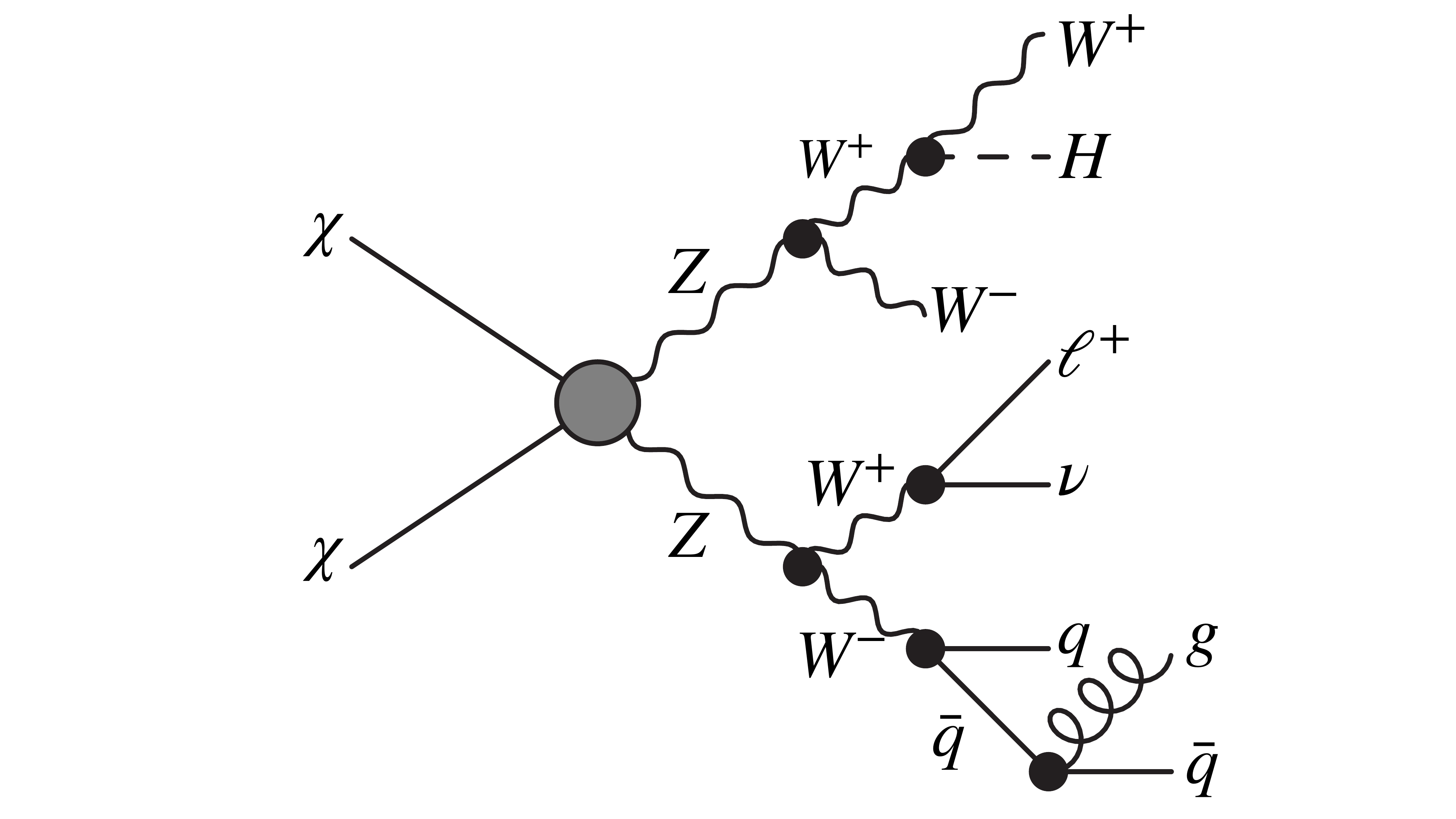}
    \caption{Feynmann diagram of an event where two DM particles ($\chi$) annihilate into two $Z$ bosons. Then, the $Z$ particles emit EW radiation, producing other bosons, which subsequently decay into fermions.}
    \label{fig:diagrams}
\end{figure*}

The approach taken in \pppc has represented the state of the art for many years. However, several new features have been added in \Pythia the meanwhile.
Indeed, in the latest \Pythia versions, the emission of $W^{\pm}$ and $Z^0$ gauge bosons off fermions is an integrated part of the ISR and FSR, and is fully interleaved with the QCD and QED emissions (see Ref. \cite{Christiansen:2014kba} for details). More recently, an implementation of a new shower algorithm has been provided in \vincia~\cite{Fischer:2016vfv}. 
It uses a $p_T$-ordered model for QCD+QED/EW helicity-dependent showers based on the Antenna formalism. Originally developed as an independent plugin, it has been incorporated into the \Pythia~8 source code since \Pythia version 8.3.

The main innovations of \vincia in relation to the standard \Pythia shower process are the following.
First, \vincia takes into account the helicity along all the shower, \ie~\vincia decomposes the shower into distinct terms for each set of contributing helicities.
Secondly, it includes the trilinear gauge boson interactions $Z^0 W^+ W^-$, $H W^+ W^-$, $H Z Z$ and $\gamma W^+ W^-$, which are neglected in the standard \Pythia shower and that can give an important contribution at high DM masses (see Appendix \ref{app:shower} for more details about the differences between the two shower options). 
As an example, in Fig.~\ref{fig:diagrams}, we show a representative Feynman diagram of an event included in \vincia but not in standard \Pythia. This event involves the production of two $Z$-bosons from DM annihilation. The two $Z$-bosons produce $W$-bosons. Then one of the four bosons subsequently radiates off a Higgs boson while the two $W$-bosons in the lower branch decay into leptons and quarks, respectively. Exemplarily, we also show gluon radiation off one of these quarks. 
In addition to the previous innovations, the authors of \Pythia have included new tuning of the model parameters in the code, which makes the code more compatible than in the past with collider data.

The main differences between \pppc and \vincia are the following. First, \pppc includes EW corrections via semi-analytical calculations using the results of Ref.~\cite{Ciafaloni:2010ti}. These corrections were considered at leading order without resummations. Second, these corrections are then matched to \Pythia to add further QCD emissions, to handle the decays of heavy resonances and to hadronise colored final-state particles. Instead, in \vincia, both the EW corrections and the other QED+QCD emissions are handled in a coherent manner. Furthermore, the decay of heavy resonances such as the $W$-boson or the top quark is fully interleaved with the rest of the shower machinery, a fact that has non-trivial impact on the kinematical distributions of the decay products. Finally, as we will demonstrate later, in the version of \Pythia used in \pppc, the photon yield for leptonic channels tends to be underestimated at low energies.

Recently, the authors of Ref.~\cite{Bauer:2020jay} (hereafter \hdms\footnote{The abbreviation derives from their public repository called HDMSpectra.}) have provided decay spectra for DM with masses significantly above the scale of EW symmetry breaking. In this approach, they evolve from the scale of the DM mass down to just above the weak scale using the DGLAP equations by adopting an implementation that considers all interactions in the unbroken SM phase, as well as a partial treatment of soft-coherence effects. The handling of the decays of the $W/Z$ bosons and the top quark is done in the unbroken phase while the decay of the SM Higgs boson is performed with \Pythia. They perform a matching by evolving across a parametrically small region through the weak scale, removing all particles with EW scale masses. Finally, these results are matched at the EW scale to \Pythia, which is then used to calculate the subsequent showering, hadronization, and light particle decays. The results of HDMS can lead to theoretical problems in the matching between the physics generated by the DGLAP formalism in the unbroken phase and the physics described by \Pythia. 
Our results are not affected by any of the issues described above because all the processes happening after the DM annihilation or decay, including hadronization, EW corrections and particle decays, are calculated internally and consistently by \vincia.  Additionally, our approach allows us to release reliable results down to DM masses much lower than \hdms, which includes spectra for values above 500 GeV only. Moreover, we use an improved version of the \Pythia code, tuning the most important parameters to LEP data.
In Refs.~\cite{Amoroso:2018qga,Jueid:2022qjg,Jueid:2023vrb} (hereafter \qcdunc), some of the authors of this paper have derived a new set of hadronization parameters (tuning) using \Pythia 8.2 through a fit to LEP and SLD data at the $Z$-boson resonance. They have also calculated a conservative set of uncertainties on the shower and hadronization model parameters. They have estimated the impact on the energy spectra of $\bar{p}$, $\gamma$ rays, $\nu$ and $e^+$, which is at most of the order of $10\%$--$20\%$ in the peak region.

In this work, we generate state-of-the-art spectra for the following cosmic particle messengers produced during DM annihilation and decay: $\bar{p}$, $\gamma$ rays, $\nu$ (in their three flavor states) and $e^+$.\footnote{We verified that the spectra of $e^+$ and $e^-$ are the same and therefore can be used for those indirect searches that do not distinguish between electrons and positrons.} We improve the existing results on this topic adopting the following strategy:
\begin{itemize}
\item We use the latest version of the \vincia algorithm, implemented in the \Pythia version 8.309, which we interface with \maddm~\cite{Backovic:2013dpa,Backovic:2015cra,Ambrogi:2018jqj}. We use \maddm for two reasons. First, it is convenient for having an event file containing all (in particular the helicity) information to be passed to \vincia. Second, we need a proper event generator for the 4-body processes as detailed in Sec.~\ref{sec:polar} since \maddm is the only event generator dedicated to DM annihilation. Thereby, we include contributions from triple gauge boson interaction in the showering and carry the helicity information throughout the showering process. We also include the effect of running quark masses. \vincia represents the state-of-the-art code for including EW corrections for masses up to hundreds of TeV.
\item We include annihilation channels with off-shell gauge bosons and new channels such as $ZH$ and $Z\gamma$.
\item We carry out an improved tuning of the hadronization parameters in \Pythia (adopting the \vincia shower as our default) to fit the available data for the production of particles from the $Z$ resonance measured at LEP. These parameters are used to make predictions for DM annihilation thanks to the jet universality.
\end{itemize}
Our results are publicly available in a \href{https://github.com/ajueid/CosmiXs.git}{github repository}.\footnote{\href{https://github.com/ajueid/CosmiXs.git}{\tt https://github.com/ajueid/CosmiXs.git}\label{fn:github}}
While we perform the simulations for annihilating DM, upon simple rescaling, the results can be used for decaying DM as well. Specifically, we cover the case of decaying scalar, pseudoscalar\footnote{In the case of pseudoscalar DM the decay into $HH$ and $W^+W^-$ and $ZZ$ are forbidden due to CP violation.}, vector and axial-vector DM\@. A dedicated study of decaying fermionic DM is left for future work. \\

The remainder of the paper is organized as follows. In Sec.~\ref{sec:had}, we describe the hadronization and EW processes that are responsible for the production of particles relevant to DM studies. The main novelties of our 
analysis are detailed in Sec.~\ref{sec:imp}. In Sec.~\ref{sec:tuning}, we report the results of the tuning of the \vincia shower algorithm and the consequent uncertainties in the final particle spectra. Finally, in Sec.~\ref{sec:results} we show the results for the particle spectra and the comparison with other reference results, and we draw our conclusions in Sec.~\ref{sec:conclusions}.

\section{Hadronization and Electroweak Model}
\label{sec:had}

In this section, we briefly discuss the physical description of the particle production from DM\@. We begin with a discussion of the general features of stable particle production from DM annihilation, which encompasses a complex sequence of phenomena including resonance decay, QED and QCD bremsstrahlung, EW showers, hadronization and hadron decays. We conclude this section with a discussion of the composition of particle spectra, with some examples of final state particles and annihilation channels.

\subsection{Emission mechanism of particles from dark matter}

Let us consider a generic annihilation process of DM particles $\chi$ into a set of final-state particles:
\begin{eqnarray}
        \chi \chi \to \underbrace{\bigg[X_{1} X_{2} \ldots X_{N} \bigg]}_{\rm Intermediate~states} \to \overbrace{\bigg(Y_{11} \ldots Y_{1a_1} \bigg) \ldots \bigg(Y_{N1} \ldots Y_{N a_N} \bigg)}^{\rm Stable~particles}.
        \label{eq:annihilation}
    \end{eqnarray}
In the case where the narrow-width approximation holds, we can factorise the whole process into a production part and a decay part. In general, the first part consists in the production of $N$ particles ($X_1, \cdots X_N$), which may be quarks, gluons, leptons or heavy resonances such as the $W/Z/H$ bosons or the top quark. These particles undergo a series of complex processes that give rise to various particles that are stable over astrophysical/cosmological scales, such as photons, positrons, neutrinos or antiprotons. The narrow width approximation holds true for particles with small decay widths, such as the Higgs boson. However, off-shell effects provide important corrections for the pair production of massive gauge bosons, especially below their production threshold. We stress that the narrow width approximation is not valid in this case since heavy particles produced either in the annihilation process or the showering process do not decay until their virtuality reaches a scale that is close to their offshellness scale. The physics modeling of stable particle production depends on the nature of the intermediate particles $X_i$ and/or their decay products ($Y_{ij}$).

\begin{figure}[!t]
    \centering
\includegraphics[width=0.70\linewidth]{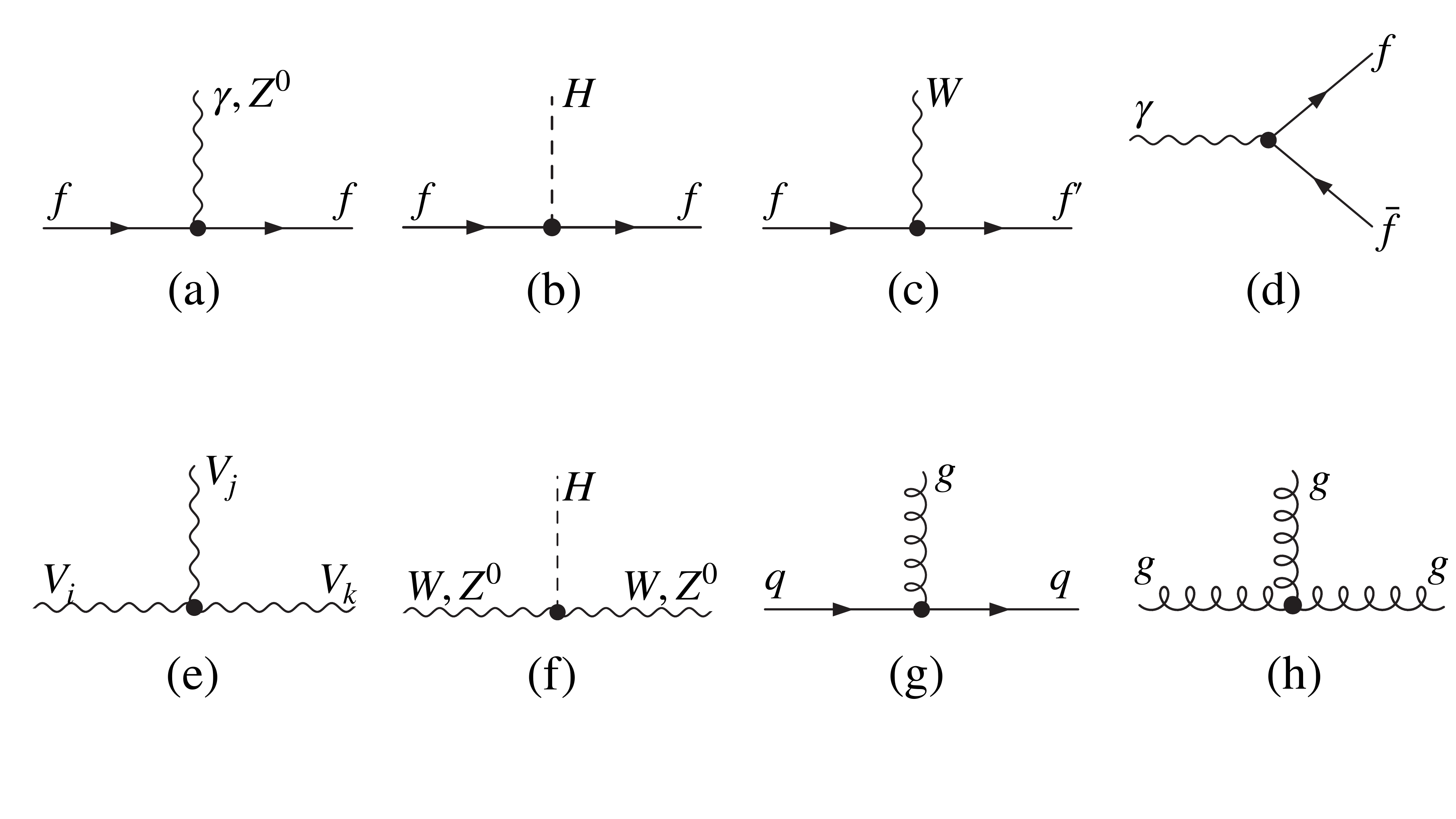}
\vspace{-0.5cm}
    \caption{Example of Feynman diagrams for the different parton-shower branchings. Here we show QED+EW emissions of gauge bosons off fermion lines (a)--(b), Higgs emission off fermion lines (c), photon splitting into $f\bar{f}$ (d), gauge boson and Higgs boson emissions off bosonic lines (e)--(f) (labeled as EWBR), gluon emission off quark lines (g) and $g\to gg$ (h).}
    \label{fig:FD:PS}
\end{figure}

QED bremsstrahlung occurs when the $X$ or the $Y$ particles are either electrically charged or include photons. In this case, additional photons and/or electrically charged particles are produced through $X_i \to X_i \gamma$ (Fig.~\ref{fig:FD:PS}a) and $\gamma \to f\bar{f}$ (Fig.~\ref{fig:FD:PS}d). The photon emission is enhanced for both soft and quasi-collinear regions.\footnote{For more details, see Section 4 of Ref. \cite{Buckley:2011ms}.} Note that the collinear photons can have very high energies ($E_\gamma \to M_\chi$) provided that the angle between the parent particle and the photon is extremely small. On the other hand, fermion pair production through $\gamma \to f\bar{f}$ can occur with subleading probabilities but enhanced at low values of photon virtualities: $(p_f + p_{\bar{f}})^2/M_\chi^2 \to 0$. If the phase space is permitting, fermions in general can emit $W/Z/H$ bosons 
(Fig.~\ref{fig:FD:PS}a--c).
The inclusion of the $W/Z$-boson emissions was included in \Pythia since version~8.176. On the other hand, due to the gauge structure of the SM, massive gauge bosons and the SM Higgs boson can undergo further weak emissions (Fig.~\ref{fig:FD:PS}e--f). Note that not all of these branchings are included in \Pythia but \vincia includes all of them through the Antenna formalism. In fact, more than 1000 Antenna functions for weak showers are implemented in \vincia. In the following, the EW radiation from bosons is called EWBR.

In the case where the $X$ or the $Y$ particles are colored particles, further colored particles are produced through QCD bremsstrahlung (Fig.~\ref{fig:FD:PS}g-h). QCD showers are treated in similar fashion as the case of QED showers, which reflects the enhancement of probabilities for both soft and collinear emissions ($X \to X g$ and $g \to q \bar{q}$) in addition to the $g \to g g$ branching. The probabilities of the QCD shower branching are controlled by the value of the strong coupling constant. 

\begin{figure}[!t]
\centering
\includegraphics[width=0.60\linewidth]{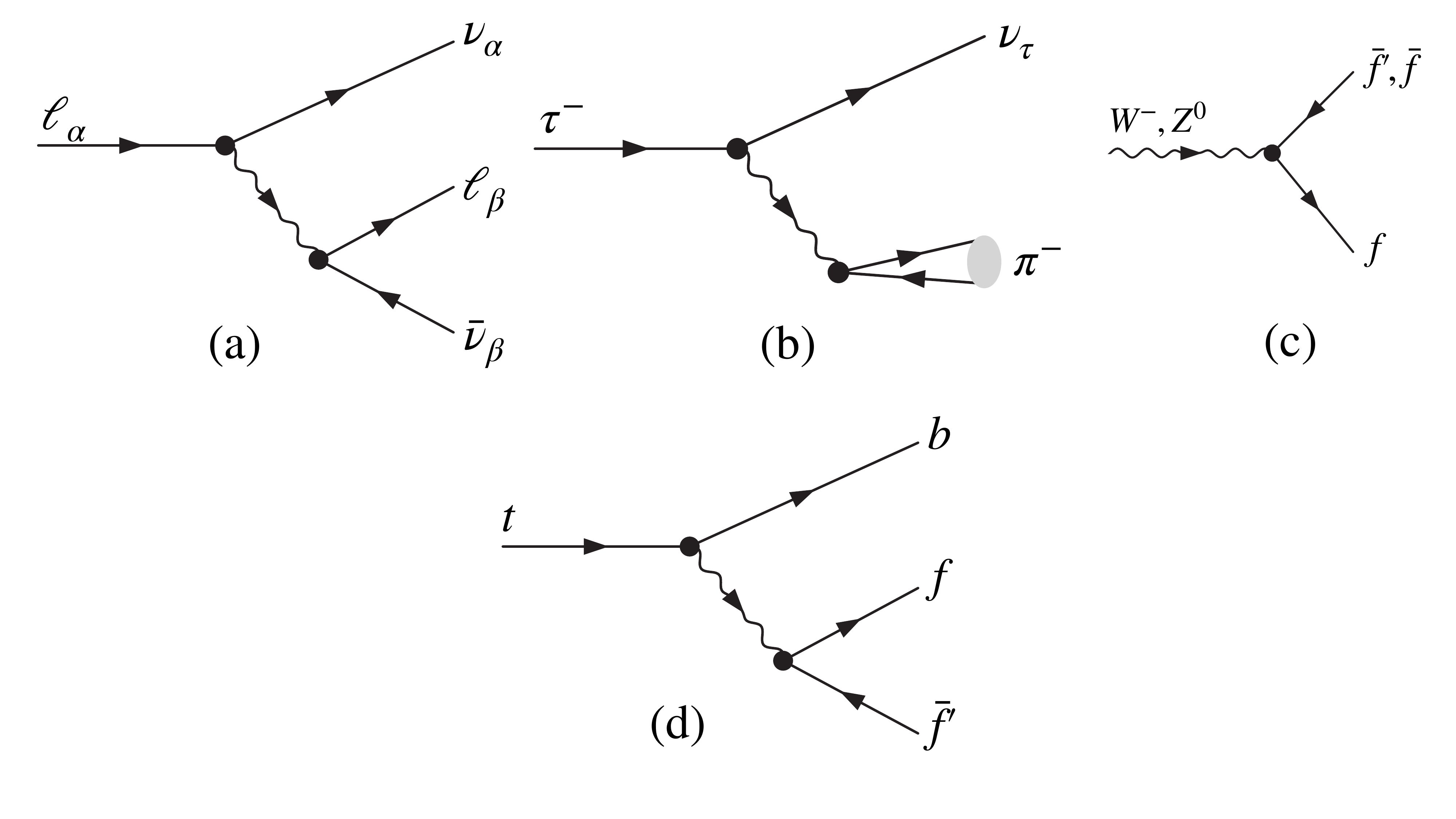}
\caption{Examples of Feynman diagrams for weak coupling driven decays of fermions and gauge bosons. We show the leptonic decays of charged leptons (a), the semi-leptonic decays of the tau lepton (b), the fermionic decays of the massive gauge bosons (c) and the three-body decay of the top quark (d).}
\label{fig:FD:resonance}
\end{figure}

The decay of short-lived particles is an important source for stable particle spectra, especially $e^+$ and $\bar{\nu}$ (Fig.~\ref{fig:FD:resonance}). The contribution of these sources to the particle spectra is dominant for the high-energy region.  We note that two annihilation/decay channels are very special. First the $t\bar{t}$ channel is the only channel where we can have resonance decays, QED+QCD showers, EW showers and hadronization. The $W^+ W^-$ channel is also very special as both electroweak/QED showers and resonant decays can occur, producing quarks and leptons which then undergo further QED+QCD showers before producing hadrons. The default treatment in \vincia is that resonance decays are interleaved with QED/QCD and EW showers. More details about the EW showers and the treatment of heavy-resonance decays in \vincia are shown in Appendix \ref{app:shower}. \\

\begin{figure}
    \centering
    \includegraphics[width=0.60\linewidth]{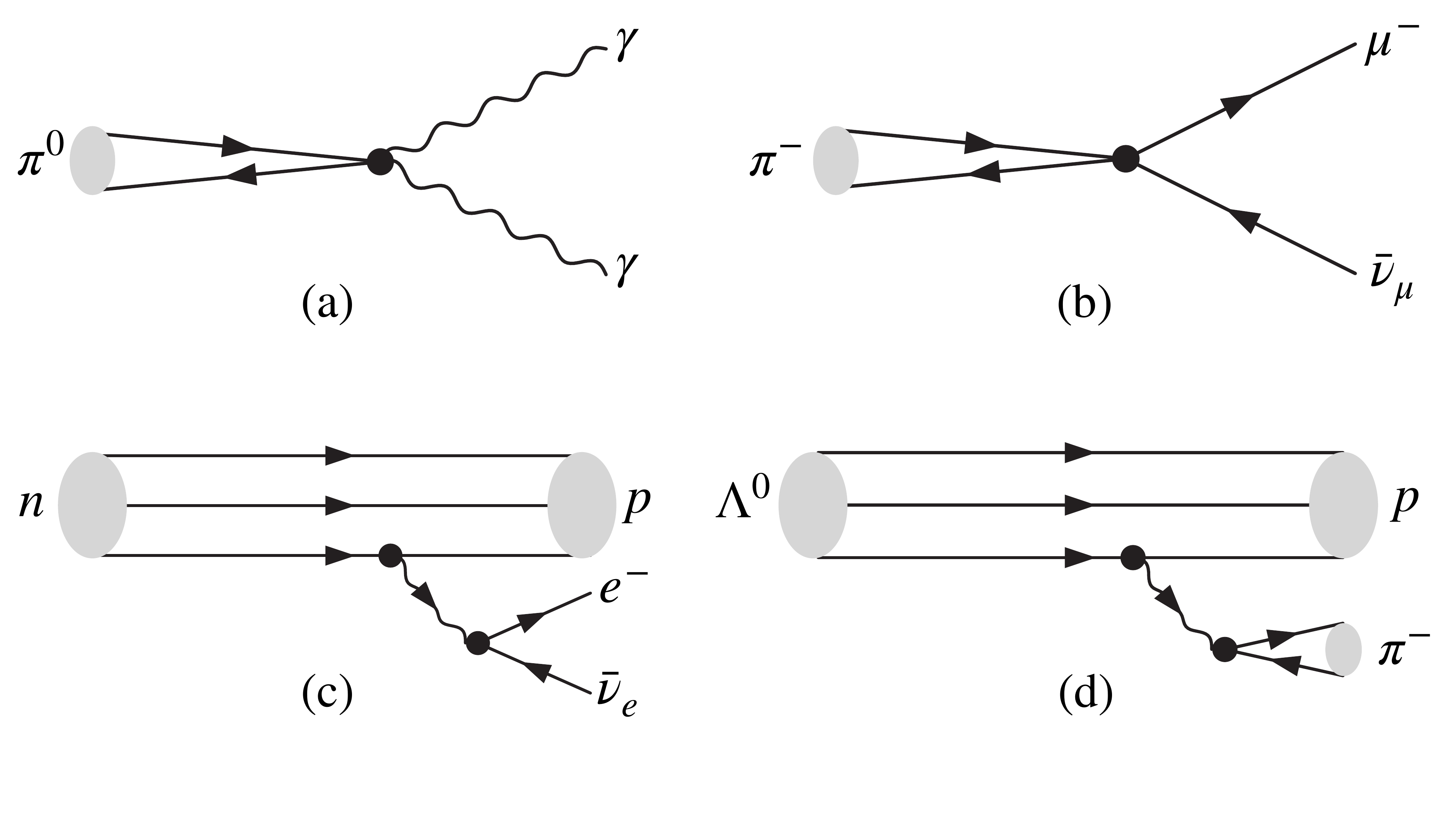}
    \caption{Examples of Feynman diagrams for the main hadron decays within QCD jets. From top left to bottom right, we show the decay of $\pi^0 \to \gamma\gamma$ (a), $\pi^- \to \mu^- \bar{\nu}_\mu$ (b), $n \to pe^- \bar{\nu}_e$ (c) and $\Lambda^0 \to p\pi^-$ (d).}
    \label{fig:FD:hadrons}
\end{figure}

Any colored particle must be confined inside colour-neutral hadrons above distance scales of order $10^{-15}$ m. This process which is called hadronization cannot be modeled using first-principles QCD but only using phenomenological models. There are two main models: string models \cite{Artru:1974hr,Sjostrand:1982fn,Andersson:1983ia,Sjostrand:1984ic} and cluster  models \cite{Webber:1983if,Winter:2003tt} which are implemented in multipurpose Monte Carlo event generators. \Pythia~8 is based on the Lund string model where the hadronization is modeled by a left-right symmetric fragmentation function $f(z)$ given by 
\begin{equation}
        f(z,m_{\perp h}) \propto N \frac{(1-z)^{a_L}}{z}\exp\left(\frac{-b_L m_{\perp h}^2}{z}\right),
\label{eq:fz}
\end{equation}
which gives the probability that a hadron $h$ gets a fraction $z \in [0, 1]$ of the remaining energy at each step of the hadronization process. In Eq.\eqref{eq:fz}, $N$ is a normalisation constant, $m_{\perp h}^2 \equiv \sqrt{m_h^2 + p_{\perp h}^2}$ is the square of the transverse mass of the hadron $h$, $a_L$ and $b_L$ are tunable parameters. This basic picture of the hadronization of the $q\bar{q}$ system is unaffected in the presence of gluons, since a gluon having a colour and an anti-colour structure can be seen as a kink in the string. Eq. \eqref{eq:fz} can also be generalized to include flavour effects: in particular, strange quarks and massive quarks at the endpoints of the string. The inclusion of heavy mass effects is achieved by the Bowler modification \cite{Bowler:1981sb,Artru:1974hr}. Baryons composed of three quarks or three antiquarks are produced similarly to mesons. The production of baryons can be achieved by breaking the strings by the production of diquarks-antidiquarks between the quark and the antiquark at the string endpoints. However, this basic picture leads to a strong correlation between the produced baryon and the antibaryon in both the flavour and the angular distributions. These correlations have been falsified by experimental measurements of $\Lambda^0$--$\bar{\Lambda}^0$ angular correlations by the \textsc{Opal} collaboration \cite{OPAL:1998tlp}. To reduce the degree of the correlation between baryons, the {\it pop-corn} mechanism was introduced \cite{Andersson:1984af, Eden:1996xi}. In this mechanism, one or more $q\bar{q}$ pairs are produced in between the diquark-antidiquark pairs, which enables the production of one or more mesons between the two baryons and therefore decreases their correlation.  With all these modifications, the string fragmentation function is given by
\begin{eqnarray}
    f(z) \propto N \frac{1}{z^{1+r_Q b_L m_Q}} z^{a_{\rm eff}} \bigg(\frac{1-z}{z}\bigg)^{a_{\rm eff}} \exp\left(\frac{-b_L m_{\perp h}^2}{z}\right),
\end{eqnarray}
with $Q=c,b$ and $a_{\rm eff} \equiv a_L + a_{\rm QQ}$. The parameters of the Lund string fragmentation function are given in Tab.~\ref{tab:ranges}. We must stress that for the purpose of this study, we do not need to tune the Bowler parameters $r_c$ and $r_b$ and we use their default values. The produced hadrons within QCD jets decay into $\gamma$ (Fig.~\ref{fig:FD:hadrons}a), $\mu^- \to e^- \bar{\nu}_e \nu_\mu$ (Fig.~\ref{fig:FD:hadrons}b), and $\bar{p}$ (Fig.~\ref{fig:FD:hadrons}c--d). The contribution of these decays dominates in the peak and the bulk regions of the spectra. For the $\bar{p}$ spectra, another important source comes from hadronization, in which case these $\bar{p}$ are called primary (see Refs.~\cite{Jueid:2022qjg,Jueid:2023vrb} for more details). 
As mentioned above, the hadronization mechanism is only solved by phenomenological models with many free parameters. Therefore, the uncertainties can be estimated based on the parameters of the hadronization model. The estimation of these uncertainties was done for the {\it first time} in Refs. \cite{Amoroso:2018qga,Jueid:2022qjg,Jueid:2023vrb,Bierlich:2023fmh}. These uncertainties were found to be of the order of $10$--$30\%$ depending on the annihilation channel, the DM mass and the final state particles. Furthermore, it was found that the impact of these uncertainties on the best-fit point of the DM mass and the annihilation cross-section can be dramatic, especially for heavy DM.

\begin{table}[!t]
\setlength\tabcolsep{2pt}
  \begin{center}
    \begin{tabular}{llcll}
\noalign{\hrule height 1pt}
      parameter & \textsc{\Pythia8} setting & Variation range & \vincia  \\
\noalign{\hrule height 1pt}
      $\sigma_{\perp}$~(GeV) & \verb|StringPT:Sigma|   & 0.0 -- 1.0 & 0.305  \\
      $a_L$              & \verb|StringZ:aLund|    & 0.0 -- 2.0 & 0.45  \\
      $b_L$              & \verb|StringZ:bLund|    & 0.2 -- 2.0 & 0.80 \\
      $a_{QQ}$ & \verb|StringZ:aExtraDiquark| & 0.0 -- 2.0 & 0.90  \\
      $r_c$   & \verb|StringZ:rFactC| & 0.0 -- 2.0 & 0.85 \\
      $r_b$   & \verb|StringZ:rFactB| &  0.0 -- 2.0 & 1.15 \\
 \noalign{\hrule height 1pt}
    \end{tabular}
  \end{center}
  \caption{\label{tab:ranges} The main parameters of the Lund fragmentation function in \Pythia~8 along with their range and their default values when using the \vincia shower algorithm.}
\end{table} 

\subsection{Composition of particle spectra}
\label{sec:composition}

\begin{figure}[!tbh]
\centering
\includegraphics[width=0.49\linewidth]{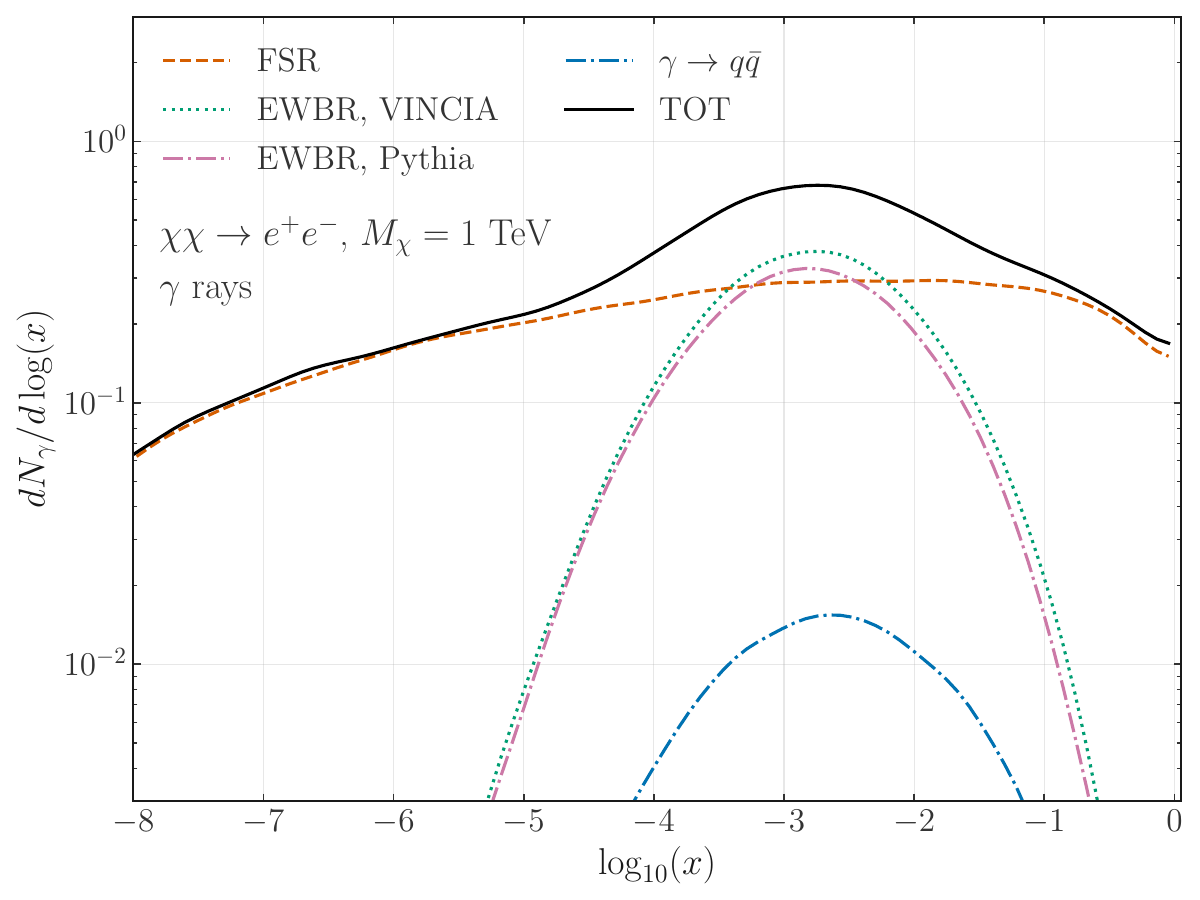}
\includegraphics[width=0.49\linewidth]{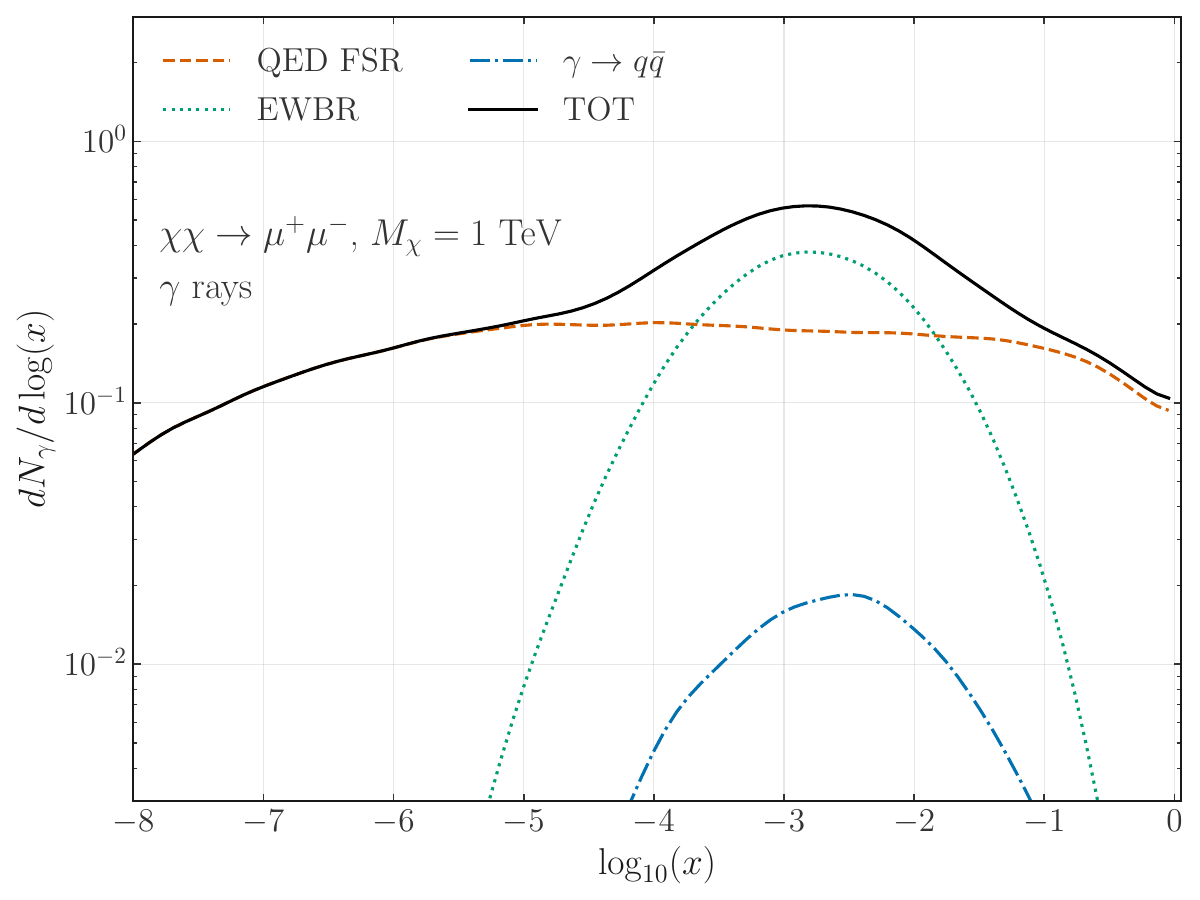}
\includegraphics[width=0.49\linewidth]{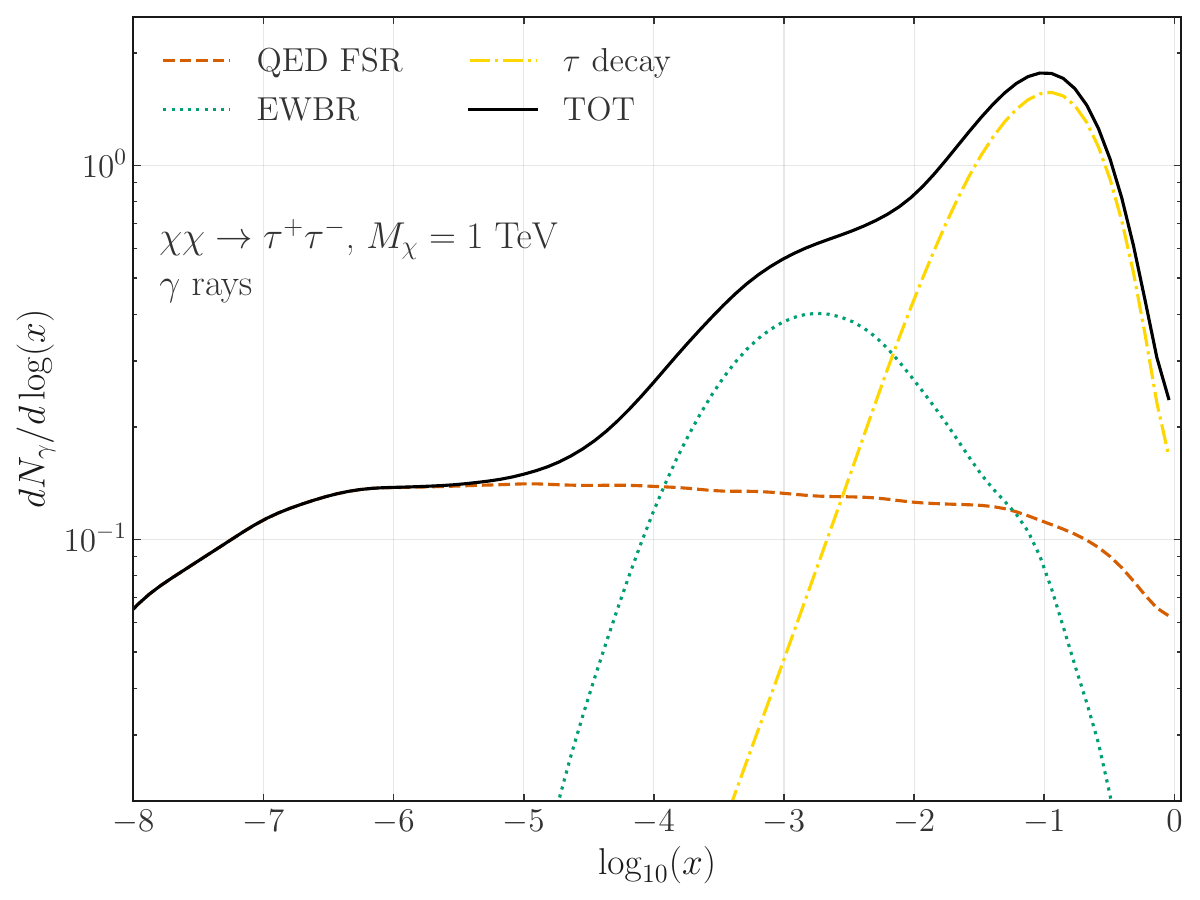}
\includegraphics[width=0.49\linewidth]{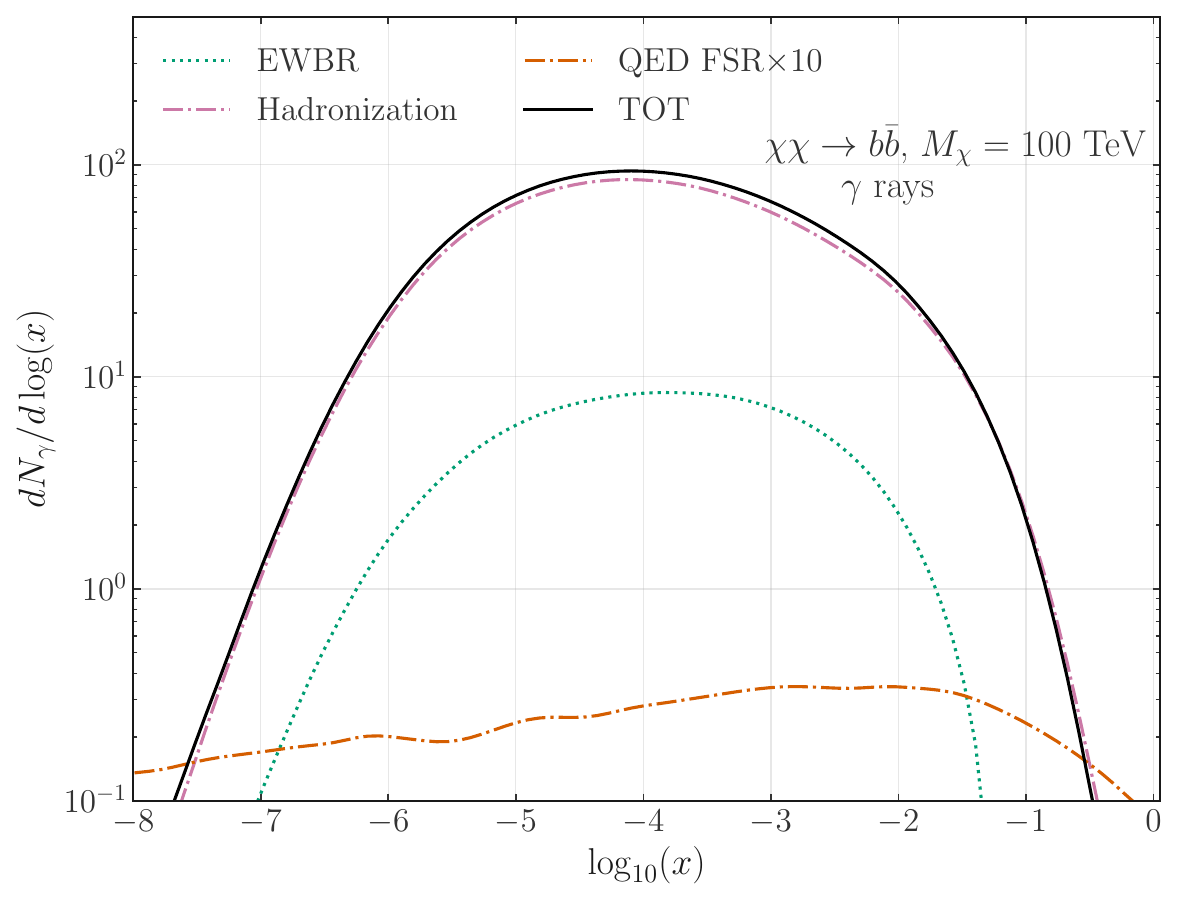}
\includegraphics[width=0.49\linewidth]{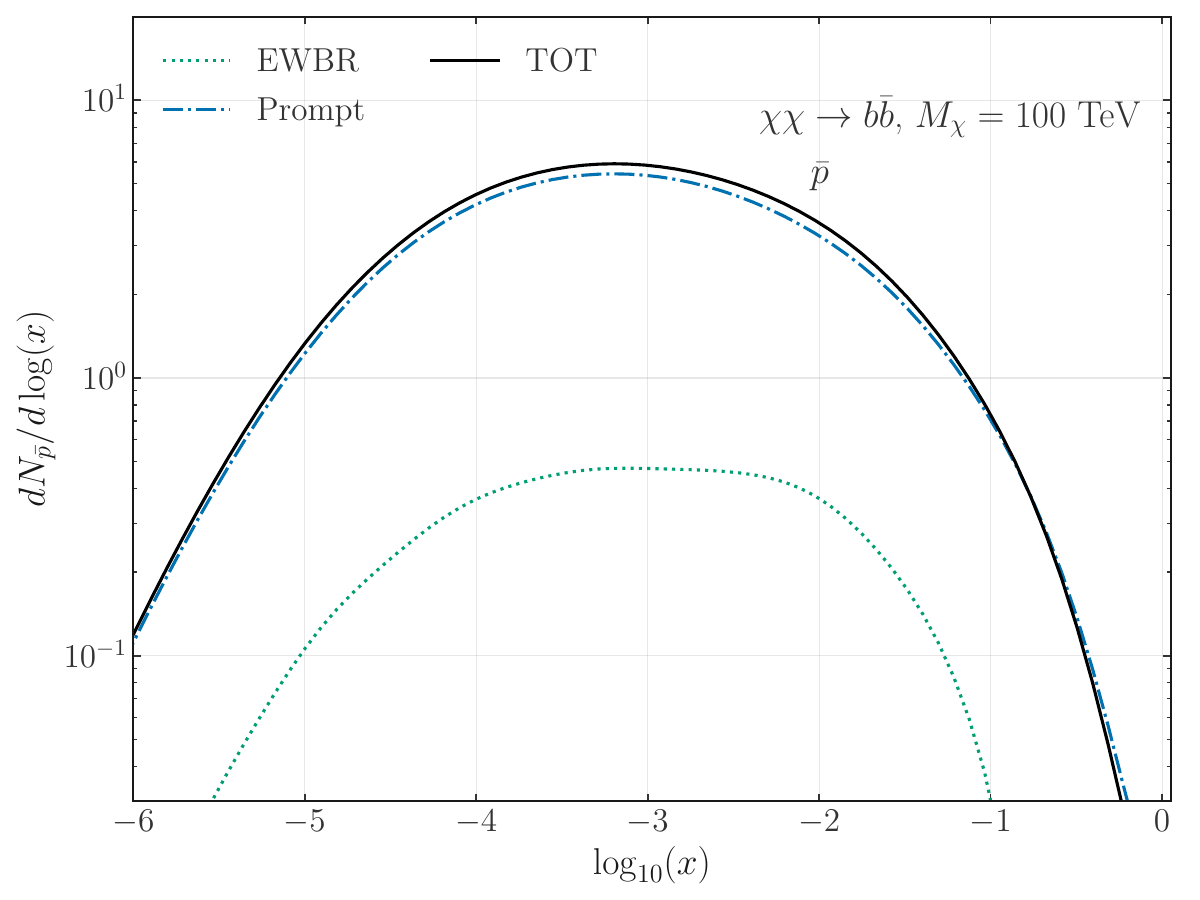}
\includegraphics[width=0.49\linewidth]{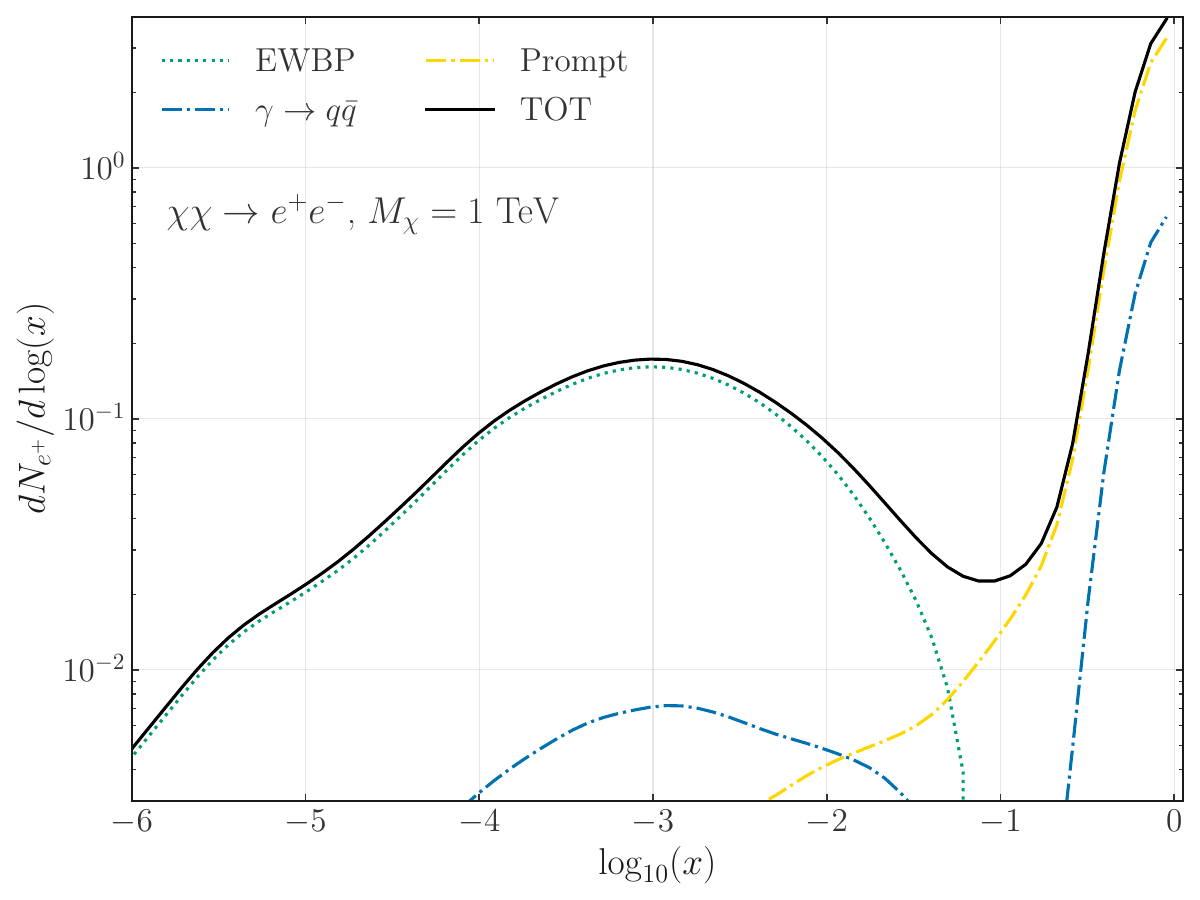}
\caption{\label{fig:spectra}
Spectra of $\gamma$ rays, positrons ($e^+$) and antiprotons ($\bar{p}$) for different annihilation channels, namely $e^{\pm}$, $\mu^{\pm}$, $\tau^{\pm}$ and $b\bar{b}$. For each plot, we report the different production mechanisms and the total spectrum in units of $\log_{10}(x)$, where $x=E/M_{\chi}$. 
We display the product of hadronization from quarks generated by DM annihilation (labeled as hadronization). The EWBR is due to the boson radiation from other bosons. This process creates quarks that hadronize into antiprotons and pions, which subsequently decay into $\gamma$ rays or $e^{\pm}$. We display also the production of photons due to QED FSR and the result of the hadronization products from quarks generated through pair production from an energetic photon (labeled as $\gamma \rightarrow q\bar{q}$). In case of $e^{\pm}$ channel we also show the direct production of positrons from the initial particles and for the $\tau$ channel the production of its decay into $\gamma$ rays through its decay into neutral pions.}
\end{figure}

In this section, we show how the different emission mechanisms explained in the previous section compose into the DM spectra.
We generate the spectra of cosmic messengers for 64 masses between 5 GeV and 100 TeV\footnote{We plan to extend the range towards larger masses in future analyses.}. We simulate 5 million annihilation events for each mass and produce the result in terms of $dN/d\log_{10}(x)$, where $x \equiv E/M_{\chi}$. $\log_{10}(x)$ is provided for 100 logarithmic values between -8 and 0. The spectra are produced for the following channels: $e^+e^-$, $\mu^+\mu^-$, $\tau^+\tau^-$, $\nu_e\nu_e$, $\nu_{\mu}\nu_{\mu}$, $\nu_{\tau}\nu_{\tau}$, $u\bar{u}$, $d\bar{d}$, $c\bar{c}$, $s\bar{s}$, $t\bar{t}$, $b\bar{b}$, $\gamma \gamma$, $g g$, $W^+W^-$, $Z Z$, $HH$, $ZH$, $\gamma Z$. In addition, for the channels $e^+e^-$, $\mu^+\mu^-$, $\tau^+\tau^-$ we also calculate separately the left-handed and right-handed spectra ($e^-_L e^+_L$ and $e^-_R e^+_R$) and for the gauge bosons $W^+W^-$, $Z Z$ the longitudinal and transverse polarization spectra ($Z_{L}Z_{L}$ and $Z_{T}Z_{T}$).
We make our results publicly available as tables in a \href{https://github.com/ajueid/CosmiXs.git}{github repository}.\footref{fn:github} 
In particular, Fig.~\ref{fig:spectra} illustrates the spectrum of $\gamma$ rays, positrons and antiprotons,  for different annihilation channels. We focus in the plot on high DM masses where EW corrections are relevant. 

\begin{figure}
\centering
\includegraphics[width=0.59\linewidth]{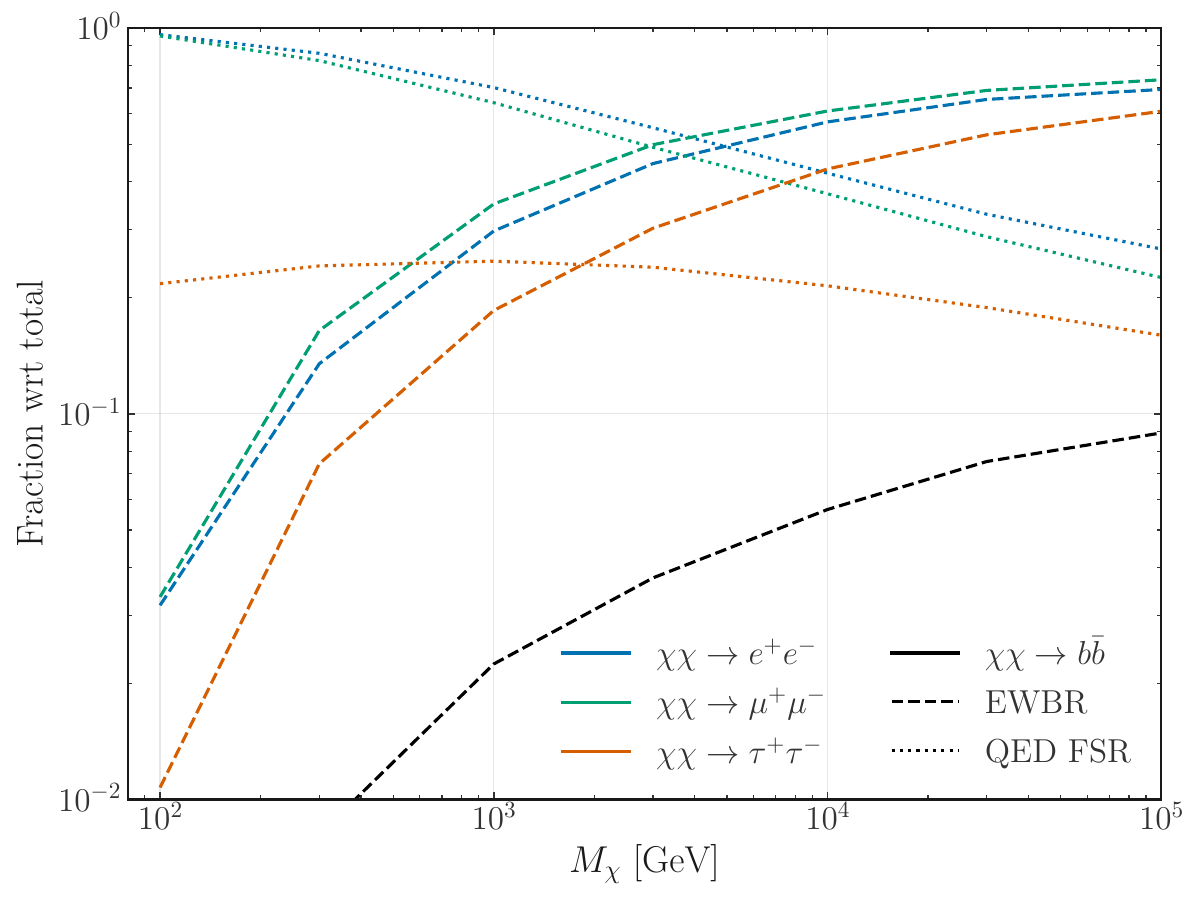}
\caption{\label{fig:frac}
Fraction between the number of photons produced through the EWBR (dashed curves) and QED FSR (dotted curves) and the total yield. We show the results for the leptonic and $b\bar{b}$ channels with different colors.}
\end{figure}

When the annihilation channel involves quarks, the main production of particles comes from the hadronization of these quarks, which emit several gluons for QCD bremsstrahlung and produce several particles in the final states.
We show in Fig.~\ref{fig:spectra} the case of $M_{\chi}=100$ TeV annihilating into $b\bar{b}$.
In particular, the production of $\gamma$ rays ($e^{\pm}$ and $\nu$) is mostly due to the hadronization process, which produces $\pi^0$ ($\pi^{\pm}$) mesons that subsequently decay into two photons (in $e^{\pm}$ and $\nu$ by muon decays). The vast majority of $\bar{p}$ are produced by the hadronization of $\bar{u}$ and $\bar{d}$ quarks. We can call this prompt or primary production. However, there is also a relevant production of secondary antiprotons, which are produced by the decay of resonances ($\Delta$-baryons), from neutrons or from hyperons, such as $\Lambda$ and $\Sigma$.
The hadronization process accounts for at least $90\%$ of the total production of particles. This process is labeled as ``Hadronization'' in Fig.~\ref{fig:spectra}.
Other two processes are responsible for the production of particles at a subleading order. These are EWBR, for which we have the radiation of bosons from other bosons (see Fig.~\ref{fig:FD:PS} d-e), and QED FSR with radiation of a photon from a fermion (see Fig.~\ref{fig:FD:PS} a). We show in Fig.~\ref{fig:frac} the evolution of the contribution of EWBR and FSR to the $\gamma$-ray total yield as a function of the DM mass. In particular, the EWBR makes at most about $10\%$ of $\gamma$ rays at 100 TeV and with a decreasing contribution going at lower DM masses. A similar result is also obtained for $\bar{p}$, $\nu$ and $e^+$. QED FSR, instead, contributes less than $1\%$ at all energies as exemplified in Fig. \ref{fig:spectra} for $b\bar{b}$. Similar results are obtained for the other quarks and for the EW and Higgs bosons.

In the case of leptonic channels, the contribution of EWBR and QED FSR becomes much more relevant. This is particularly true for the production of $\gamma$ rays and antiprotons.
Fig.~\ref{fig:spectra} shows the spectra of $\gamma$ rays for the $e^{\pm}$, $\mu^{\pm}$ and $\tau^{\pm}$ channels for a DM mass of 1 TeV. We also show the production of positrons for the $e^{\pm}$ channel. This result is representative of all the energies above a few hundreds of GeV where EWBR becomes relevant. In particular, for $e^{\pm}$ and $\mu^{\pm}$ cases the production of photons is dominated by FSR at all energies except for $\log_{10}(x)\sim [-5,-1]$ where the EWBR gives the most important contribution.
In the case of the production of positrons, electrons and neutrinos, EWBR makes the largest contribution at $\log_{10}(x)<-2$. Instead, at higher energies, these particles are produced directly from the initial leptons (referred to as  ``Prompt'' in the figure). There is another important process that produces some of these particles, namely the production of some quarks from very energetic photons (see Fig.\ref{fig:FD:PS} c). These quarks hadronize and produce mesons and hadrons, which subsequently decay into $e^{\pm}$ and $\nu$. This process, called $\gamma\rightarrow q\bar{q}$, accounts for at most $10\%$ of the total yield.
Finally, antiprotons are mainly produced by the processes EWBR and $\gamma\rightarrow q\bar{q}$.
In the case of the $\tau^+ \tau^-$ channel, there is another important process that produces $\gamma$ rays, positrons and neutrinos, namely the decay of the $\tau$ lepton into charged and neutral pions, which has a branching ratio of $65\%$. This process contributes mainly at $\log_{10}(x)>-2$.

We show the variation of the contribution of EWBR to the $\gamma$-ray spectrum with respect to the total spectrum as a function of DM mass in Fig. \ref{fig:frac}. For the leptonic channels, FSR dominates the spectrum for the $\mu$ and $e$ channels, while its contribution decreases to $20-30\%$ at 100 TeV. In the case of the $\tau$ channel, on the other hand, the FSR yield remains roughly constant with the DM mass, with a contribution between $15\%$ and $25\%$. In contrast, the EWBR contribution becomes more important with increasing DM mass, reaching between $60\%$ and $80\%$ for $M_\chi =100$ TeV.

\section{Main novelties of our analysis}
\label{sec:imp}

In this section, we elaborate on the novelties of our model and assumptions with respect to the state-of-the-art literature.

\subsection{Polarization and off-shell contribution}
\label{sec:polar}

The method used in \pppc and \qcdunc to generate the DM spectra is based on \Pythia and the \emph{resonance approach}. 
This method assumes that a spinless resonance with a center-of-mass energy of twice the DM mass is produced and subsequently decays into the specified final state (\ie~annihilation) channel.\footnote{An example of this method is present in the example {\tt main07.cc} of the \Pythia code.} 
For example, if the annihilation channel is $b\bar{b}$, \Pythia simulates an $e^+e^-$ collision of energy $E_{\rm beam} = M_{\chi}$, which generates a resonance ${\cal R}$ with energy of $2 M_{\chi}$. Then, $\mathcal{R}$ decays into $b\bar{b}$ with an isotropic distribution emission with respect to the polar angle, \ie~the angle of the incoming $e^{\pm}$. After being produced, the pair of the SM particles, \ie~$b\bar{b}$, undergo the QED and QCD shower that produce the final particles.

The resonance approach implemented in \Pythia and used in \pppc and \qcdunc does not take into account some important aspects of the particle physics theory of DM annihilation and particle production:
\begin{itemize}
\item {\bf Off shell contribution:} The contribution for the annihilation into a pair of massive gauge bosons, $VV$ ($V= {W^{\pm}, Z}$), is zero if the DM mass is smaller than the boson's pole mass, as the resonance is restricted to decay into a pair of on-shell particles only. 
However, gauge bosons can be produced off-shell (see, \eg,~\cite{Cuoco:2016jqt}) \ie~with an effective mass different from its pole mass, since their decay is modeled with a Breit-Wigner function with a width $\Gamma$ that is 2.09  (2.50) GeV for W$^{\pm} (Z)$ (see \cite{Erler:2019hds} for a recent review).

\item {\bf Polarization of the gauge bosons:} The resonance approach does not take into account the spin of the massive gauge bosons produced by the DM annihilation. Instead, the method works as if the annihilation channel is through the Higgs boson, which is a scalar. Massive gauge bosons are spin-1 particles, and they carry three possible polarizations (two transverse states ($\pm 1$) and one longitudinal state). The boson polarization affects the kinematics (angular and momentum distribution) of the fermions produced after their decay \cite{Stirling:2012zt}. This effect cannot be taken into account with the generic resonance from which all the particles produced are unpolarized.
\end{itemize}

The above two problems can be solved by using an external code that calculates the matrix elements in a given BSM model. In this way, the spin of the particles is included in the hard process and the kinematics of the particles are generated according to the polarization of the bosons and their off-shell contribution.
We use \mg~\cite{Alwall:2014hca} and its wrapper \maddm~\cite{Backovic:2013dpa,Backovic:2015cra,Ambrogi:2018jqj}. 
Specifically, we use the LTS version of \mg (currently 2.9.16) while for \maddm we use a custom version that incorporates \vincia and our new tuning. This version of \maddm is not yet public, but will be made available to the reader upon request. In Appendix \ref{app:spectrum}, we describe the \maddm commands we use to generate the spectra.

For annihilation into off-shell vector bosons, we generate the full four-fermion process  with \maddm. Here, we consider annihilation via a $s$-channel scalar mediator, specifically the SM Higgs, $\chi \chi \rightarrow H \rightarrow VV$. To this end, we employ the Singlet Scalar model with a Higgs portal (see, \eg, \cite{DiMauro:2023tho}). 
We generate the off-shell process by including the $V$ decays in the hard process:
\begin{equation}
\label{eq:4B}
\chi \chi \rightarrow H \rightarrow VV \rightarrow 4f ,
\end{equation}
where $4f$ indicates the production of 4 fermions from the $V$ decay. We will call the diagrams with four fermions in the final state of the hard process the four-body diagram.
In this process, the gauge bosons are virtual particles. Therefore, they can have a mass below the pole, meaning that the channel is kinematically open also for $M_{\chi}<M_V$. We consider the four-body process up to 100 GeV. 
For higher masses, off-shell effects are fully negligible.
In fact, for a mass of 100 GeV, we have checked that the spectra of the four-body processes agree with the ones of the on-shell production processes, $\chi \chi \rightarrow VV $, within Monte Carlo uncertainties.
The method of using the four-body diagrams calculated by \maddm together with the \vincia shower algorithm provides a consistent framework to take into account the off-shell contribution and the helicity information of the massive bosons. 

\begin{figure}
\centering
\includegraphics[width=0.65\linewidth]{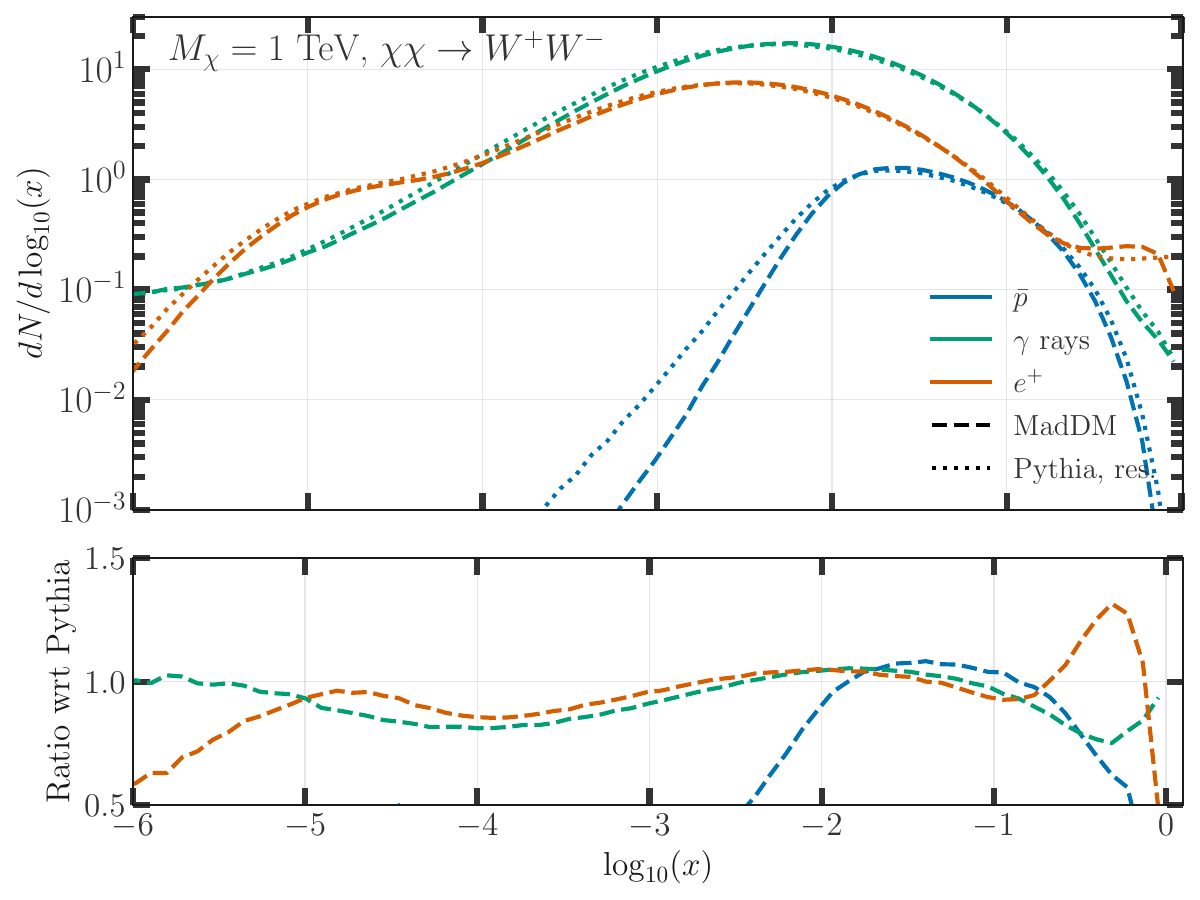}
\caption{\label{fig:pol}
Comparison between the spectra generated with the \Pythia resonance method (``\Pythia, res", dotted curves) and with \maddm using the Singlet scalar model (dashed curves) for the $W^+W^-$ channel. We show the spectra of $\gamma$ rays (green), $\bar{p}$ (blue) and positrons (red).}
\end{figure}

We now turn to the discussion of polarization of massive gauge bosons.  
As we use \maddm for the generation of the hard process, the spin information of all final state particles are stored in the Les Houches Event File ({\tt LHEF}) \cite{Alwall:2007mw}.
Then the {\tt LHEF} is passed onto \Pythia which produces the showering taking into account the polarization information of the massive bosons.
In Fig.~\ref{fig:pol}, we show the effect of the polarization for a DM mass of 1~TeV\@. We consider the $W^+W^-$ annihilation channel and  calculate the spectrum for the production of $\gamma$ rays, $\bar{p}$ and $e^+$.
The figure reveals that the largest differences in the spectra occur in the regions below and above the peak of the distribution. This is true for all particles messengers. For $\gamma$ rays and $e^+$ spectra the difference can be up to $20$--$30\%$, while for $\bar{p}$ the deviation can also reach $50\%$ but at the tails of the distribution.
At the peak of the spectra, the results obtained with and without taking boson polarisation into account are very similar in the range of a few $\%$.
The effect of the polarization of the gauge bosons becomes more and more relevant the larger the DM mass is.

The difference in the spectra between the case with and without polarization information is due to the different kinematics of the quarks and leptons produced in the decay of the $W^{\pm}$ in the two cases.
To check this, we produced the spectra of the quarks for the four-body diagram with the Higgs bosons channel $\chi \chi \rightarrow H H \rightarrow 4f$ (named as 4-body, $HH$), the two-body case for the $W$ channel $\chi \chi \rightarrow W^+ W^-$ and with the default spin-0 resonance method of \Pythia (spin0, res). Again, we exemplarily consider a DM mass of 1 TeV. All three methods lead to the same result for the quark spectra: the energy distribution is flat, \ie~the probability of producing a quark with any energy between 0 and the DM mass is the same. In all three cases, the polarization of the generated boson is not taken into account.
Instead, in the four-body process with the $W$ channel, there is a peak at about half the DM mass and the distribution is not uniform.
This effect is due to the $V-A$ structure of the charge and neutral current interactions in the SM, which generates asymmetries, called the forward-backward and left-right asymmetries. The forward-backward asymmetry has been studied in $e^+e^-$ collisions at the $Z$ resonance, which subsequently decays into a pair of fermions (see, \eg, \cite{DELPHI:2000wje}). The fermions are not produced symmetrically with respect to the polar angle, which is the angle relative to the electron beam. This asymmetry is typically 0 for $\sqrt{s}=m_Z$. The left-right asymmetry instead requires polarized beams and is related to the asymmetry in the cross-section for the production of fermions with the two chiralities.\footnote{In some studies related to the top quark, it was found that the energy spectra of the charged leptons was strongly correlated to their angular distributions. The latter is a direct probe of the spin of the top quark (see for example Refs. \cite{PrasathV:2014omf,Jueid:2018wnj,Arhrib:2018bxc} for more details).}

Finally, we want to mention that for large DM masses, higher order corrections in the hard process can become significant. When including these contributions, a matching scheme between the radiation from the fixed-order calculation and showering algorithm has to be employed. Note that radiation in the hard process is model-dependent as, in addition to FSR, it can involve IB and ISR diagrams. In general, in a fixed-order calculation the inclusion of all diagrams of a given order in perturbation theory is vital to ensure gauge invariance and, hence, physically meaningful results. In this work, we refrain from including higher-order corrections in the hard process, to support applicability of the generated spectra to a wide range of models. Note that for intermediate mass scales, the effect on the resulting spectra is small.  Considering, for instance, the Singlet Scalar model and a DM mass of 1 TeV, the spectra from the four-body processes $\chi \chi \rightarrow VV \rightarrow 4f, \; \chi \chi\rightarrow VV \rightarrow2f\, 2V, \; \chi \chi\rightarrow VV \rightarrow4V$ and the two-body process $\chi \chi \rightarrow  V V$ (with showering performed by \vincia) agree within less than $5\%$ in the relevant energy range.
\subsection{Effect due to the choice of the BSM model}

The resonance method implemented in \Pythia is a widely used method in the literature that accounts for the production of a pair of SM particles from the decay of a spin-0 resonance. Other physical assumptions leading to the annihilation channel and associated with a specific BSM model could produce important differences in the final DM spectra. In particular, the choice of a DM model plays an important role in the annihilation channels with very relevant EW corrections. For example, DM simplified models~\cite{Abdallah:2015ter} have couplings between new mediators and fermions that are similar to the $V-A$ structure of the EW processes in the SM\@. The exact values of the $V-A$ coupling parameters $g^{V}$ and $g^{A}$ can produce a preference for the production of certain helicity states that have different particle spectra \cite{Cirelli:2010xx}.

As an example, we consider a DM simplified model with a mediator $Y_1$ of spin-1 and a Dirac fermion $\chi$ as DM particle.
The additional Lagrangian added to the SM one is given by:
\begin{equation}
\label{eq:DMsimp1}
\bar{\chi} \gamma_{\mu} (g_{\chi}^V + \gamma_5 g_{\chi}^A) \chi Y_1^{\mu} +  
\sum^{N_f}_i  \bar{F}_i \gamma_{\mu} (g^{V}_{ij} + \gamma_5 g^{A}_{ij}) F_j Y^{\mu}_1,
\end{equation}
where $F_i$ is the fermionic multiplet and $g^{V}_{ij}$ and $g^{A}_{ij}$ are the vector and axial couplings between $F_i$ and $Y^{\mu}_1$, which resemble the $V-A$ structure of the neutral current in the SM\@.
Therefore, there are coupling parameters $g^{V}_{ij}$ and $g^{A}_{ij}$ for each of the fermionic states written above.
Typically, these models are taken with only the vector structure, \ie~by assuming that $g^A=0$ \cite{Arina:2018zcq} (see for caveats \eg~here~\cite{Kahlhoefer:2015bea,Englert:2016joy,Ellis:2017tkh}). 
This implies that the annihilation of DM particles produce both left-handed and right-handed fermions, \ie~$\sum^{N_F}_i (g^{V}_{L,ij} \bar{F}_{L,i} \gamma_{\mu} F_{L,j} +
g^{V}_{R,ij} \bar{F}_{R,i} \gamma_{\mu} F_{R,j} ) Y^{\mu}_1$, where $F_{L,i}$ and $F_{R,i}$ are the left-handed and right-handed components of the fermions.

However, this can change drastically if different assumptions are made for the values of $g^A$ and $g^V$.
Lets for example assume that $g^A = -g^V$. In this case the Lagrangian associated to the production of fermions is given by:
\begin{equation}
\sum^{N_f}_i g^{V}_{ij} \bar{F}_i \gamma_{\mu} (1 - \gamma_5 ) F_j Y^{\mu}_1 = \sum^{N_f}_i  g^{V}_{ij} \bar{F}_{L,i} \gamma_{\mu} F_{L,j} Y^{\mu}_1.
\end{equation}
This implies that, as in SM charged current interactions between the $W$ boson and the fermions, the mediator $Y_1$ couples only to the left-handed fermions and there is no coupling with the right-handed fermions.
The model thus produces completely polarized fermions which have left-handed chirality and negative helicity. The opposite applies for antifermions.
Instead, if $g^A = g^V$ the mediator $Y_1$ couples only to right-handed fermions and there is no coupling with left-handed fermions.
The model thus produces completely polarized fermions, which have right-handed chirality and positive helicity. The opposite applies for antifermions.

Instead, in the case of scalar and pseudoscalar couplings between the mediator and SM fermions, the messenger particle spectra do not change when different values of the couplings are considered.
This can be demonstrated considering the DM simplified model with a mediator $Y_0$ of spin-0 and a Dirac fermion $\chi$ as DM particle. This model is described by the following interaction term in the Lagrangian:
\begin{equation}
\label{eq:DMsimp0}
\bar{\chi} (g_{\chi}^S + i \gamma_5 g_{\chi}^{P}) \chi Y_0 +  
\sum^{N_f}_i  \bar{F}_i (g^{S}_{ij} + i \gamma_5 g^{P}_{ij}) F_j Y_0,
\end{equation}
where $g^{S}_{ij}$ and $g^{P}_{ij}$ are the scalar and pseudoscalar couplings, respectively, between $F_i$ and $Y_0$. Regardless the values of the scalar and pseudoscalar couplings ($g^{S}_{ij}=0$ or $g^{P}_{ij}=0$ or $g^{P}_{ij}=g^{S}_{ij}$ or $g^{P}_{ij}=-g^{S}_{ij}$), this model does not select any specific chirality of the fermions.

Note that the same conclusions hold for decaying DM considering its interaction with SM fermions to be described by the second terms in  Eqs.~\eqref{eq:DMsimp1} and~\eqref{eq:DMsimp0}, respectively, for spin-1 and spin-0 DM\@. In this case the mediator is replaced by the respective DM particle.

As explained in Sec.~\ref{sec:composition}, the EW corrections with the emission of a $W^{\pm}$ gauge boson can be a very important production mechanisms in DM spectra.
When we consider models with mediators of spin 0 or mediators spin 1 with $g^V$ or $g^A=0$, the fermions produced from the mediator contains both left and right helicities with the same probability. Therefore, the production of $W^{\pm}$ and $Z$ is turned on.
When, $g^A=g^V$ the mediator produces only right-handed fermions which do not produce $W^{\pm}$ for EWBR, instead $Z$ for EWBR are still produced with the same rate as in the previous case.
Finally, when $g^A=-g^V$ the mediator produces only left-handed fermions which produce both $W^{\pm}$ and $Z$ for EWBR\@. Since in this case fermions have only the left-handed helicity the production of $W^{\pm}$ is enhanced by a factor of two with respect to the case with spin 0, $g^V$ or $g^A=0$, which contain both helicity states.

\begin{figure}[!t]
\centering
\includegraphics[width=0.49\linewidth]{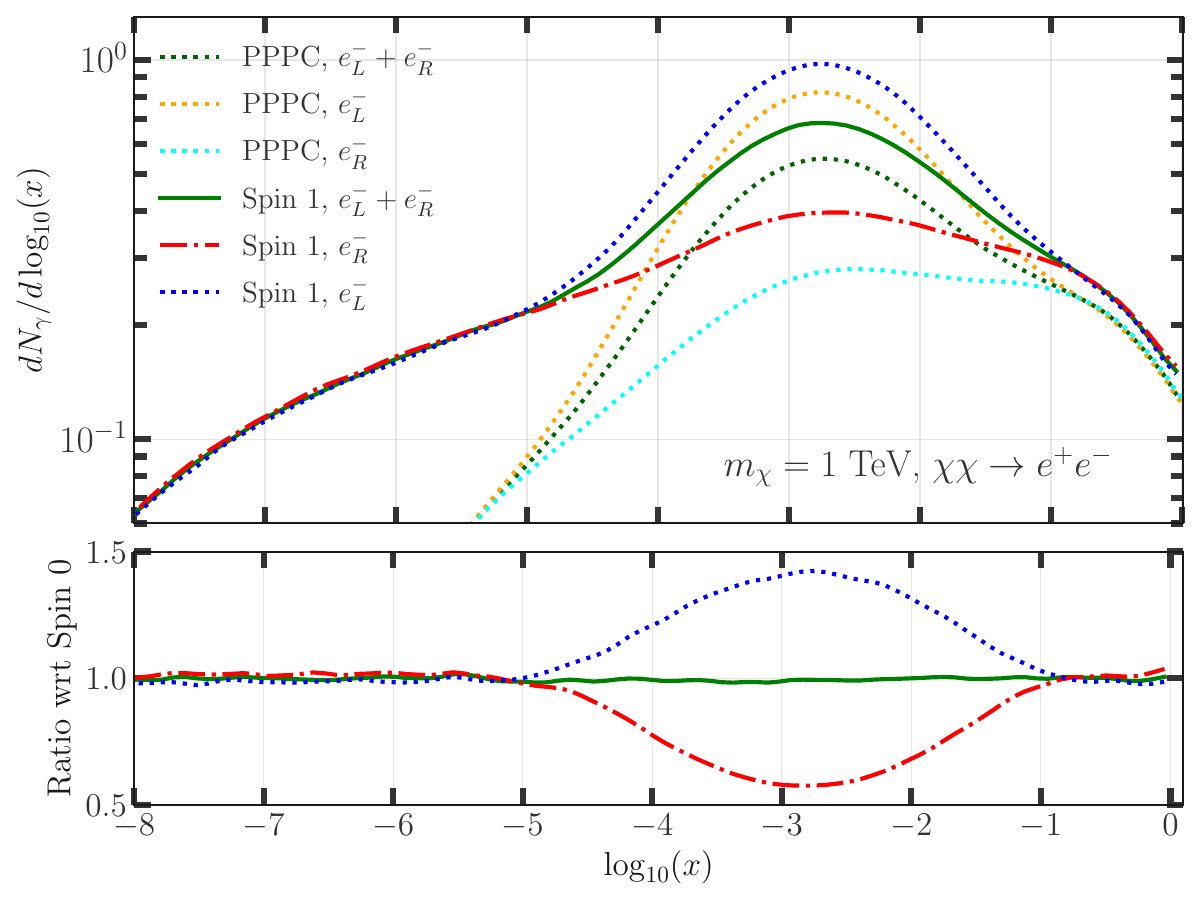}
\includegraphics[width=0.49\linewidth]{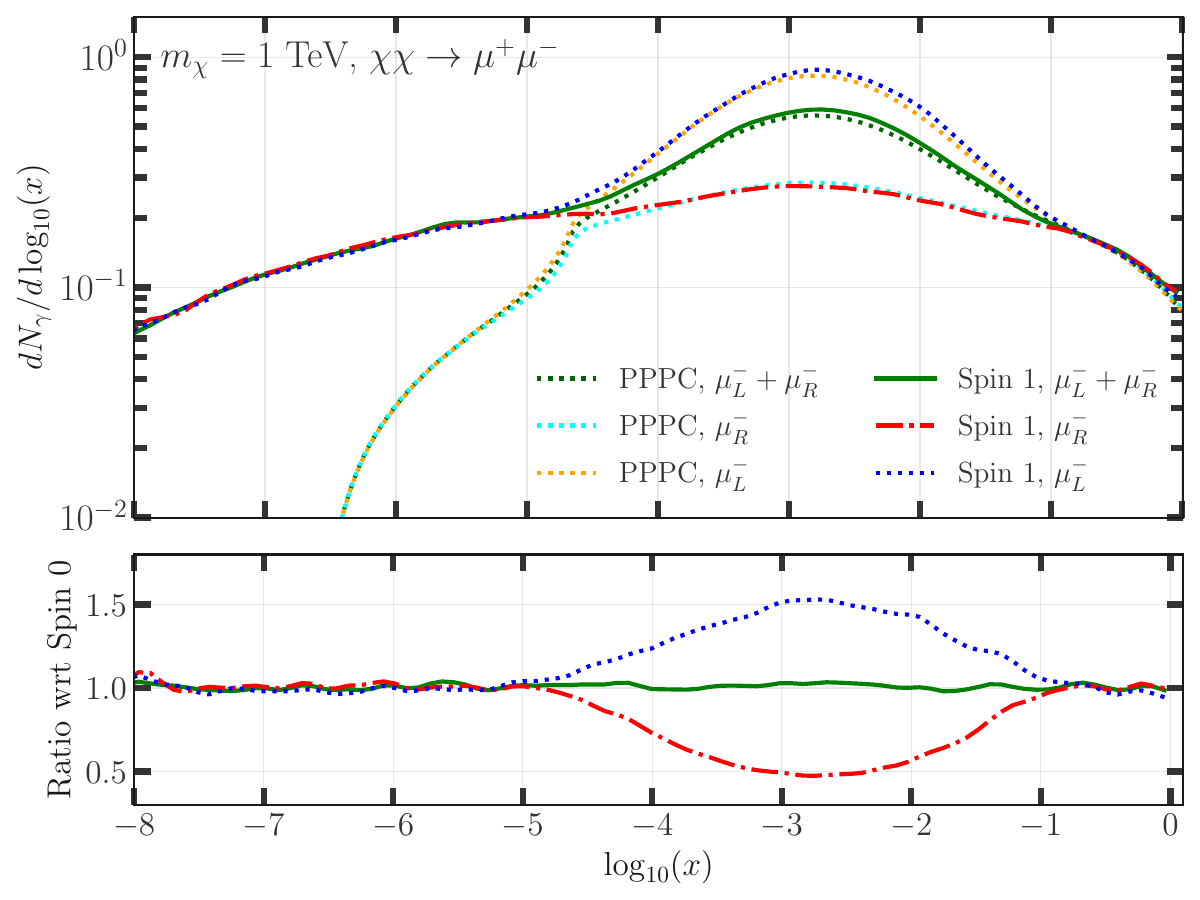}
\includegraphics[width=0.49\linewidth]{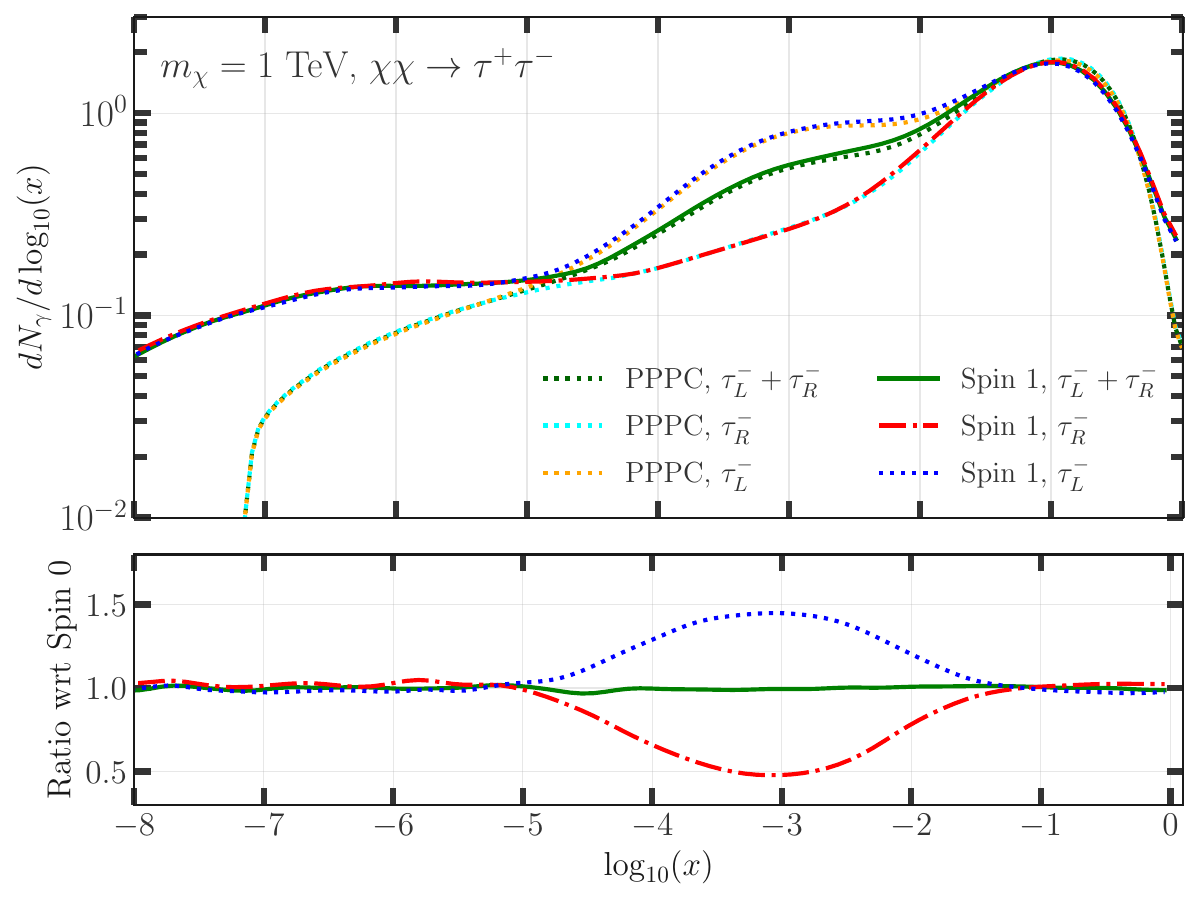}
\includegraphics[width=0.49\linewidth]{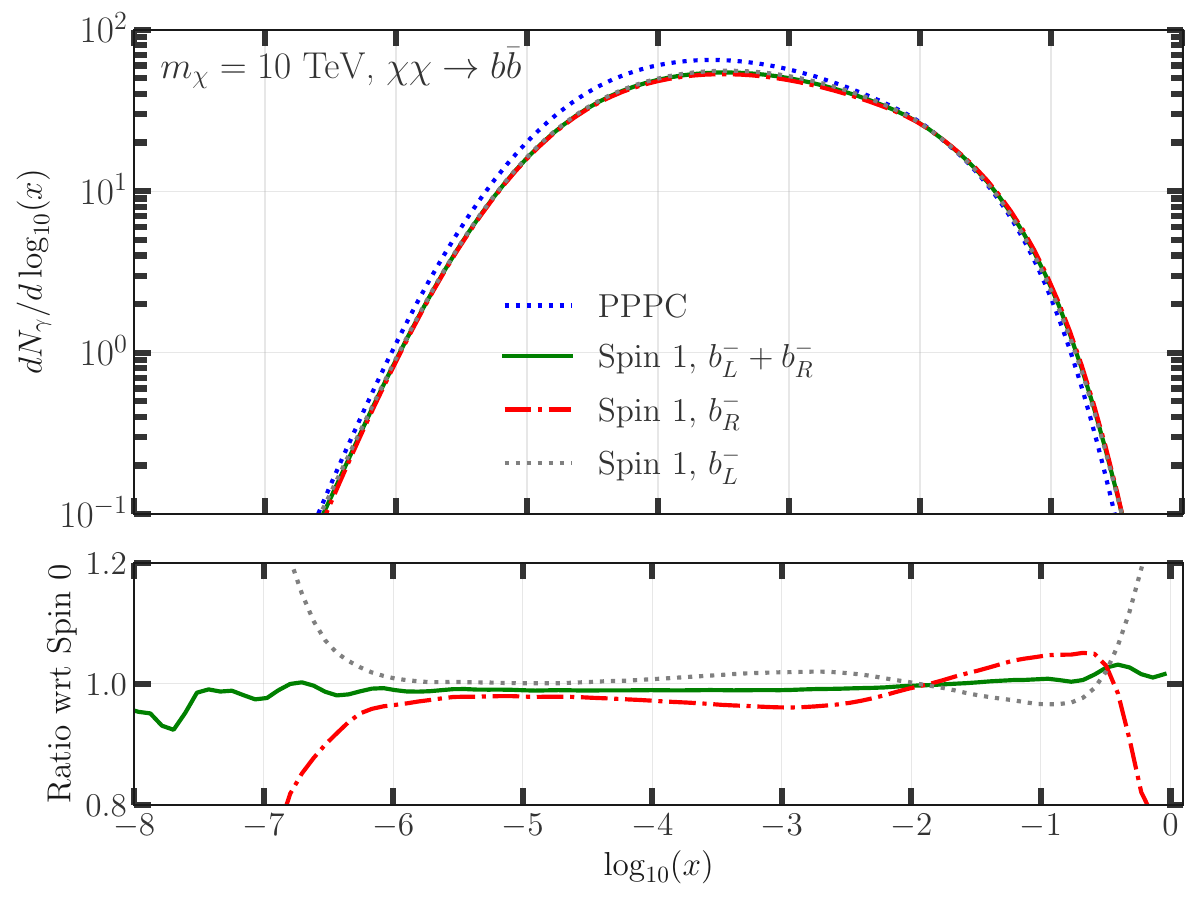}
\caption{\label{fig:spectrahel}
Effect of the mediator spin and coupling parameters when considering the DM simplified model in the spectra of $\gamma$ rays. We show the results obtained for the $e^-e^-$, $\mu^+\mu^-$, $\tau^+\tau^-$ and $b\bar{b}$ annihilation channels (from top left to bottom right). In each figure we report the results obtained when using $g^A=0$ which correspond to the case for which we have both left and right-handed helicity states (Spin 1, $F_L+F_R$), $g^A=g^V$ ($g^A=-g^V$), which corresponds to the case where only right-handed (left-handed) fermions are produced, Spin 1, $F_R$ (Spin 1, $F_L$ ). Results obtained with \pppc are shown for comparison.}
\end{figure}

In Fig.~\ref{fig:spectrahel}, we show the difference between the DM spectra obtained for different choices of coupling parameters for the $e^+e^-$, $\mu^+\mu^-$, $\tau^+\tau^-$ and $b\bar{b}$ annihilation channels.
We tested the following cases
\begin{itemize}
\item SHP model, spectra for both helicity states (labeled as Spin 0).
\item DMsimp model spin 1 ($g^V=0$ or $g^A=0$), spectra for both helicity states (Spin 1, $F_L+F_R$).
\item DMsimp model spin 1 ($g^V=-g^A$), spectra for left-handed fermions (Spin 1, $F_L$).
\item DMsimp model spin 1 ($g^V=g^A$), spectra for right-handed fermions (Spin 1, $F_R$).
\end{itemize}

As expected, the results obtained with the spin 0 case are the same as the spin 1 case, $F_L+F_R$, , within the statistical errors. Therefore, in Fig.~\ref{fig:spectrahel} we decide not to show the Spin 0 spectra.
In case of the $b\bar{b}$ channel, and for all hadronic channels, the differences between the tested cases are at most at the level of $5-10\%$. This is due to the fact that the main production process of particles is hadronization generated from pairs of quarks or gauge bosons produced from DM annihilation (see, Sec.~\ref{sec:composition}). Instead, the EWBR for hadronic channels contribute, through a secondary hadronization process, at most with $10\%$ of the total yield for $\gamma$ rays (see, Fig.~\ref{fig:frac}).

For leptonic channels, the EWBR can provide a very important contribution and thus the choice of the coupling parameters of the BSM model can have a much larger effect with respect to what found for the hadronic channels.
In case of leptonic channels, for $g^A=g^V$, which selects only right-handed fermions, is the one with the lowest spectra at the peak. This is due to the fact that the mediator produces only right-handed fermions, which cannot couple with $W^{\pm}$. Therefore, the EWBR with $W^{\pm}$ is not present and does not contribute to the spectrum, giving thus a reduced yield of final particles. Instead, the case with Spin 0 or Spin 1 with either $g^A=0$ or $g^V=0$ contains fermions both right and left-handed. Therefore, when the left-handed fermions are produced, which happens on average for half of the events, the coupling with $W^{\pm}$ is present and the EWBR with these bosons contribute to the spectrum.
Finally, when $g^A=-g^V$ only left-handed fermions are produced, so the production of $W^{\pm}$ is twice as large as the Spin 0 case and the spectrum is much larger at the peak.
Since the variation in the spectra between the tested cases is due to the contribution of EWBR the differences in the spectra are present at values of $\log_{10}(x)=[-5,-1]$ where this process gives the largest contribution (see, Sec.~\ref{sec:composition}).

\pppc provides the spectra for the individual helicity states. The authors have published this result only for leptons since for quarks, the difference is minimal. We compare our results for left and right-handed helicity states spectra with the \pppc spectra in Fig.~\ref{fig:spectrahel}.
We obtain very similar results for the $\mu^+ \mu^-$ and $\tau^+ \tau^-$ at energies $\log_{10}(x)>-5$. Instead, at smaller energies, our results are systematically larger.
We will discuss more extensively this in Sec.~\ref{sec:results}.
Instead, for the $e^{\pm}$ channel the differences are not only at low energies but also at the peak of the distribution where the contribution of the EWBR is the largest.

\subsection{Improved shower algorithm}

The final very important improvement we introduce in our paper is the fact that we use the \vincia shower algorithm and a specific tuning of the model parameters.
We already discussed this in Sec.~\ref{sec:had} so here we will report the key points.
The first advantage of using \vincia is that it takes into account the helicity of the particles during all the shower algorithm. The standard \Pythia shower algorithm does not take into account the helicity, and thus it misses important effects as the ones we showed in the previous section, which are due to the $V-A$ structure of the SM EW interactions. 
The second improvement is related to the fact that \vincia takes into account the trilinear diagrams of the massive bosons, which starts to be important at DM masses above 1 TeV. These diagrams are not taken into account in the standard \Pythia shower algorithm.
The third important new aspect is that we perform a tuning of the main \vincia model parameter in order to match the LEP data for the production of particles at the $Z$ boson resonance, which we expect to be similar with the DM annihilation process.
This is a very significant improvement with respect to, \pppc which was generated with \Pythia version 8.135. This version of \Pythia, using the default parameters, did not have a specific tuning to collider data.
Using the latest \Pythia version has remarkable effects, as we will discuss later in the paper. The most important one is for the photons produced for QED FSR, which was highly underestimated in \pppc at low energies.

\section{Tuning of the hadronization model}
\label{sec:tuning}

In this section, we discuss the tuning of the Lund hadronization model parameters to a set of measurements performed by the experimental collaborations of LEP at the $Z$--boson pole. We first discuss the technical setup we have adopted in the fits and then present our results.

\begin{table}[!tbp]
\setlength\tabcolsep{10pt}
  \begin{center}
    \begin{tabular}{lcccc}
\noalign{\hrule height 1pt}
      Parameter & \textsc{Monash} & \vincia~(default) & \Pythia~ \cite{Jueid:2022qjg,Jueid:2023vrb} & This work \\
\noalign{\hrule height 1pt}
      $a_L$ &  $0.68$ & $0.45$ & $0.601$ & $0.337\pm 0.015$  \\
      $b_L$ & $0.98$ & $0.80$ & $0.897$ & $0.784\pm 0.020$ \\
      $\sigma_\perp~({\rm GeV})$  & $0.335$ & $0.305$ & $0.307$ & $0.296 \pm 0.003$  \\
      $a_{QQ}$ & $0.97$ & $0.90$ & $1.671$ & $1.246 \pm 0.082$ \\ [0.8ex]
\noalign{\hrule height 1pt}
      $\chi^2/N_{\rm df}$ & $1034.52/852$ & $786.11/852$ & $676.69/852$ & $660.21/852$ \\
\noalign{\hrule height 1pt}
      \end{tabular}
  \end{center}
  \caption{\label{tab:tunes:results} Tuning results of the parameters of the hadronization model in \Pythia~8 using the \vincia shower plugin. For comparison, we show the results in the baseline \textsc{Monash} tuning in both \Pythia and \vincia as well as the results of Refs. \cite{Jueid:2022qjg, Jueid:2023vrb}.}
\end{table}

\begin{figure}[!t]
\centering
\includegraphics[width=0.73\linewidth]{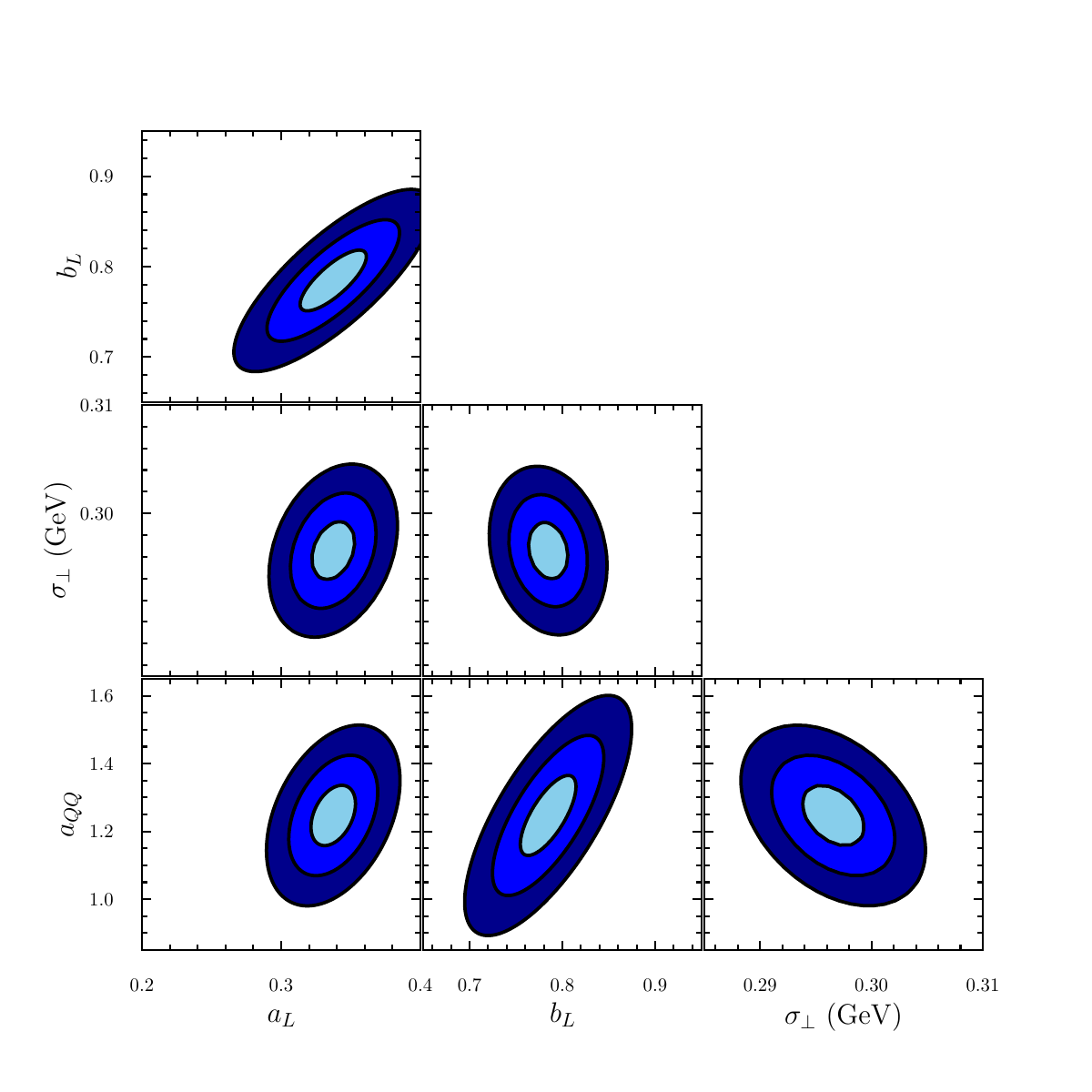}
\caption{Tuning results projected on the different parameters and using the same measurements as in Refs. \cite{Jueid:2022qjg, Jueid:2023vrb}. $68\%$, $95\%$ and $99.5\%$ confidence level contours are shown in turquoise, blue, and dark blue respectively.}
\label{fig:results:tunes}
\end{figure}

\subsection{Technical setup}

We use \Pythia version 8.309 \cite{Bierlich:2022pfr} to generate samples at the particle level and with \vincia shower plugin as the default. The \textsc{Monash} tuning is used as our baseline for further tuning \cite{Skands:2014pea}. We use \textsc{Rivet} version 3.1.7 for the implementation of the different measurements at the particle level \cite{Bierlich:2019rhm}. \textsc{Professor} version 2.3.3 is used to perform the tuning of the parameters \cite{Buckley:2009bj}. In this work, we only tune four parameters which are $a_L$, $b_L$, $a_{\rm QQ}$ and $\sigma_\perp$. The default values of these parameters in \vincia along with their allowed ranges are given in Tab.  \ref{tab:ranges}. The \textsc{Professor} toolkit uses analytical expressions that model the physical dependence of the observables on the different parameters. This dependence is derived by fitting the Monte Carlo predictions to a set of points in the four-dimensional parameter space. Then the best-fit points are derived by a standard $\chi^2$ minimisation using \textsc{Minuit} \cite{James:1975dr} and which is implemented in \textsc{Professor} as the default option. To assess the quality of the tuning we estimate the goodness-of-fit (GoF) defined as 
\begin{equation}
 \chi^2 = \sum_{\mathcal{O}} \sum_{b\in \mathcal{O}} \bigg(\frac{f_{(b)}(\{p_i\}) - {\rm Data}_{(b)}}{\Delta_b}\bigg)^2,
\label{eq:GoF}
\end{equation}
where ${\rm Data}_{(b)}$ is the experimental value of the observable ${\cal O}$ at a bin $b$, $f_{(b)}(\{p_k\})$ is the value of the analytical expression of the response function that models the theory prediction and is cast as a polynomial of the parameters (defined in equation \ref{eq:interp}) $\{p_k\}\equiv \{a_L, b_L, a_{\rm QQ}, \sigma_\perp\}$ and $\Delta_b$ is the total error on the observable ${\cal O}$ at a bin $b$. There are three types of errors on the observable ${\cal O}$ at bin $b$: experimental errors on the measurements, MC errors due to the limitation of the size of the MC samples and the theory errors. Given that we have simulated 2 million events for each point in the parameter space, we find that the MC errors are the smallest contribution to the error budget. In our analysis, we do not assume any correlation between the different errors, as this information was not provided by the experimental collaborations. On the other hand, following Ref. \cite{Skands:2014pea} we add a flat $5\%$ uncertainty for each bin so that we can avoid overfitting effects and as a sanity limit for the accuracy in theory predictions for both the perturbative and the non-perturbative effects. The total error is thus given by
\begin{eqnarray}
\Delta_b = \sqrt{\sigma_{b, \rm exp}^2 + \sigma_{b, {\rm MC}}^2 + [0.05 \times f_{(b)}(\{p_i\})]^2},
\end{eqnarray}

The number of degrees-of-freedom ($N_{\rm df}$) is defined as the total number of bins minus the number of independent parameters
\begin{eqnarray}
N_{\rm df} = \sum_{\mathcal{O}} |b \in \mathcal{O}| - N_{\rm params}.
\label{eq:NDF}
\end{eqnarray}
A good fit implies that $\chi^2/N_{\rm df}$ per number of degrees-of-freedom to be $\approx 1$. 
The polynomial dependence of the true MC response is cast as a fourth-order polynomial:

\begin{eqnarray}
    f_{(b)}(\{p_i\}) = \alpha_0^{(b)} + \sum_{i=1}^4 \beta_i^{(b)} p_i + \sum_{i,j=1}^4 \gamma_{ij}^{(b)} p_i p_j + \sum_{i,j,k = 1}^4 \delta_{ijk}^{(b)} p_i p_j p_k + \sum_{i,j,k,\ell=1}^4 \epsilon_{ijk\ell}^{(b)} p_i p_j p_k p_\ell,
    \label{eq:interp}
\end{eqnarray}
with $\alpha, \beta, \gamma,~{\rm and}~ \delta$ being the polynomial coefficients determined in the fit and $\{p_i\}$ are the parameters of the  Lund hadronization model. To compute the response function, we have randomly generated 500 MC samples that correspond to the four-dimensional parameter space. The order of the polynomial function plays a crucial role in both the quality of the fits and the consistency of the interpolated results with the true MC response at the minimum of the model parameters. Performance of the interpolation procedure is determined through the estimate of the residuals, which are defined as 
\begin{eqnarray}
    {\rm Residuals} \equiv \sum_i \frac{f_i(\{p_k\})  - {\rm MC}_i}{{\rm MC}_i},
    \label{eq:residuals}
\end{eqnarray}
where the sum runs over all the bins and for all the measurements and ${\rm MC}_i$ represents the true MC response. We have checked that a four-order polynomial response function is good enough to model the true response, since that $95\%$ of the residual distribution is within $0.02$.

\begin{table}[!tbp]
\setlength\tabcolsep{3pt}
\begin{center}
\begin{tabular}{lcc !{\vrule width 1pt} lcc}
\noalign{\hrule height 1pt}
 Measurement    & Experiment &  $\chi^2/N_{\rm bins}$  &  Measurement    & Experiment &  $\chi^2/N_{\rm bins}$ \\
\noalign{\hrule height 1pt}
 $1 - T$    &  \textsc{Aleph} \cite{ALEPH:1996oqp} & 0.13   &   $C$--parameter & \textsc{Aleph} \cite{ALEPH:1996oqp} & 0.39 \\  
 $\log(1/x_p)$ & \textsc{Aleph} \cite{ALEPH:1996oqp} & 0.19 & $\langle N_{\rm ch} \rangle$ & \textsc{Aleph} \cite{ALEPH:1996oqp} & 0.028 \\
 $\langle N_{\rm ch} \rangle~(|Y| < 0.5)$ & \textsc{Aleph} \cite{ALEPH:1996oqp} & 0.012 & $\langle N_{\rm ch} \rangle~(|Y| < 1.0)$ & \textsc{Aleph} \cite{ALEPH:1996oqp} & 0.028 \\
 $\langle N_{\rm ch} \rangle~(|Y| < 1.5)$ & \textsc{Aleph} \cite{ALEPH:1996oqp} & 0.030  &  $\langle N_{\rm ch} \rangle~(|Y| < 2.0)$ & \textsc{Aleph} \cite{ALEPH:1996oqp} & 0.040 \\ 
 $\pi^\pm$ spectrum  & \textsc{Aleph} \cite{ALEPH:1996oqp}  & 0.67 & $\pi^0$ spectrum  & \textsc{Aleph} \cite{ALEPH:1996oqp}  & 0.24 \\ 
 $\Lambda^0$ spectrum & \textsc{Aleph} \cite{ALEPH:1999udi} & 1.24  & $\Lambda^0$ spectrum ($2$--jet events) & \textsc{Aleph} \cite{ALEPH:1999udi} & 1.31 \\ 
 Thrust  & \textsc{Aleph} \cite{ALEPH:2003obs} & 0.097 & $C$--parameter &  \textsc{Aleph} \cite{ALEPH:2003obs} & 0.35 \\ 
 \noalign{\hrule height 1pt}
 $N_{\rm ch}$~($y_{\rm cut} = 0.01$) & \textsc{Delphi} \cite{DELPHI:1992sar} & 5.99 &  $N_{\rm ch}$~($y_{\rm cut} = 0.02$) & \textsc{Delphi} \cite{DELPHI:1992sar} & 4.88 \\ 
 $\Lambda^0$ spectrum & \textsc{Delphi} \cite{DELPHI:1993vpj} & 1.34 & $\langle N_{\Lambda^0} \rangle$ & \textsc{Delphi} \cite{DELPHI:1993vpj} & 0.53 \\
 $\pi^0$ momentum & \textsc{Delphi} \cite{DELPHI:1995kfu} & 0.41 & $\log(1/x_p)$ & \textsc{Delphi} \cite{DELPHI:1995kfu} & 0.33 \\ 
 $1 - T$ & \textsc{Delphi} \cite{DELPHI:1995kfu} & 0.18 & $C$--parameter & \textsc{Delphi} \cite{DELPHI:1995kfu} & 0.34 \\ 
 $\langle N_{\rm ch} \rangle$ & \textsc{Delphi} \cite{DELPHI:1995kfu} & 0.031 &  $\langle N_{\pi^\pm} \rangle$ & \textsc{Delphi} \cite{DELPHI:1995kfu} & 0.063 \\ 
 $\langle N_{\pi^0} \rangle$ & \textsc{Delphi} \cite{DELPHI:1995kfu} & 0.39 & $\langle N_{\rho} \rangle$ & \textsc{Delphi} \cite{DELPHI:1995kfu} & 3.40 \\ 
  $\langle N_{p} \rangle$ & \textsc{Delphi} \cite{DELPHI:1995kfu} & 2.30 &   $\langle N_{\Lambda^0} \rangle$ & \textsc{Delphi} \cite{DELPHI:1995kfu} & 1.54 \\  
  $\langle N_{\rm ch} \rangle$ & \textsc{Delphi} \cite{DELPHI:1998cgx} & 0.005 & $\langle N_{\rm \pi^\pm} \rangle$ & \textsc{Delphi} \cite{DELPHI:1998cgx} & 0.10 \\ 
  $\langle N_{p} \rangle$ & \textsc{Delphi} \cite{DELPHI:1998cgx} & 0.05 & $N_{p/\bar{p}}/N_{\rm ch}$ & \textsc{Delphi} \cite{DELPHI:1998cgx} & 0.27 \\
  $\pi^\pm$ momentum & \textsc{Delphi} \cite{DELPHI:1998cgx} & 0.46 & $p/\bar{p}$ momentum & \textsc{Delphi} \cite{DELPHI:1998cgx} & 0.43 \\ 
  \noalign{\hrule height 1pt}
  Thrust (udsc events) & \textsc{L3} \cite{L3:2004cdh} & 0.34 & $C$--parameter (udsc events) & \textsc{L3} \cite{L3:2004cdh} & 0.22 \\ 
  Charged multiplicity & \textsc{L3} \cite{L3:2004cdh} & 3.39 & $\log(1/x_p)$ & \textsc{L3} \cite{L3:2004cdh} & 0.96 \\ 
  $x_p$ (udsc events) & \textsc{L3} \cite{L3:2004cdh} & 0.78 \\ 
  \noalign{\hrule height 1pt}
  $\langle N_{\rm ch} \rangle$ & \textsc{Opal} \cite{OPAL:1991lzj}  & 0.37 & $\pi^\pm$ spectrum & \textsc{Opal} \cite{OPAL:1994zan} & 0.25 \\ 
$\Lambda^0$ scaled energy & \textsc{Opal} \cite{OPAL:1996gsw}  & 1.49 & $\pi^0$ scaled momentum & \textsc{Opal} \cite{OPAL:1998enc} & 0.12 \\ 
  All events $\log(1/x_p)$ & \textsc{Opal} \cite{OPAL:1998arz}  & 0.38 & $\langle N_{\rm ch} \rangle$ & \textsc{Opal} \cite{OPAL:1998arz} & 0.16 \\ 
  $1-T$ & \textsc{Opal} \cite{OPAL:2004wof} & 0.10 & $C$--parameter & \textsc{Opal} \cite{OPAL:2004wof} & 0.35 \\ 
  \noalign{\hrule height 1pt}
\end{tabular}     
\end{center}
\caption{Contributions to the $\chi^2/N_{\rm df}$ per each measurement that was included in the tuning.}
\label{tab:mean:contribution}
\end{table}

\subsection{Results of the tuning}

In this section, we discuss the results of the fits of the hadronization-function parameters. In order to have a good model of hadronization, we not only include the spectra of photons, neutral and charged pions, and baryons but also the measurements for the event shapes (in particular the Thrust and the $C$-parameter), mean identified particle multiplicities, charged multiplicities and the charged momentum distributions. We have included measurements performed by \textsc{Aleph} \cite{ALEPH:1996oqp,ALEPH:1999udi,ALEPH:2003obs}, \textsc{Delphi} \cite{DELPHI:1992sar,DELPHI:1993vpj,DELPHI:1995kfu,DELPHI:1998cgx}, \textsc{L3} \cite{L3:2004cdh} and \textsc{Opal} \cite{OPAL:1991lzj,OPAL:1994zan,OPAL:1996gsw,OPAL:1998enc,OPAL:1998arz,OPAL:2004wof} Collaborations (more details can be found in Appendix A of Ref. \cite{Jueid:2023vrb}). In total, we have included 47 measurements containing 856 bins. To ensure that this fit has a good convergence behavior, we have used a total number of 100 scans for the minimization. The results of the tuning are shown in Tab. \ref{tab:tunes:results}, where we also show the results of the \textsc{Monash} tuning with the \Pythia~8 shower plugin \cite{Skands:2014pea}, with the \vincia  shower plugin (Antenna shower) and with the tuning presented in Refs. \cite{Jueid:2022qjg, Jueid:2023vrb}. 
We also see that the tuning of this work leads to a very good, $\chi^2/N_{\rm df}$ which is slightly better than previous \textsc{Monash} tuning.

We display the $68\%$, $95\%$ and $99\%$ CL contours projected on the full parameter space in Fig. \ref{fig:results:tunes}. The figure shows some degree of correlations between the parameters. For instance, one can see that $a_L$, $b_L$ and $a_{QQ}$  are highly correlated as expected, while $\sigma_\perp$ has a positive correlation with $a_L$ and negative correlation with $b_L$ and $a_{QQ}$. To assess the quality of our tuning we calculate the mean contribution to the total GoF defined as $\chi^2/N_{\rm bins}$ for all the distributions in Tab. \ref{tab:mean:contribution}. We can see that for most of the measurements -- especially those directly connected to the stable particle spectra -- the  model at the best-fit point yields a very good, $\chi^2$ which is of the order 1 or less. \\

\begin{table}[!th]
 \setlength\tabcolsep{14pt}
 \begin{center}
\begin{adjustbox}{max width=\textwidth}
  \centering
    \begin{tabular}{l c c c c}
    \toprule
    \toprule
Tuning & $a_L$ & $b_L$ & $\sigma_\perp~({\rm GeV})$	& $a_{QQ}$ \\
\toprule
Central	& $0.337$	& $0.784$ &	$0.296$	& $1.246$ \\
\toprule
\multicolumn{5}{l}{{$1\sigma$ eigentunes }} \\
\toprule
Variation $1^+$	& $0.345$ &	$0.803$	&  $0.295$ &	$1.345$ \\
Variation $1^-$	& $0.329$ & $0.766$ &  $0.297$ &	$1.149$ \\
Variation $2^+$	& $0.968$ &	$1.444$ &  $0.339$ &	$1.074$ \\
Variation $2^-$	& $0.047$ &	$0.482$ &  $0.277$ &	$1.326$ \\
Variation $3^+$	& $0.207$ &	$0.907$	&  $0.255$ &    $1.233$ \\
Variation $3^-$	& $0.476$ &	$0.652$ &	$0.340$ &	$1.262$ \\
Variation $4^+$	& $0.327$ &	$0.791$ &	$0.346$	 & $1.247$ \\
Variation $4^-$	& $0.346$ &	$0.778$ &	$0.250$ &	$1.246$ \\
\bottomrule
\end{tabular}
\hspace{0.2cm}
\end{adjustbox}
\end{center}
    \caption{Hessian variations corresponding to $1\sigma$, defined as $\Delta \chi^2 /N_{\rm df} = 1$.}
    \label{tab:eigentunes}
\end{table}

\begin{figure}[!t]
    \centering
    \includegraphics[width=0.495\linewidth]{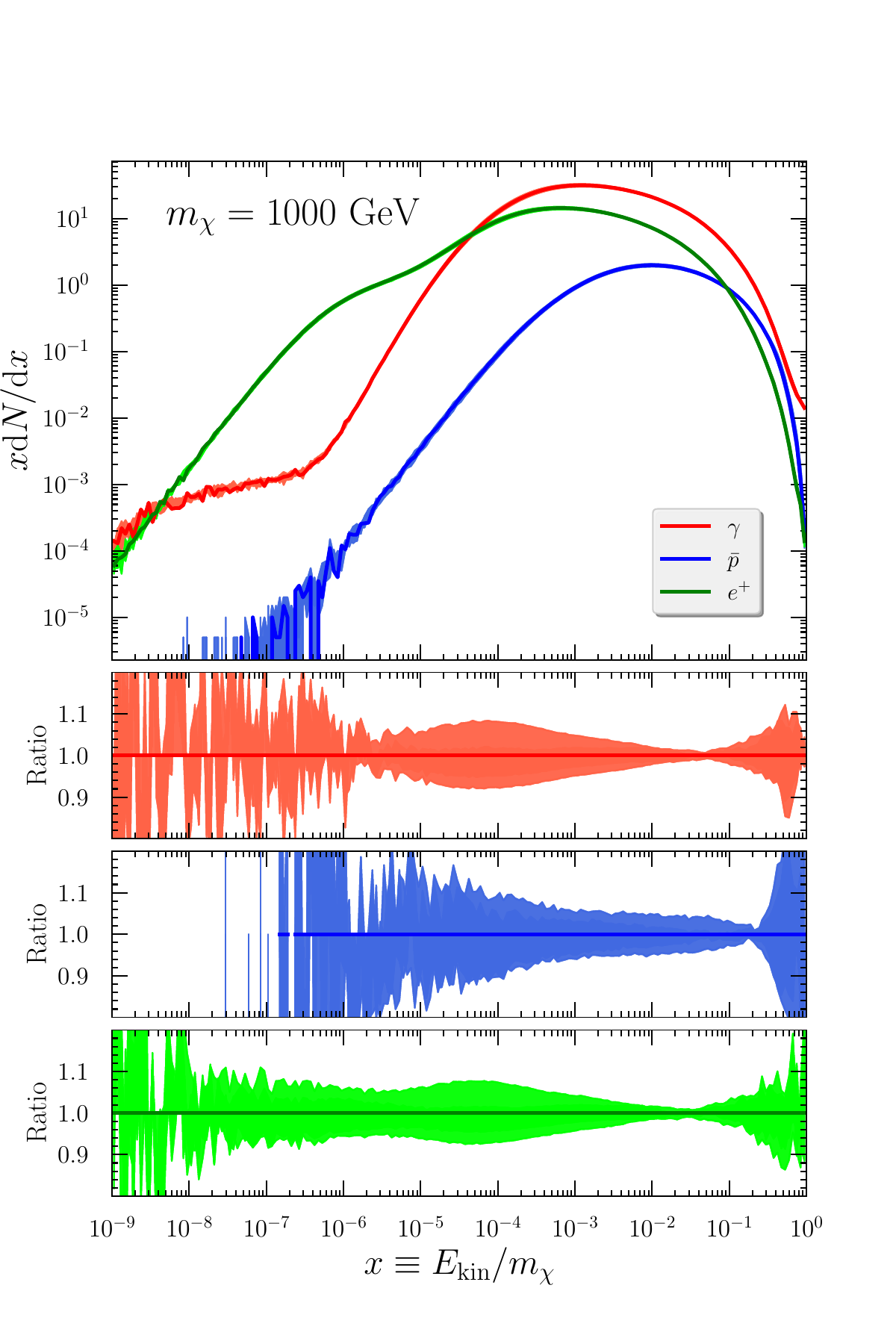}
    \hfill
    \includegraphics[width=0.495\linewidth]{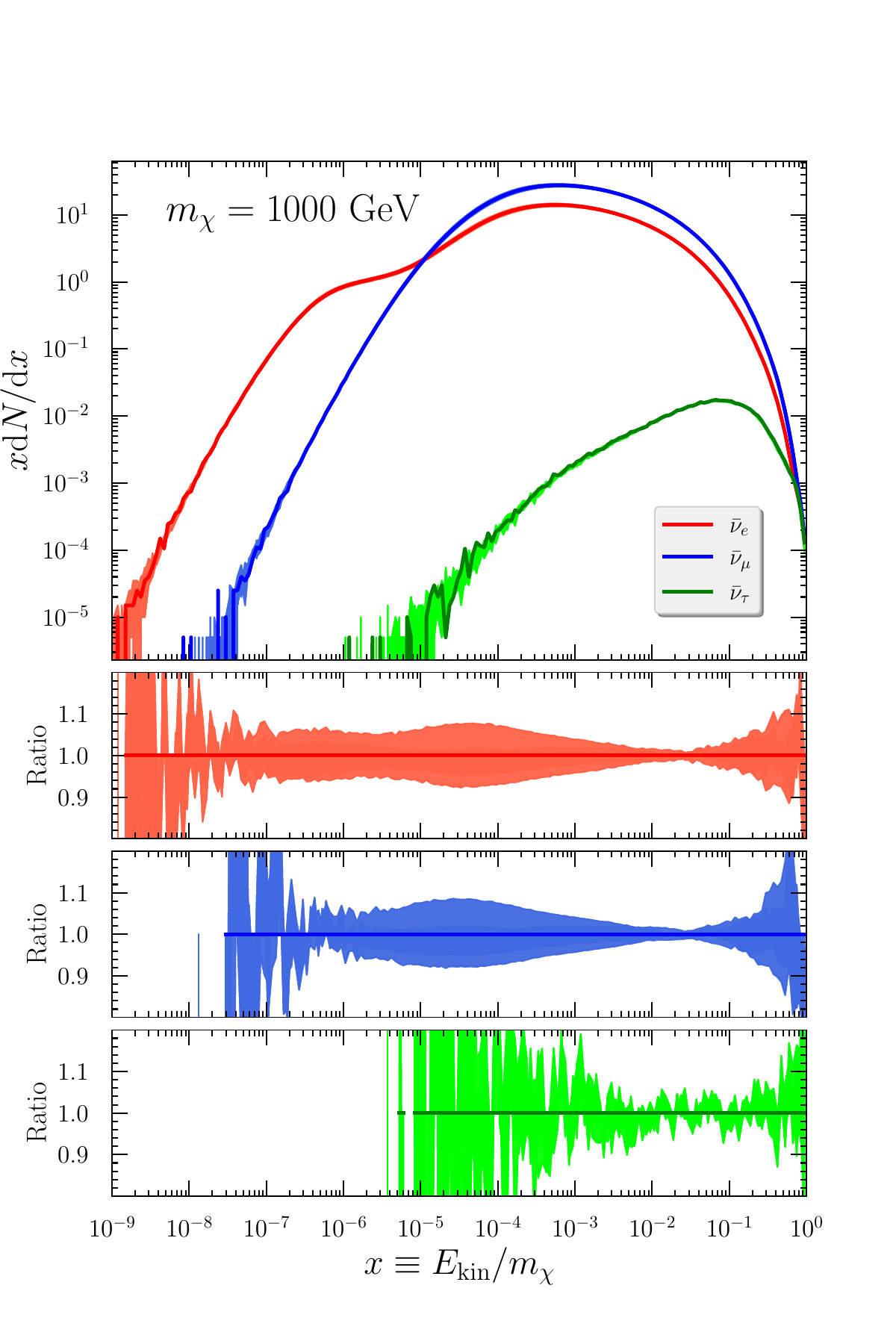}
    \caption{Spectra of stable particles in DM annihilation into $q\bar{q}$ where $q$ refers to all the quarks except the top quark and having universal couplings for $m_\chi = 1000$ GeV. We show the spectra of $\gamma$, $\bar{p}$, $e^+$ (left panel) and of $\bar{\nu}$ (right panel). For each particle species, we also estimate the $1\sigma$ uncertainty bands from QCD hadronization. To have a better visibility, we also show these uncertainties in the ratio-to-nominal subplots.}
    \label{fig:MDM:1000:uncertainties}
\end{figure}

We close this section with a brief discussion of the uncertainties that are related to the hadronization model. While in principle a comparison between different MC event generators such as \textsc{Herwig}~7 \cite{Bellm:2015jjp} and \textsc{Sherpa}~2 \cite{Gleisberg:2008ta} with the results of \Pythia~8 can yield a gross estimate of the theory systematics, we have found that this is not the case. In Ref. \cite{Amoroso:2018qga}, we have shown that the envelope spanned by the three MC event generators, for distributions that are relevant for \eg~$\gamma$-rays, cannot define a systematic and conservative estimate of the uncertainties that are allowed by the data. This is a result that has already been discussed some time ago in Ref. \cite{Cembranos:2013cfa} using slightly older MC event generators. The situation is completely different for the case of antiproton spectra. In Refs. \cite{Jueid:2022qjg, Jueid:2023vrb} it was shown that the relative differences between the different multipurpose MC event generators can be extremely large and can reach up to $50$--$60\%$, especially at the extremes of the baryon spectra. Those differences cannot be defined as uncertainties on the hadronization model. Therefore, estimating uncertainties within the same model seems to be the correct method for these analyses \cite{Amoroso:2018qga, Jueid:2022qjg, Jueid:2023vrb}. Here, we only briefly discuss the size of these uncertainties assuming the \vincia shower algorithm as our default option. We do not provide the uncertainties on the tabulated spectra, but the interested reader can find them in the \href{https://github.com/ajueid/CosmiXs.git}{github repository}.\footref{fn:github}
The \textsc{Professor} tuning allows for an estimate of the uncertainties on the parameters using the Hessian method which is widely used in the PDF community, see \eg~Ref. \cite{Pumplin:2001ct} for more details. The method consists of diagonalising the $\chi^2$ covariance matrix near the minimum of the parameter space. The variation around the minimum of the $\Delta \chi^2$ can be expanded as: 
\begin{eqnarray}
    \Delta \chi^2 = \sum_i K_{i}(x_i) (x_i - x_i^0) + \sum_i \sum_j H_{ij}(x_i, x_j) (x_i - x_i^0)(x_j - x_j^0),
\end{eqnarray}
where $K_{i}(x_i)$ is the first derivative of the $\chi^2$ which vanishes near the minimum and $H_{ij} = \partial^2 \chi^2/\partial x_i \partial x_j$ is the Hessian matrix. The sum is over all the parameters of the model. The diagonalisation of $H_{ij}$ leads to the so-called prinicipal directions (eigenvectors) and the corresponding eigenvalues. We get a set of $2 \times N_{\rm params}$ variations, which in our case correspond to $8$ variations. Imposing a constraint on the maximum variation with maximum radius equal to the confidence level of the departure from the minimum, one obtains a one-sigma variation if $\Delta \chi^2 / N_{\rm df}= 1$, two-sigma variation if $\Delta \chi^2 / N_{\rm df} = 4$ and so on. The results of the Hessian method for the $1\sigma$ eigentunes\footnote{The eigentunes or the Hessian variations correspond to the principal directions obtained from the diagonalization of the matrix $H_{ij}$  near the minimum. In other words, they are obtained as corresponding to a fixed change in the goodness-of-fit measure which is found by imposing a constraint on the maximum variation which is defined on an hypersphere (called the tolerance $T$). A one-sigma eigentunes corresponds to $\Delta\chi^2/N_{\rm df} = 1$, a two-sigma eigentunes corresponds to $\Delta\chi^2 / N_{\rm df} = 4$ and so on.} 
are shown in Tab. \ref{tab:eigentunes}. The impact of these uncertainties on the particle spectra is shown in Fig. \ref{fig:MDM:1000:uncertainties} where we give an example for a $1000$ GeV DM annihilating into $q\bar{q}$. We can see that the uncertainties range from $10\%$--$30\%$ depending on the energy region. These results are in a good agreement with the findings of Refs. \cite{Jueid:2022qjg,Jueid:2023vrb}.  

\section{Results for the particle spectra}
\label{sec:results}

\subsection{Overview of the particle spectra}
\begin{figure}
    \centering
    \includegraphics[width=0.32\linewidth]{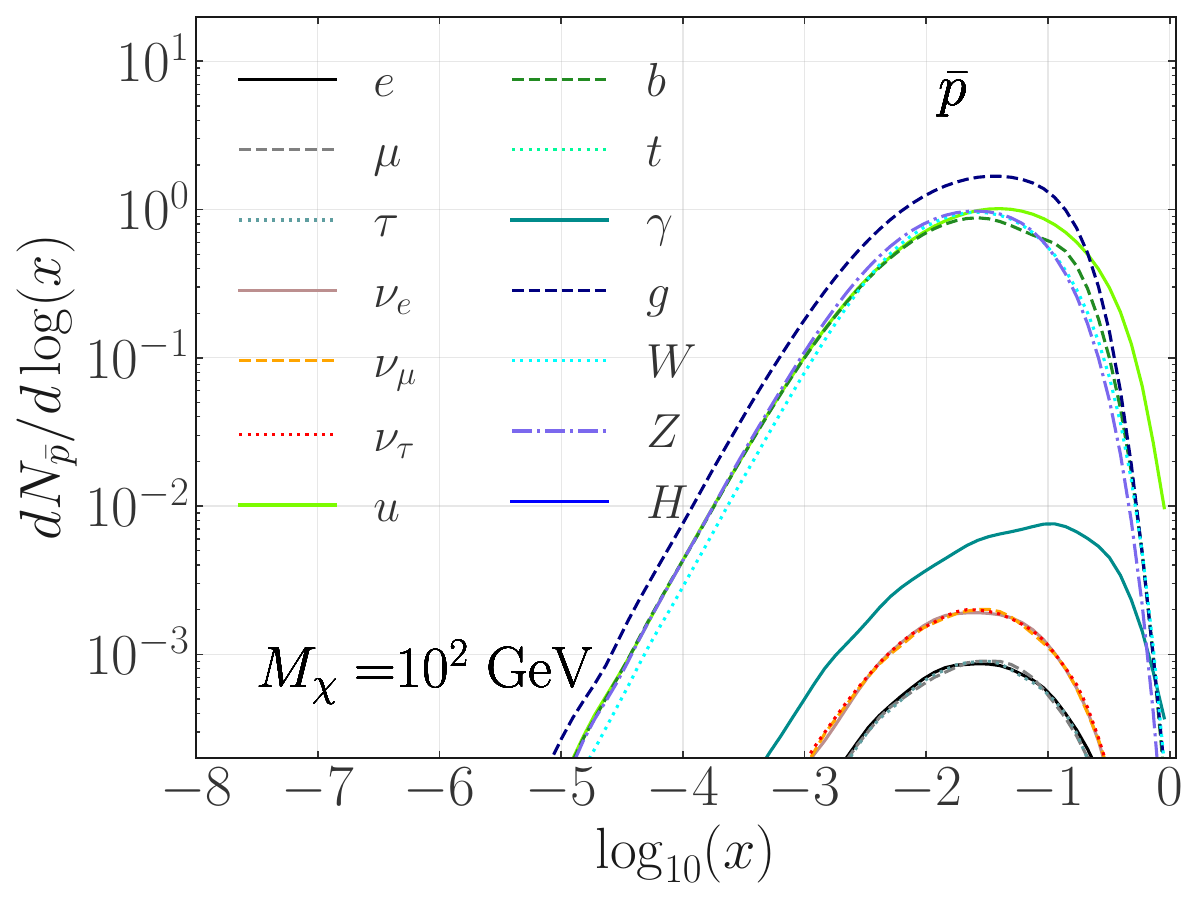}
    \includegraphics[width=0.32\linewidth]{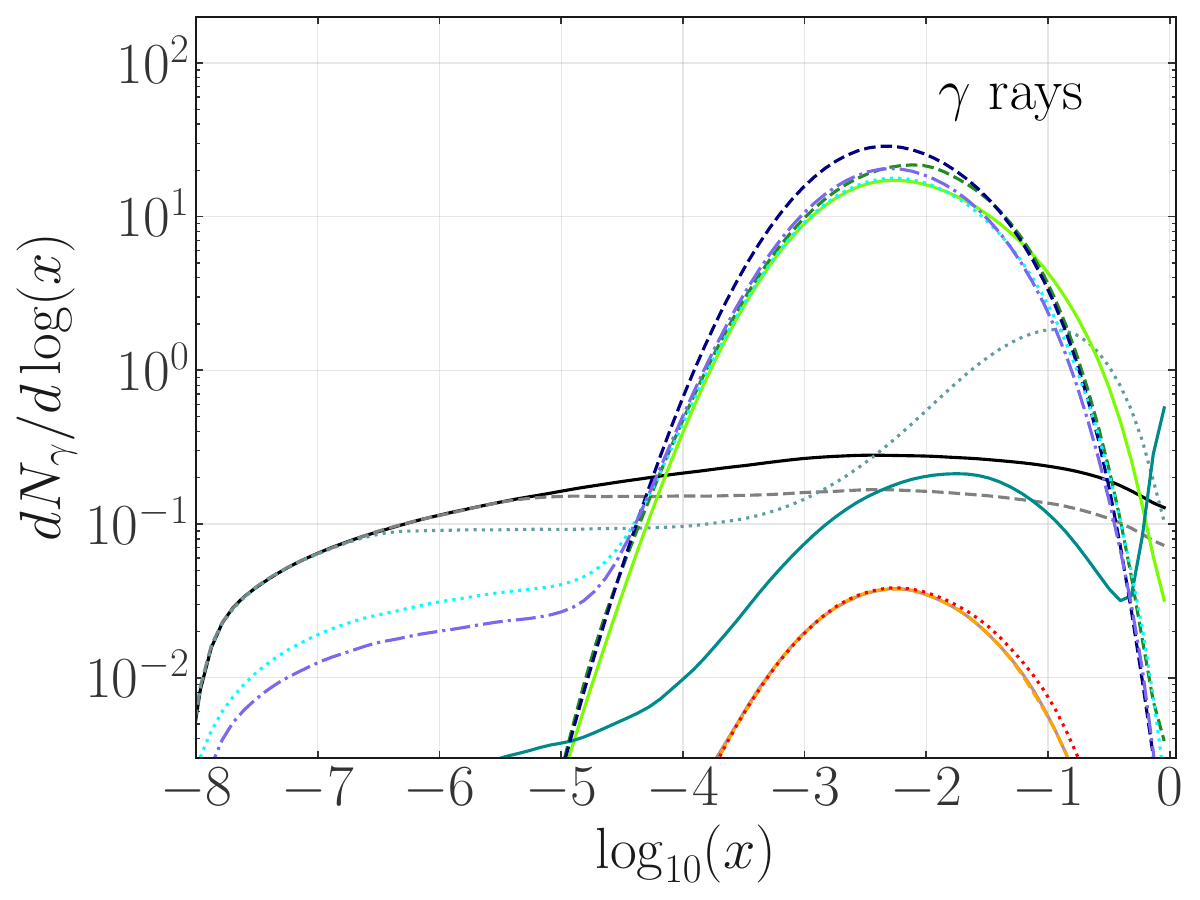}
    \includegraphics[width=0.32\linewidth]{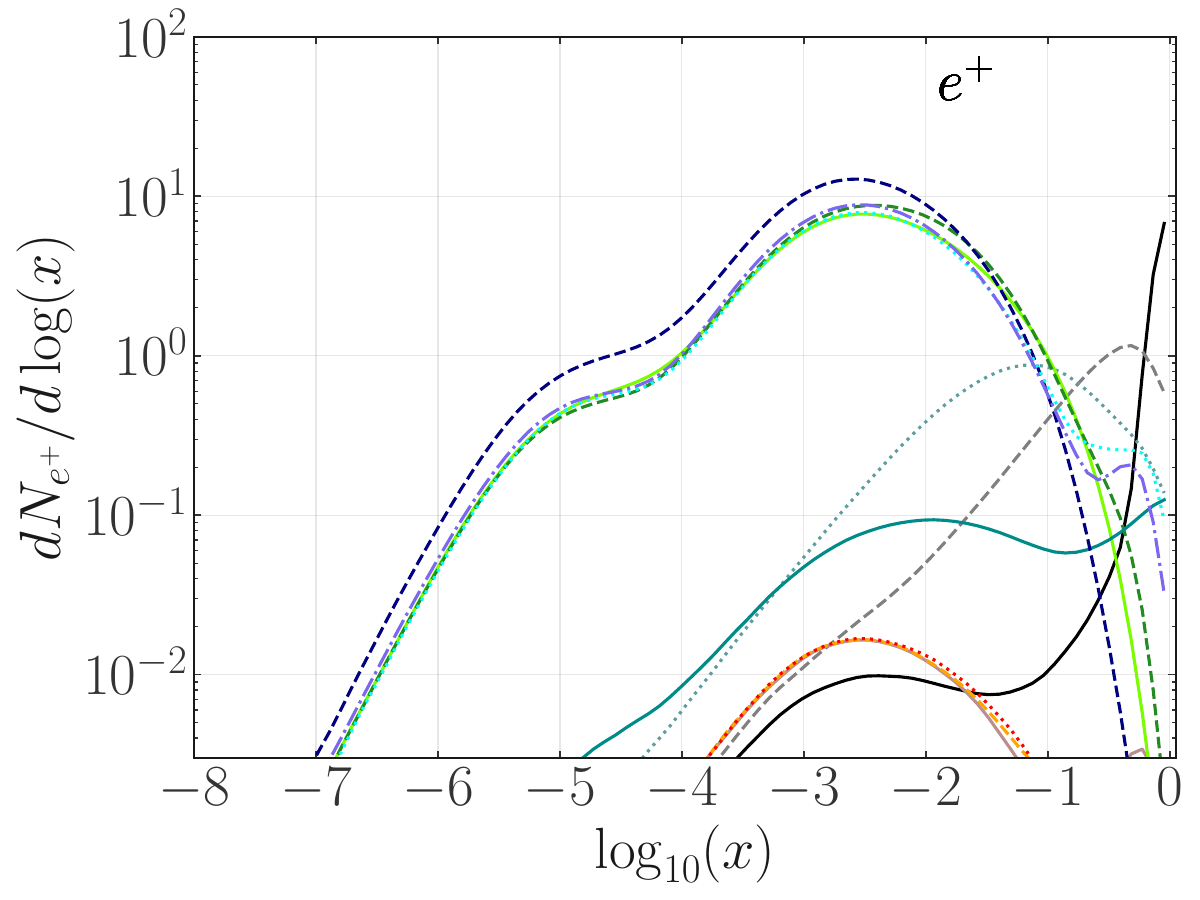}
    \includegraphics[width=0.32\linewidth]{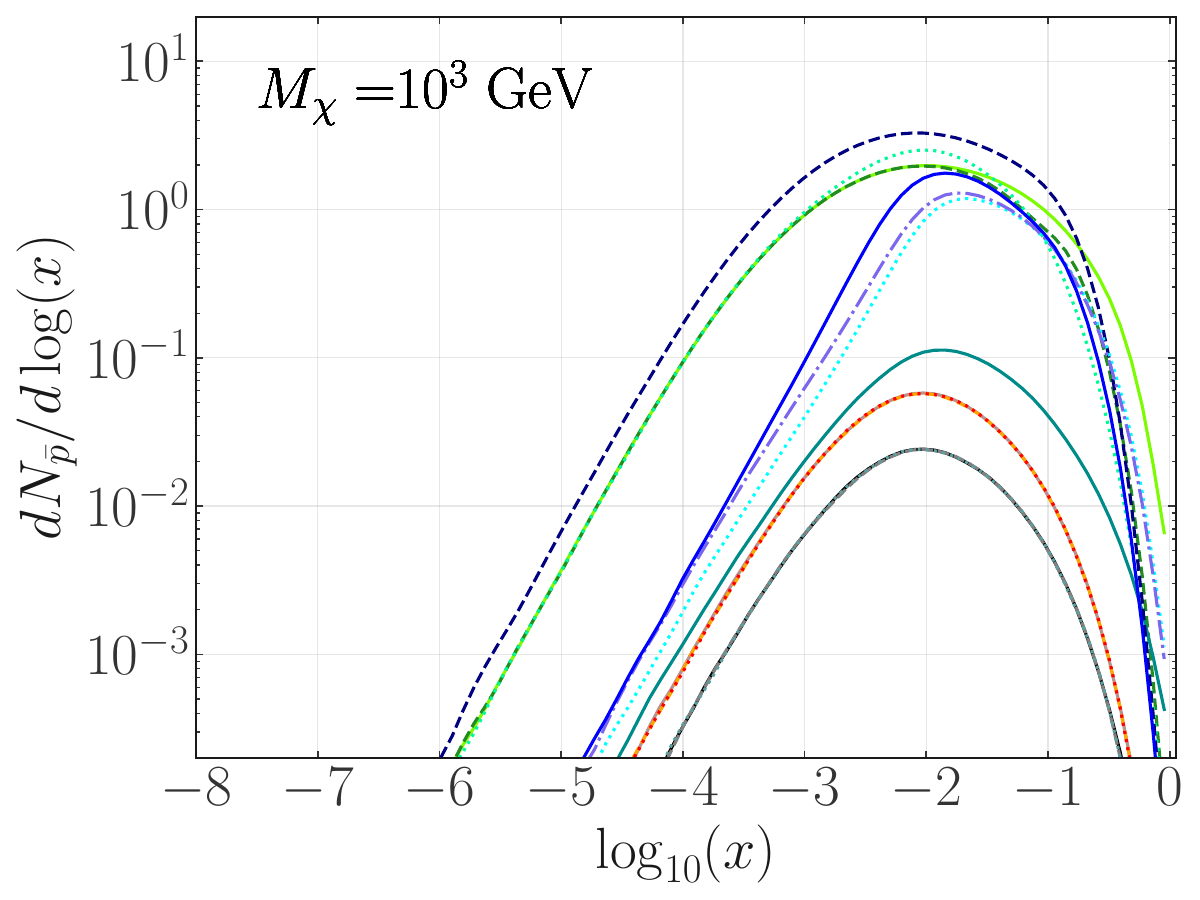}
    \includegraphics[width=0.32\linewidth]{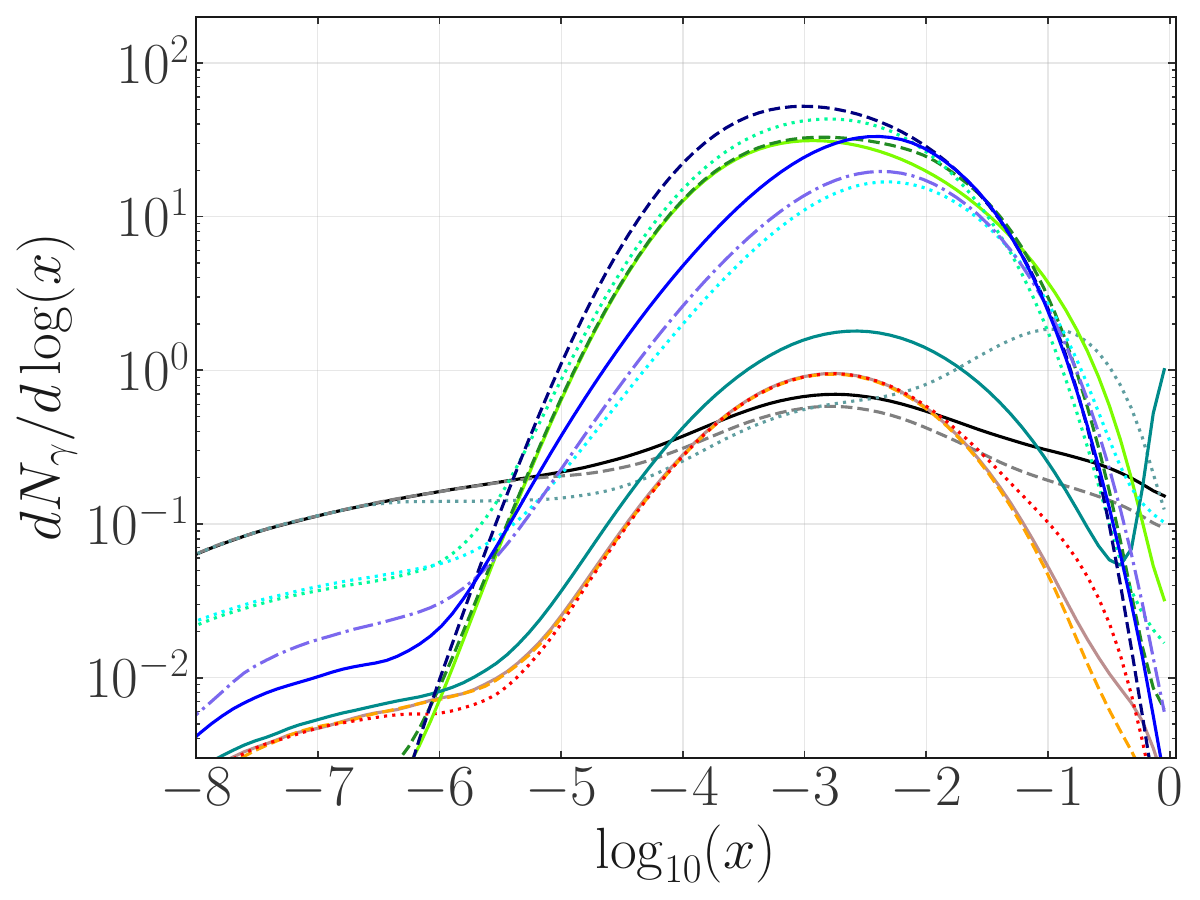}
    \includegraphics[width=0.32\linewidth]{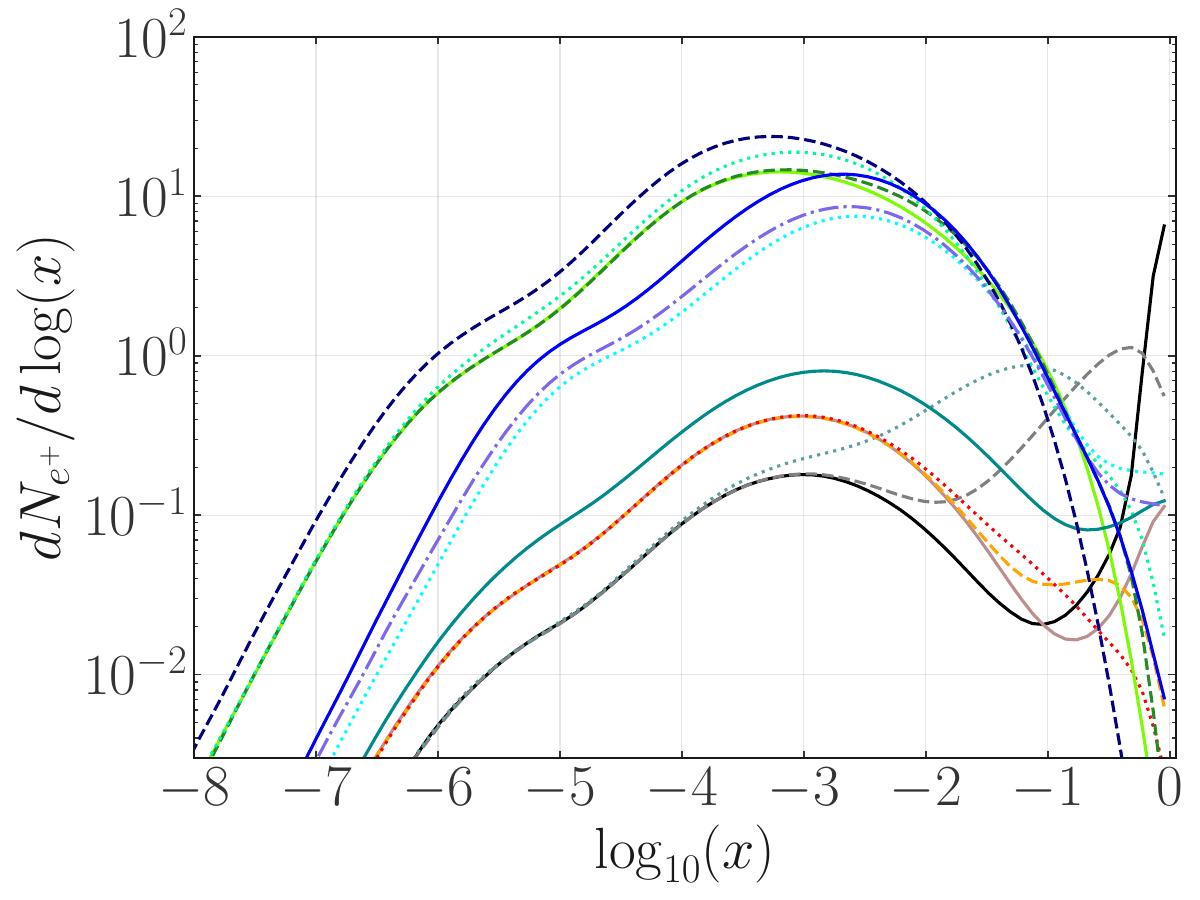}
    \includegraphics[width=0.32\linewidth]{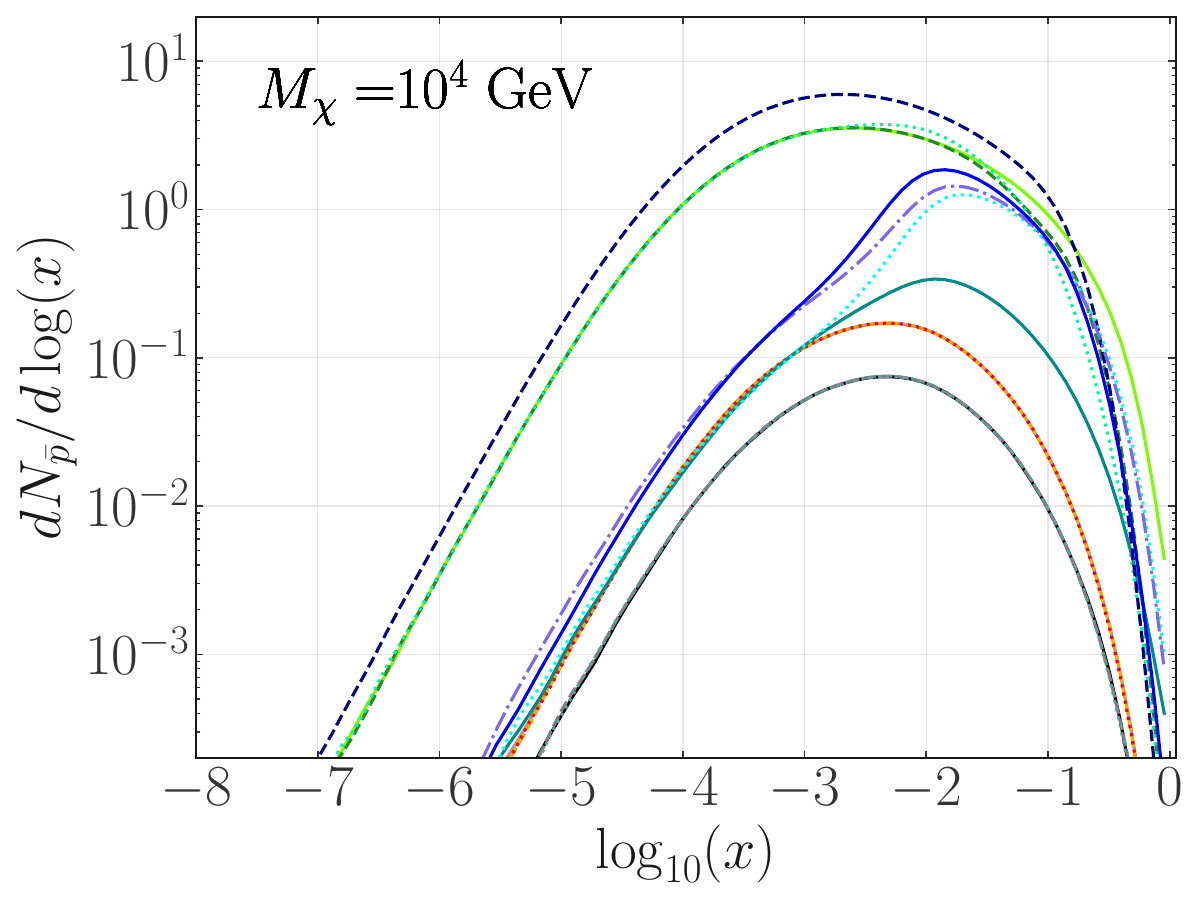}
    \includegraphics[width=0.32\linewidth]{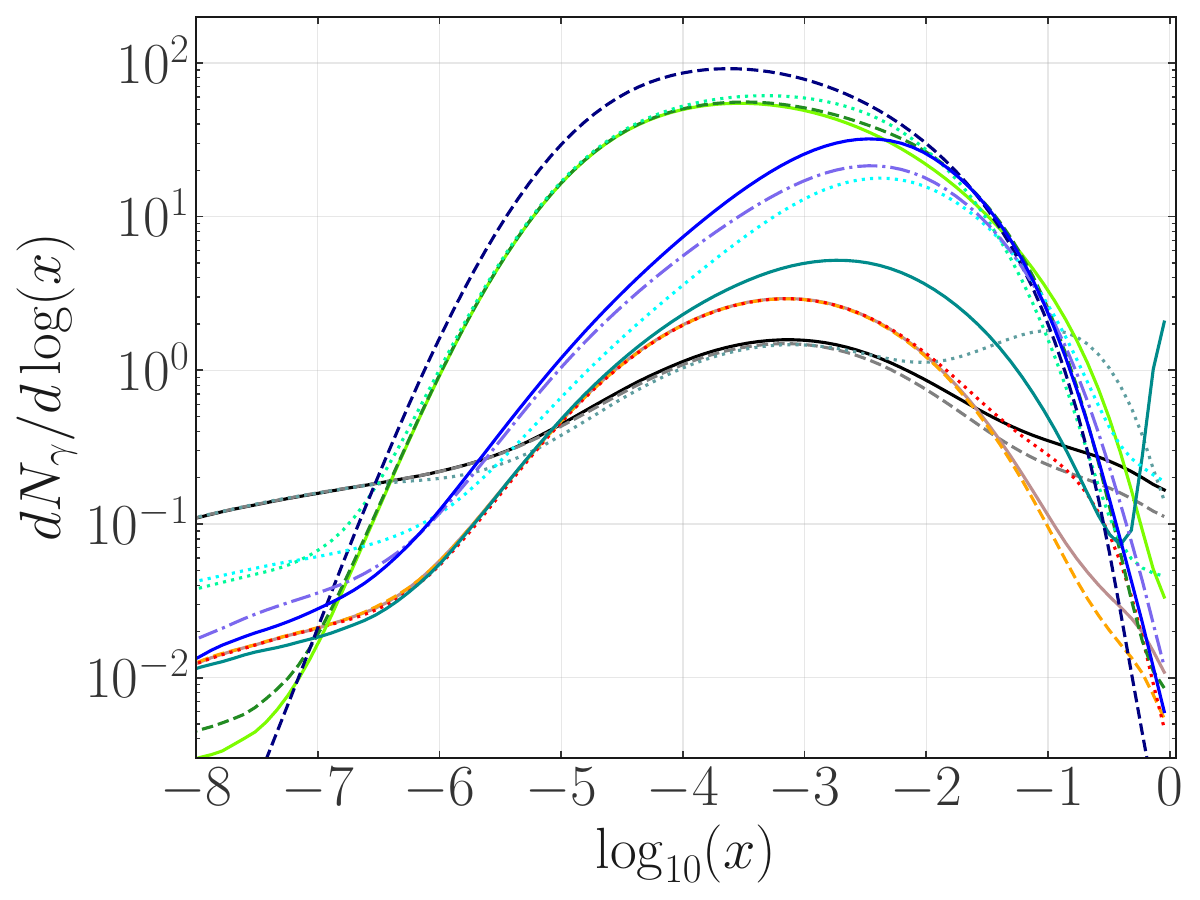}
    \includegraphics[width=0.32\linewidth]{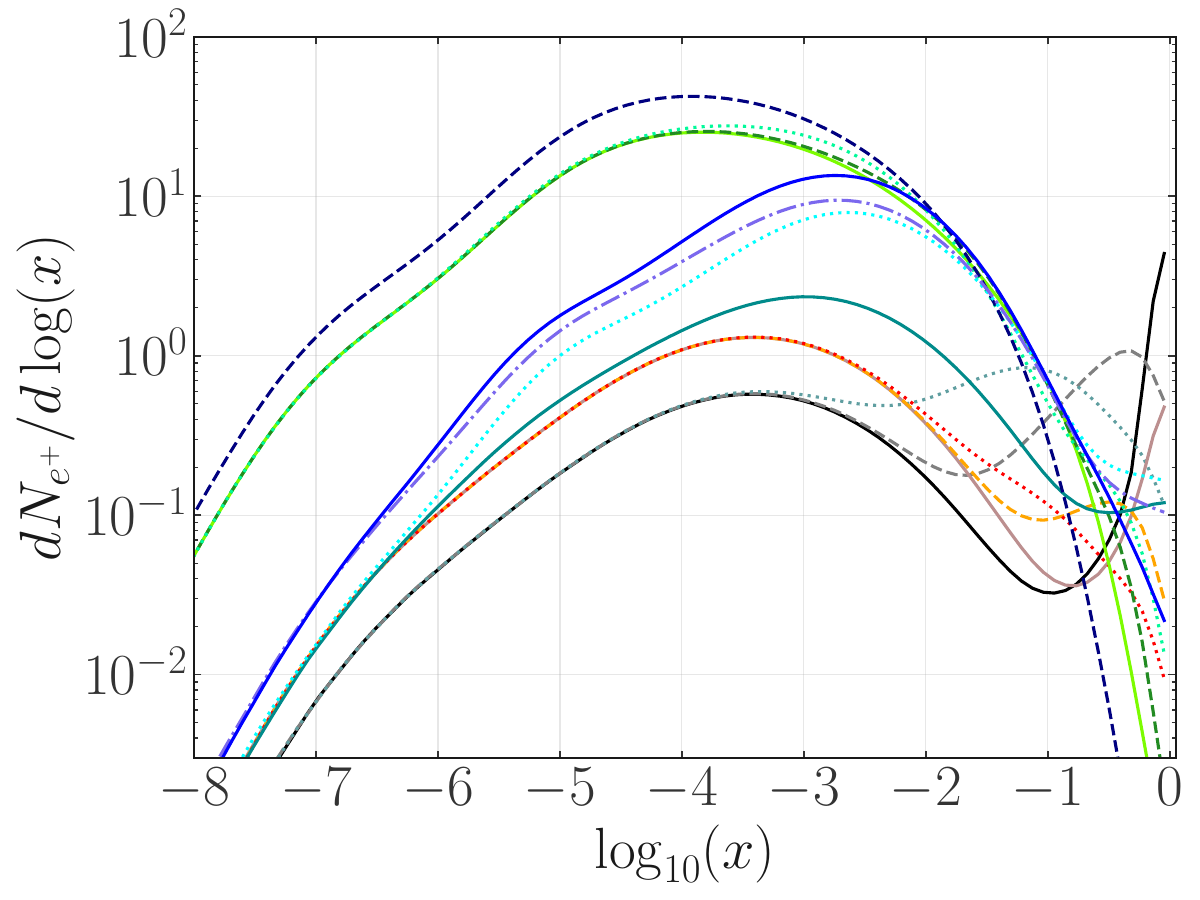}
    \includegraphics[width=0.32\linewidth]{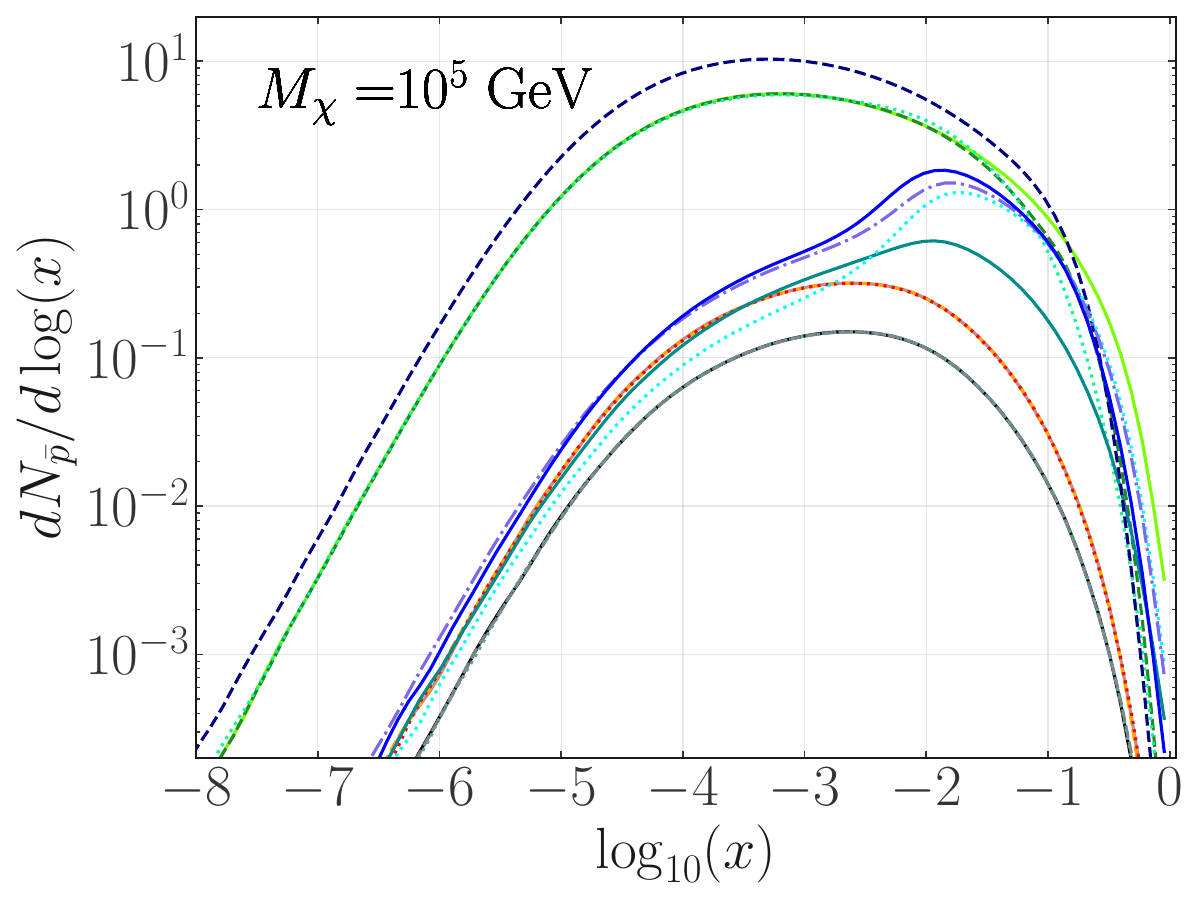}
    \includegraphics[width=0.32\linewidth]{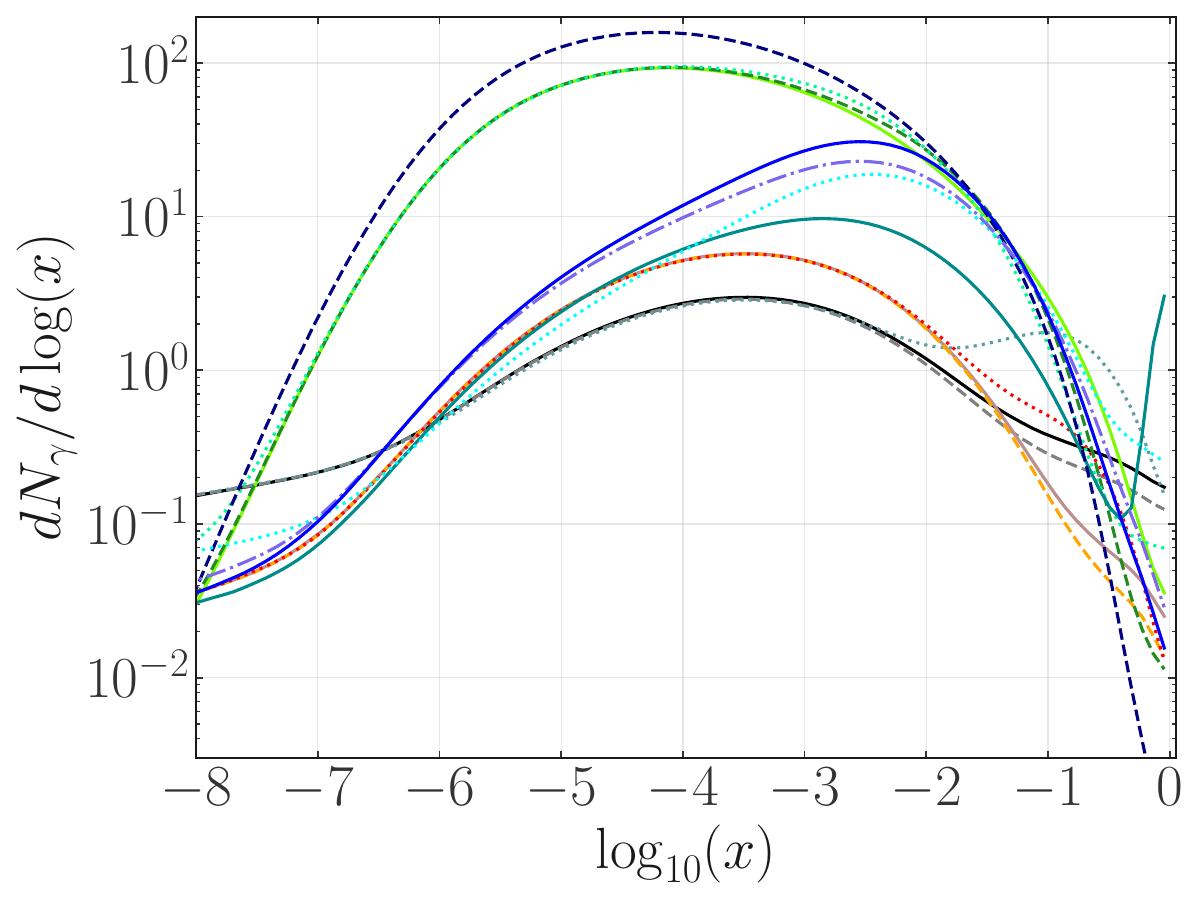}
    \includegraphics[width=0.32\linewidth]{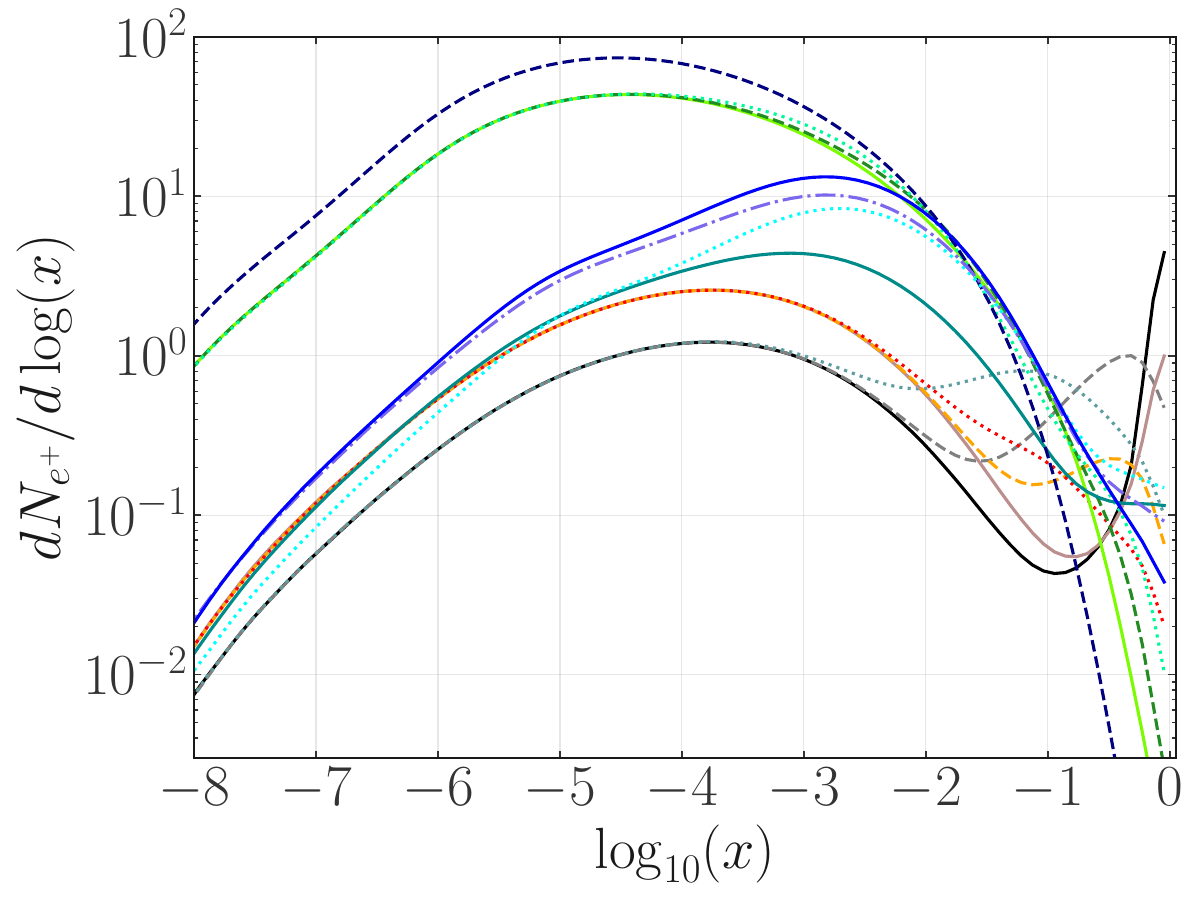}
    \caption{Energy spectra of cosmic messengers produced from DM annihilation. We show from top to bottom the results obtained for DM mass of 100 GeV, 1 TeV, 10 TeV and 100 TeV. We display from left to right the spectra for $\bar{p}$, $\gamma$ rays and $e^+$.}
    \label{fig:spectraresults}
\end{figure}

\begin{figure}
    \centering
    \includegraphics[width=0.32\linewidth]{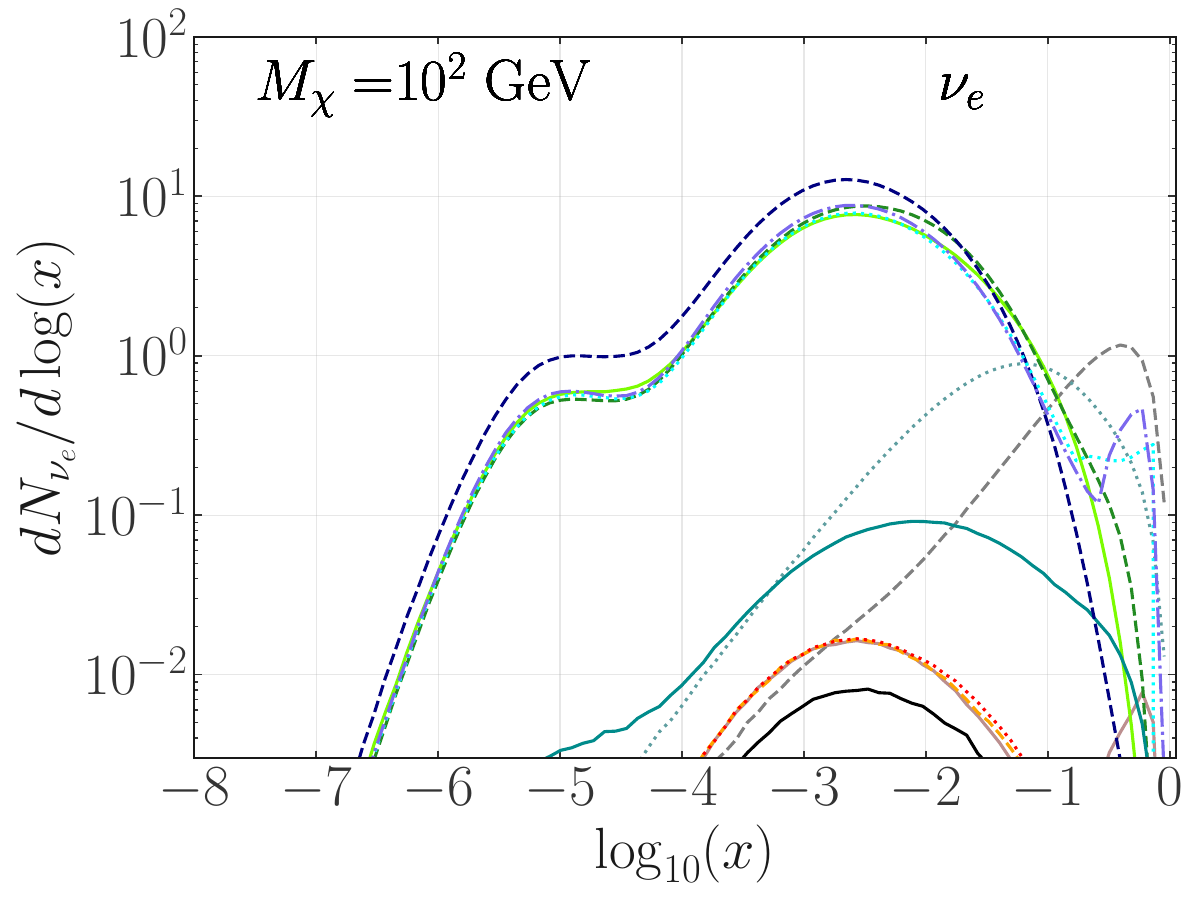}
    \includegraphics[width=0.32\linewidth]{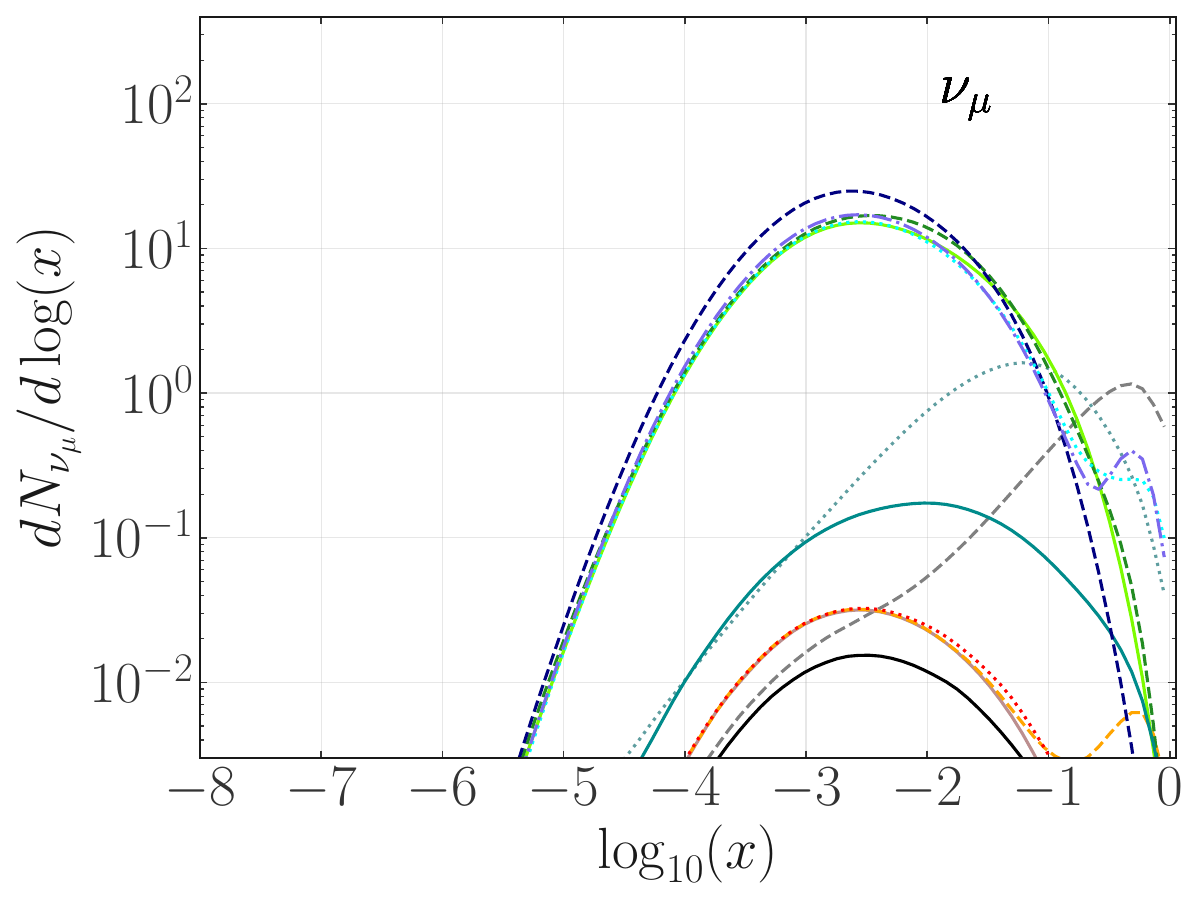}
    \includegraphics[width=0.32\linewidth]{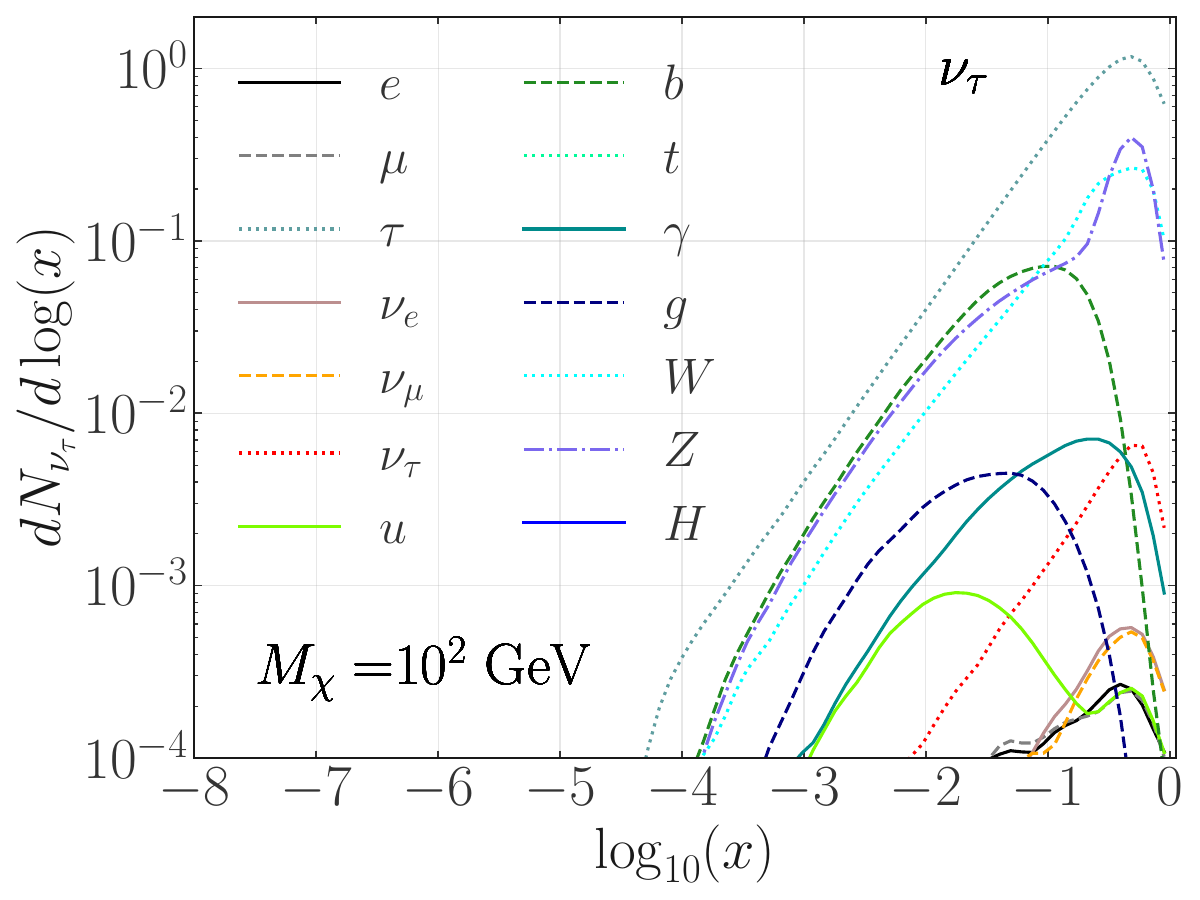}
    \includegraphics[width=0.32\linewidth]{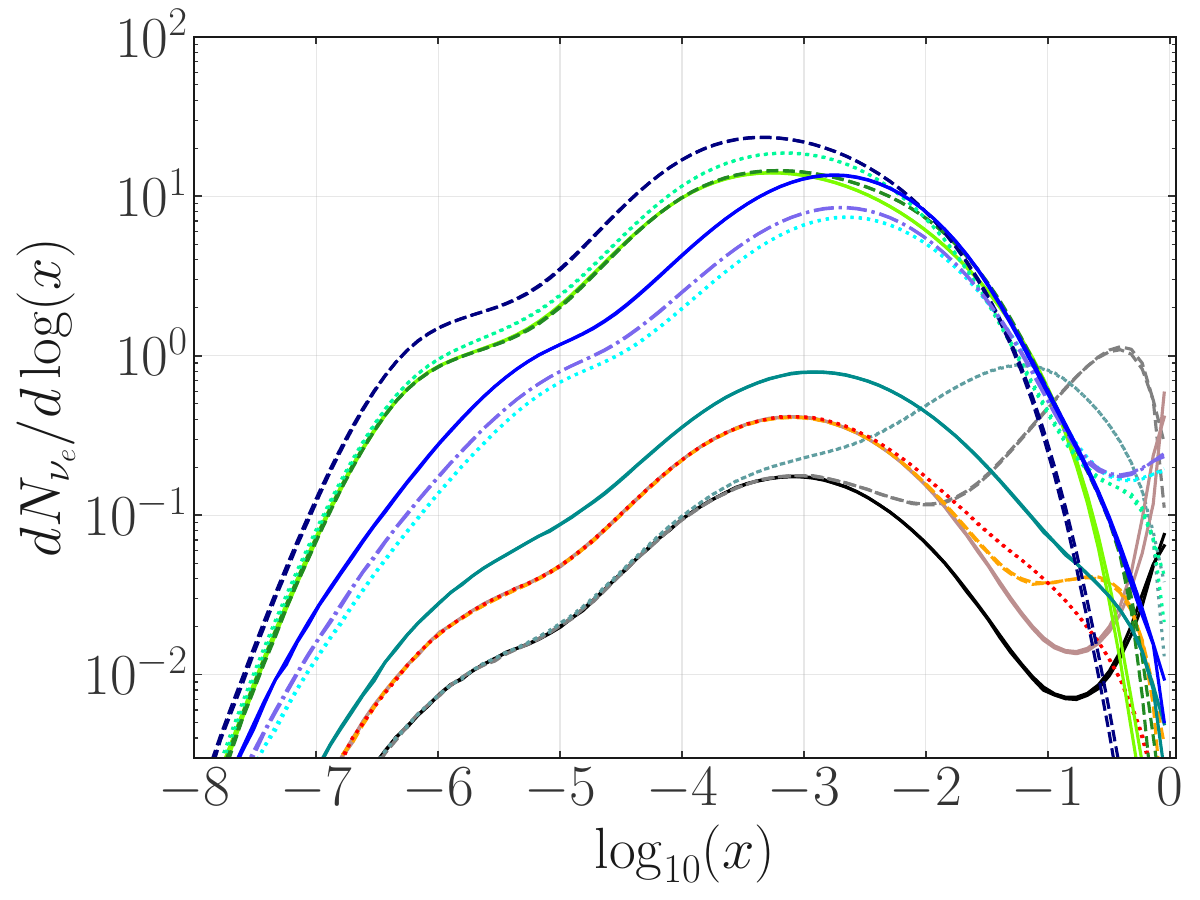}
    \includegraphics[width=0.32\linewidth]{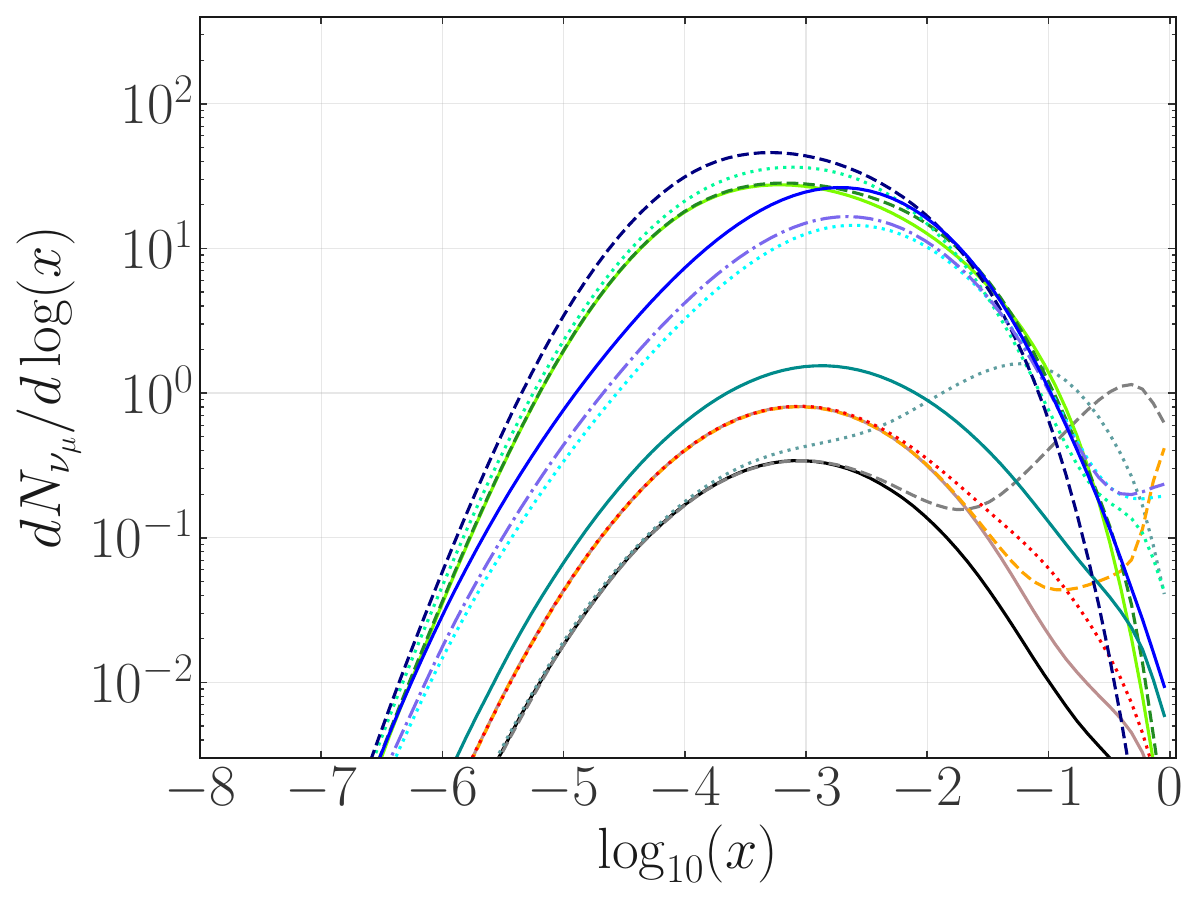}
    \includegraphics[width=0.32\linewidth]{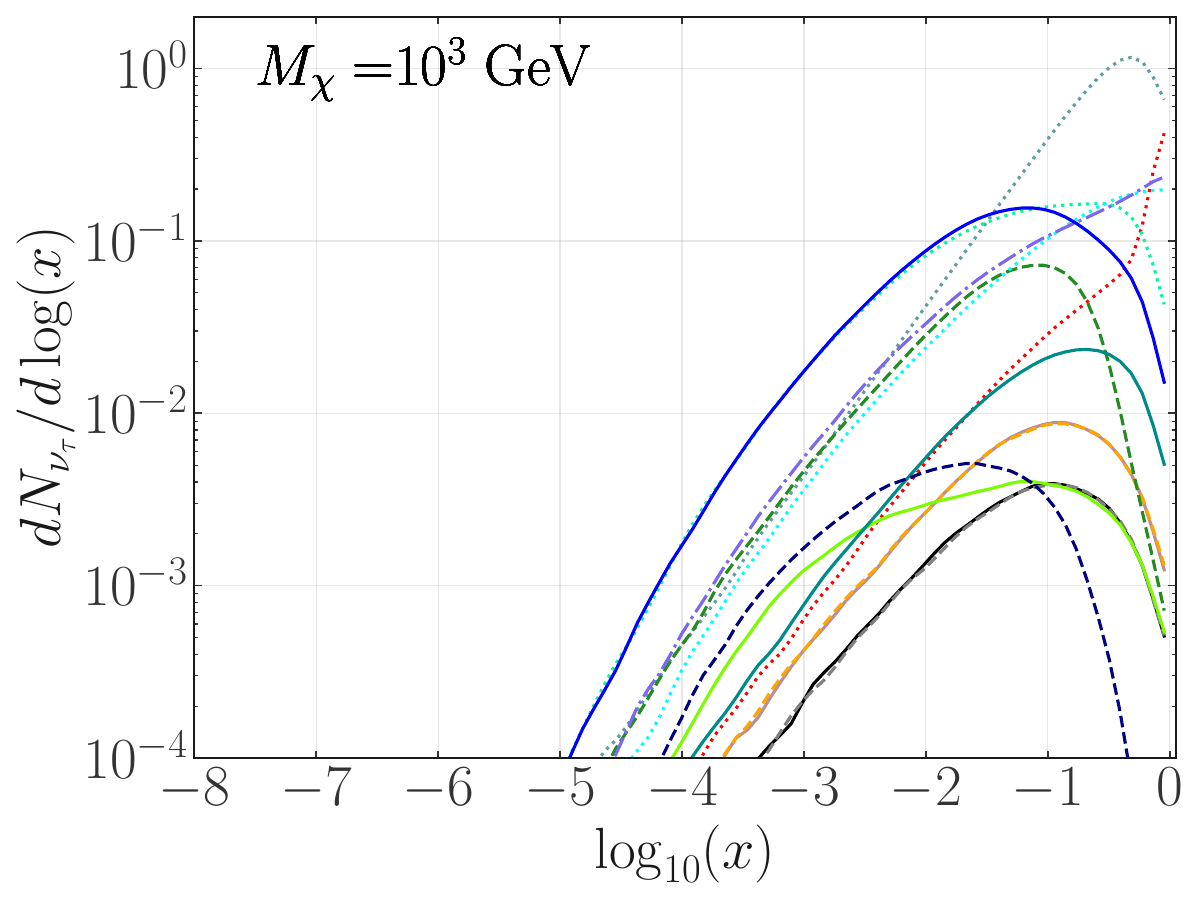}
    \includegraphics[width=0.32\linewidth]{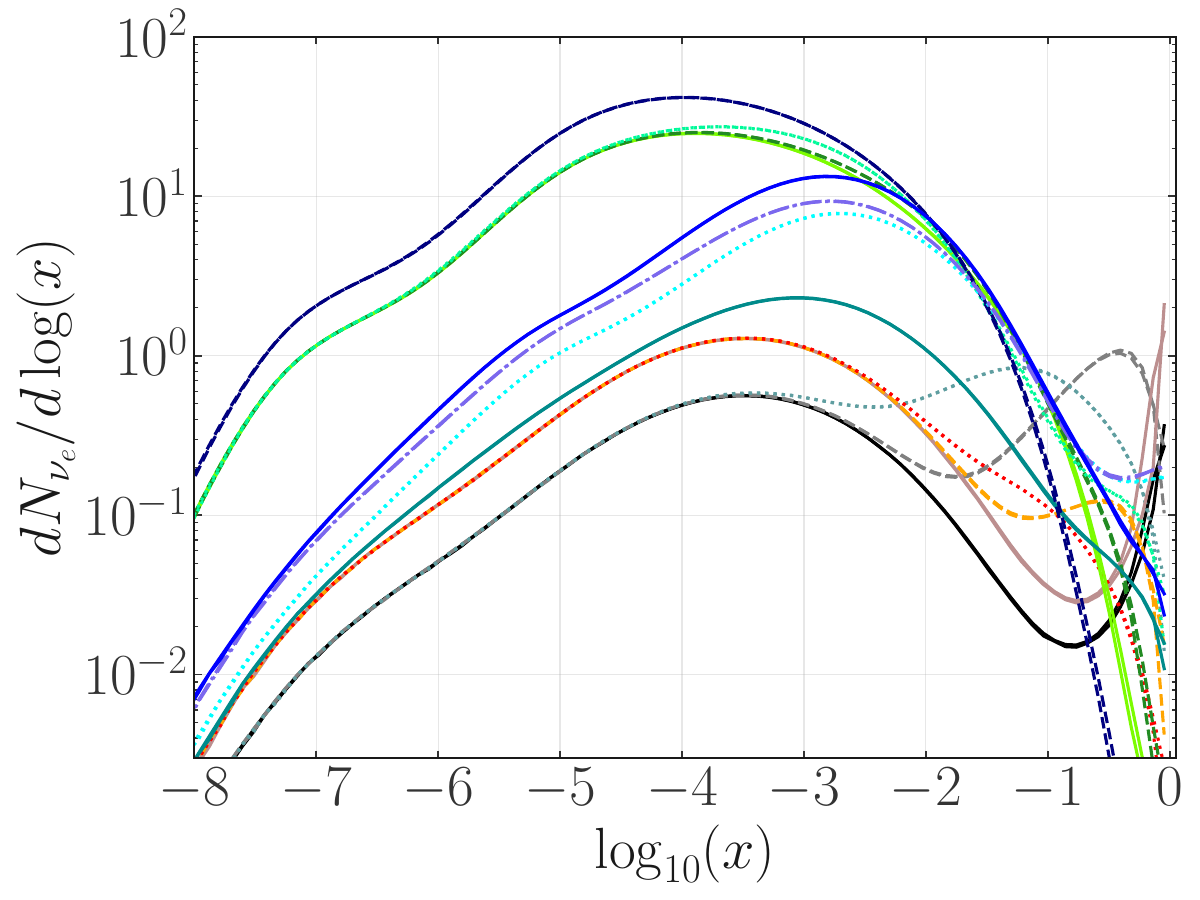}
    \includegraphics[width=0.32\linewidth]{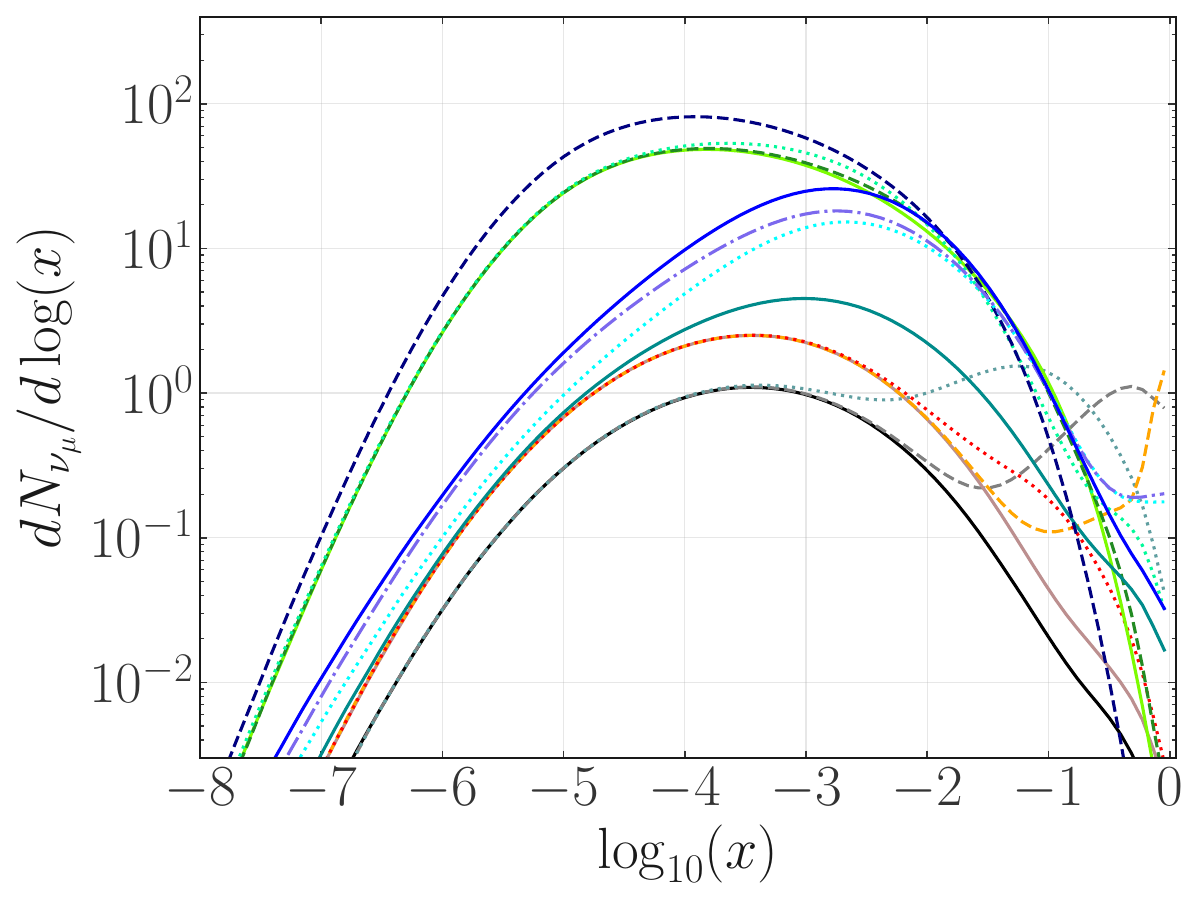}
    \includegraphics[width=0.32\linewidth]{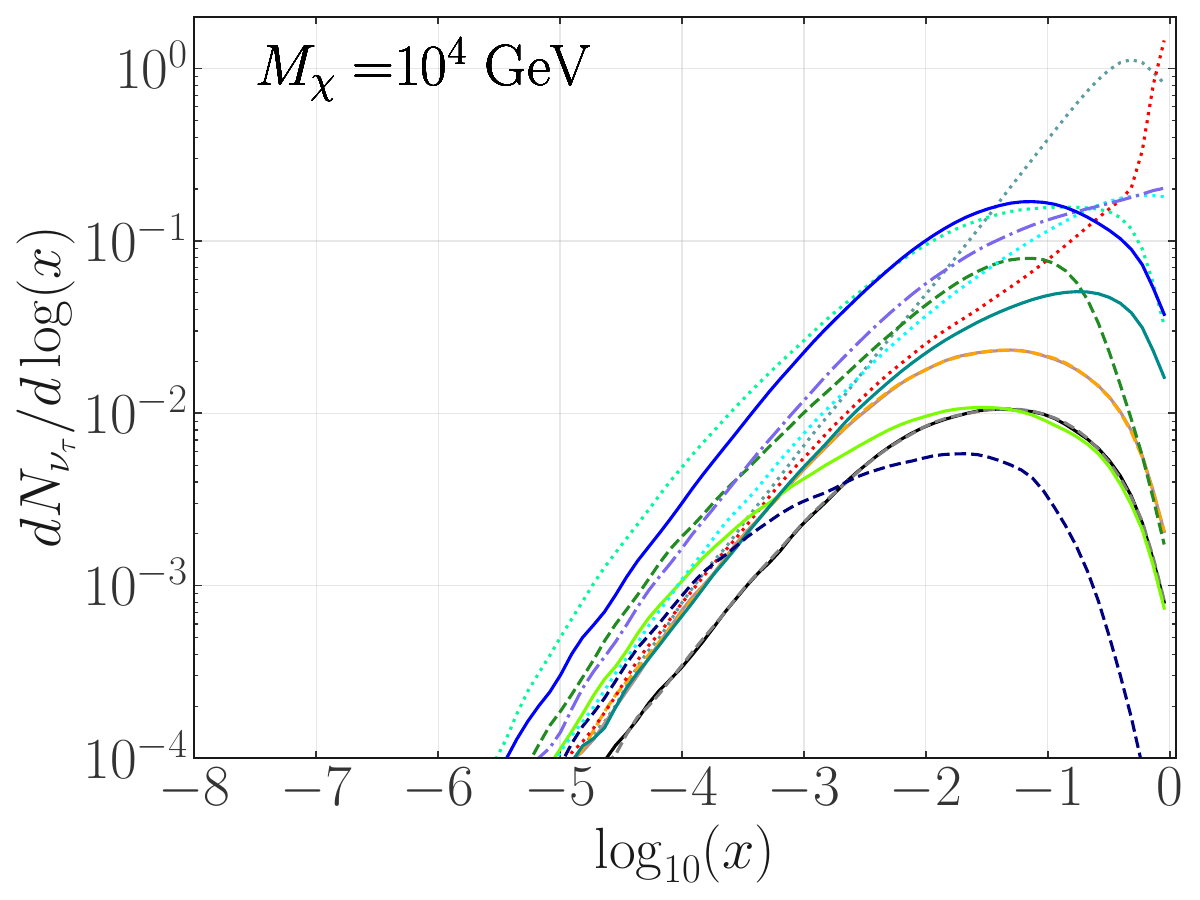}
    \includegraphics[width=0.32\linewidth]{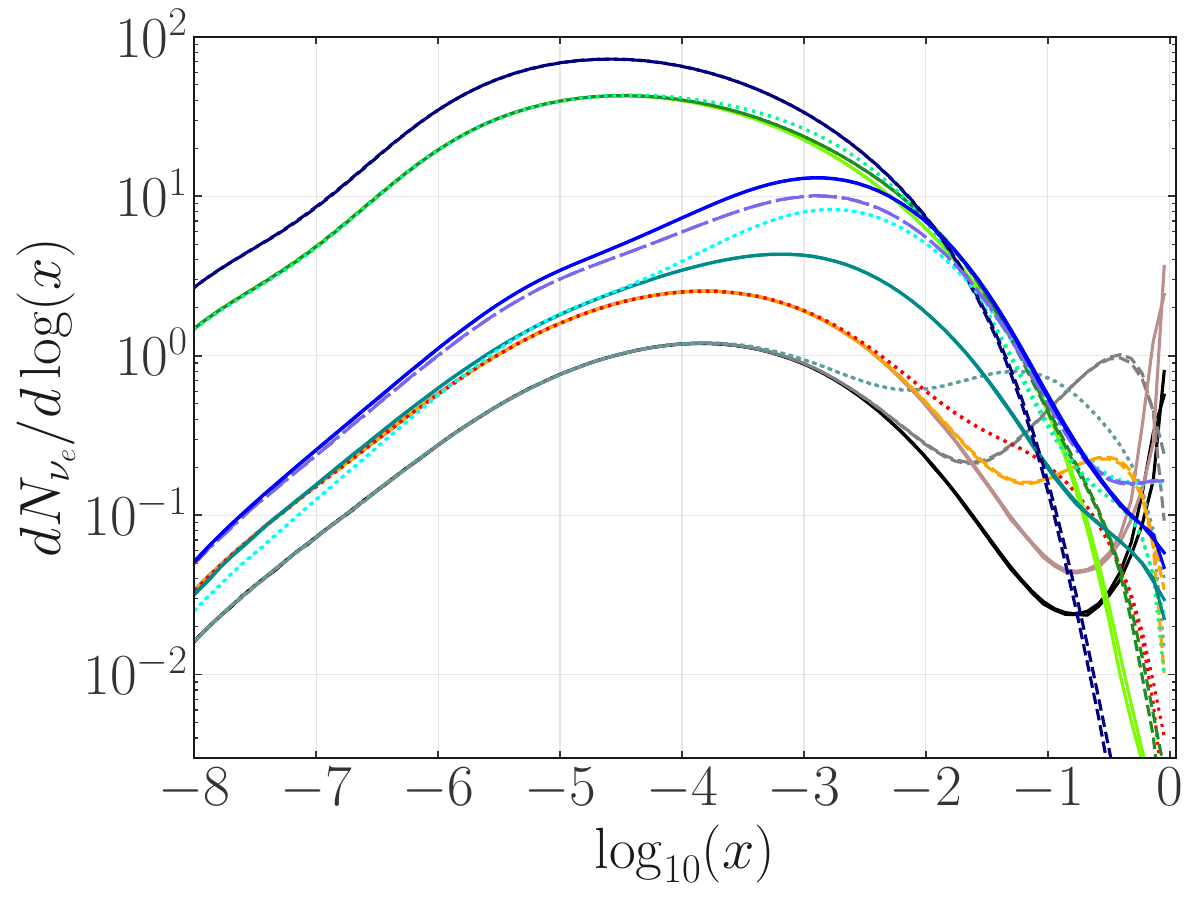}
    \includegraphics[width=0.32\linewidth]{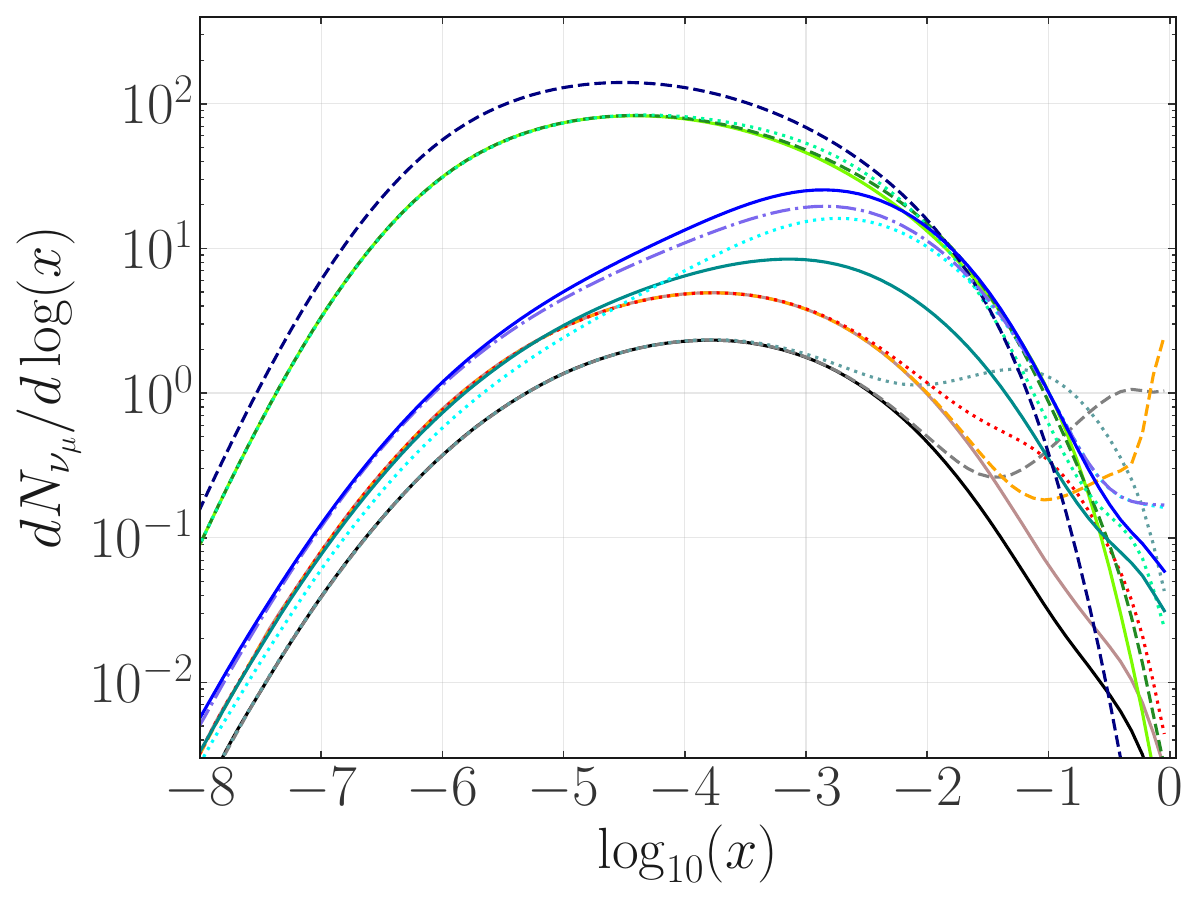}
    \includegraphics[width=0.32\linewidth]{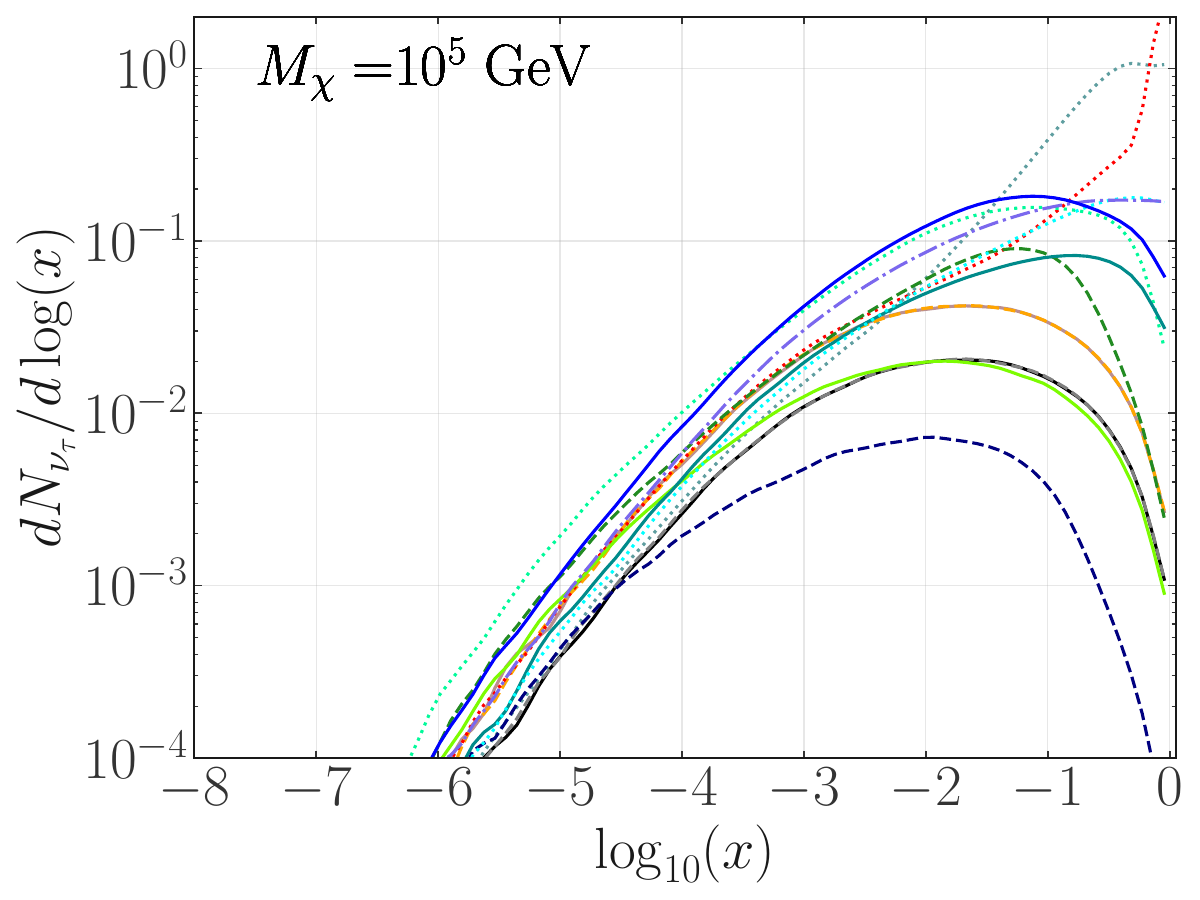}
    \caption{Same as Fig.~\ref{fig:spectraresults} for the spectra of the three flavors of $\nu$.}
    \label{fig:spectraresultsnu}
\end{figure}

Here we show a selection of results for the particle spectra we obtain.
In particular, in Figs. \ref{fig:spectraresults} and \ref{fig:spectraresultsnu}, we show the spectra obtained for all channels considered and focusing on DM masses between 100 GeV and 100 TeV.
We show the results obtained for the annihilation of DM particles. However, for a scalar, pseudoscalar, vector and axial vector DM particle 
\footnote{Due to the conservation of the spin quantum number, fermionic decaying DM particles produce an odd number of SM fermions in the final state. Such channels are not included in our current work and are deferred to future work.}
our results can also be used for decaying DM upon simple rescaling. Specifically, the annihilation spectra for $m_{\rm DM}^{\rm ann}$ correspond to the decay spectra of $2m_{\rm DM}^{\rm dec}$ 
while rescaling the $x$ variable accordingly.
We have explicitly checked that the spectra of DM annihilation via a mediator with a given spin and coupling structure is equal to the ones of decaying DM particle with the same properties as the mediator. Therefore, our tables can be used for DM decays that select a specific lepton chirality or gauge boson polarization.

\begin{figure}
\centering
\includegraphics[width=0.32\linewidth]{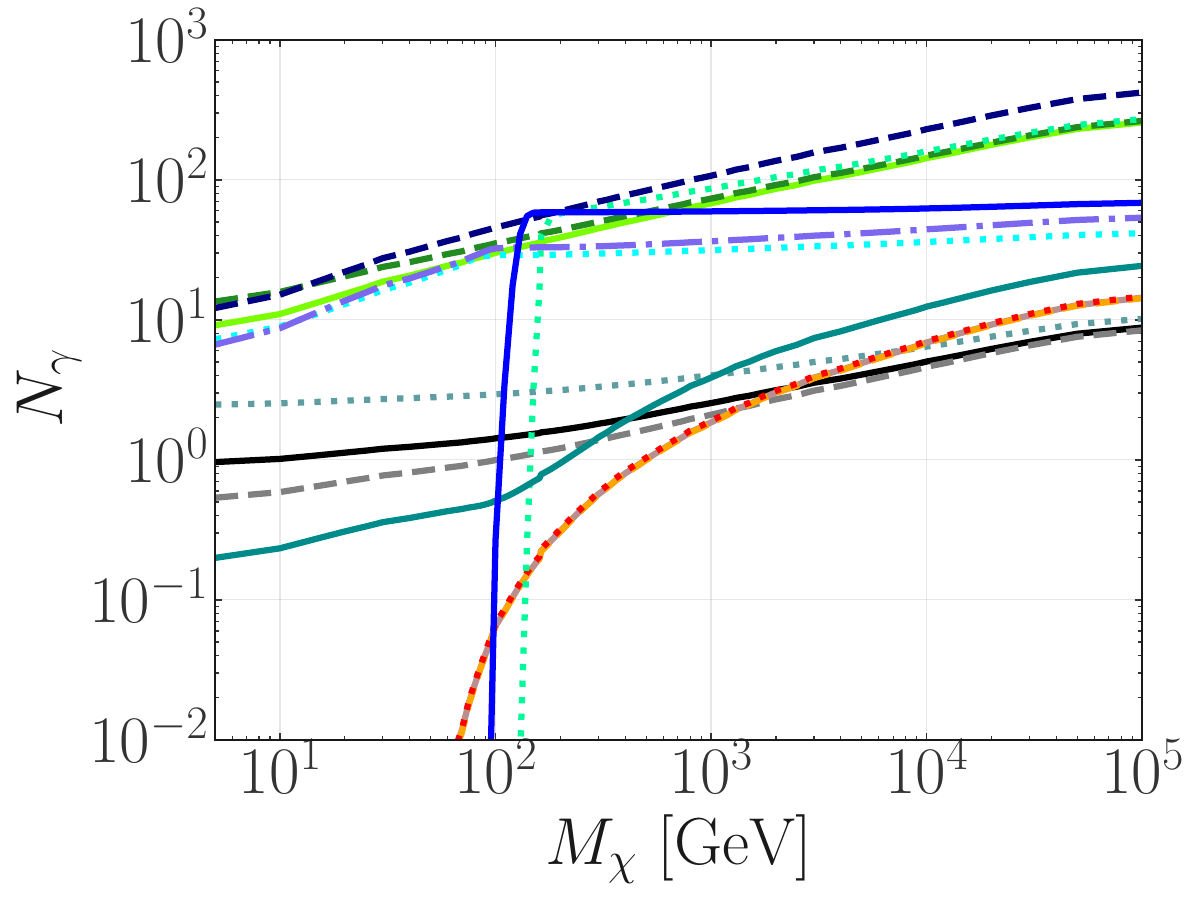}
\includegraphics[width=0.32\linewidth]{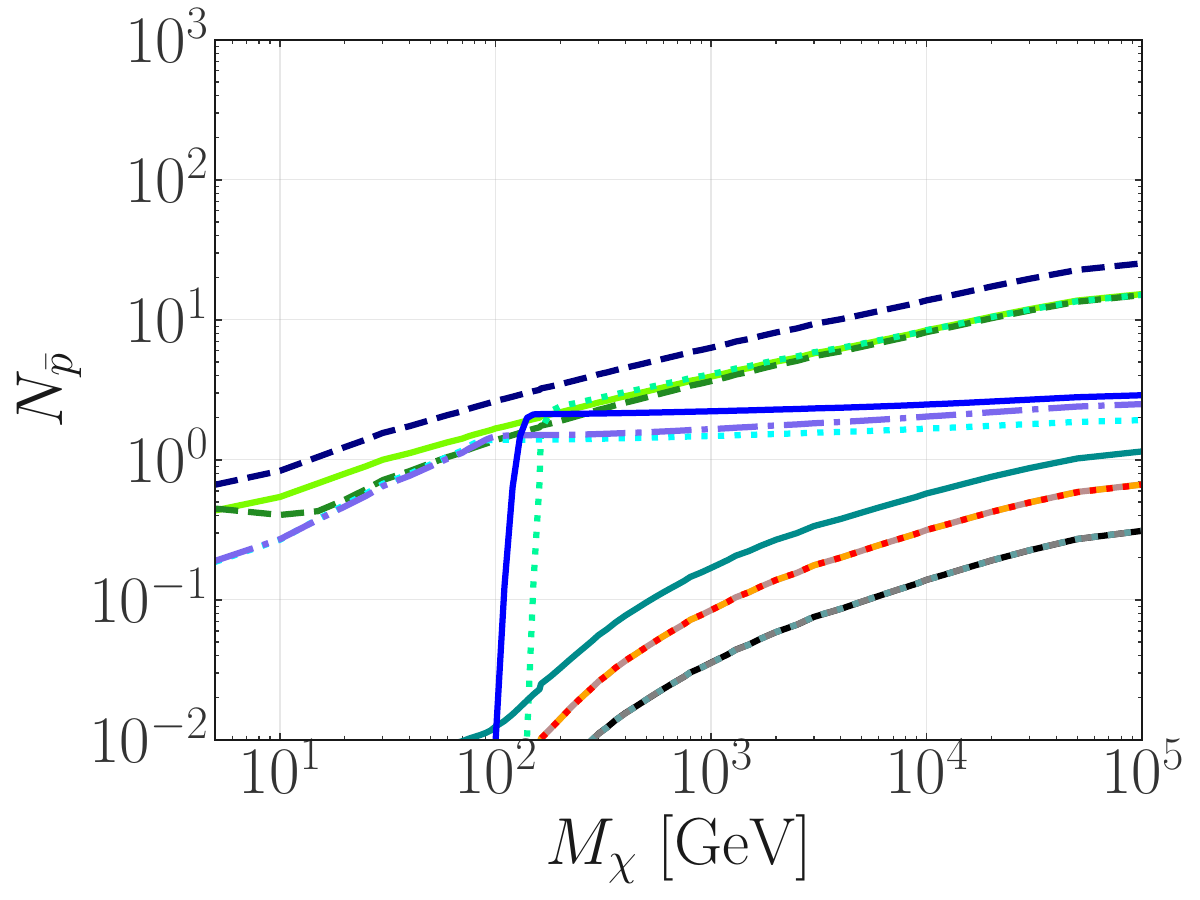}
\includegraphics[width=0.32\linewidth]{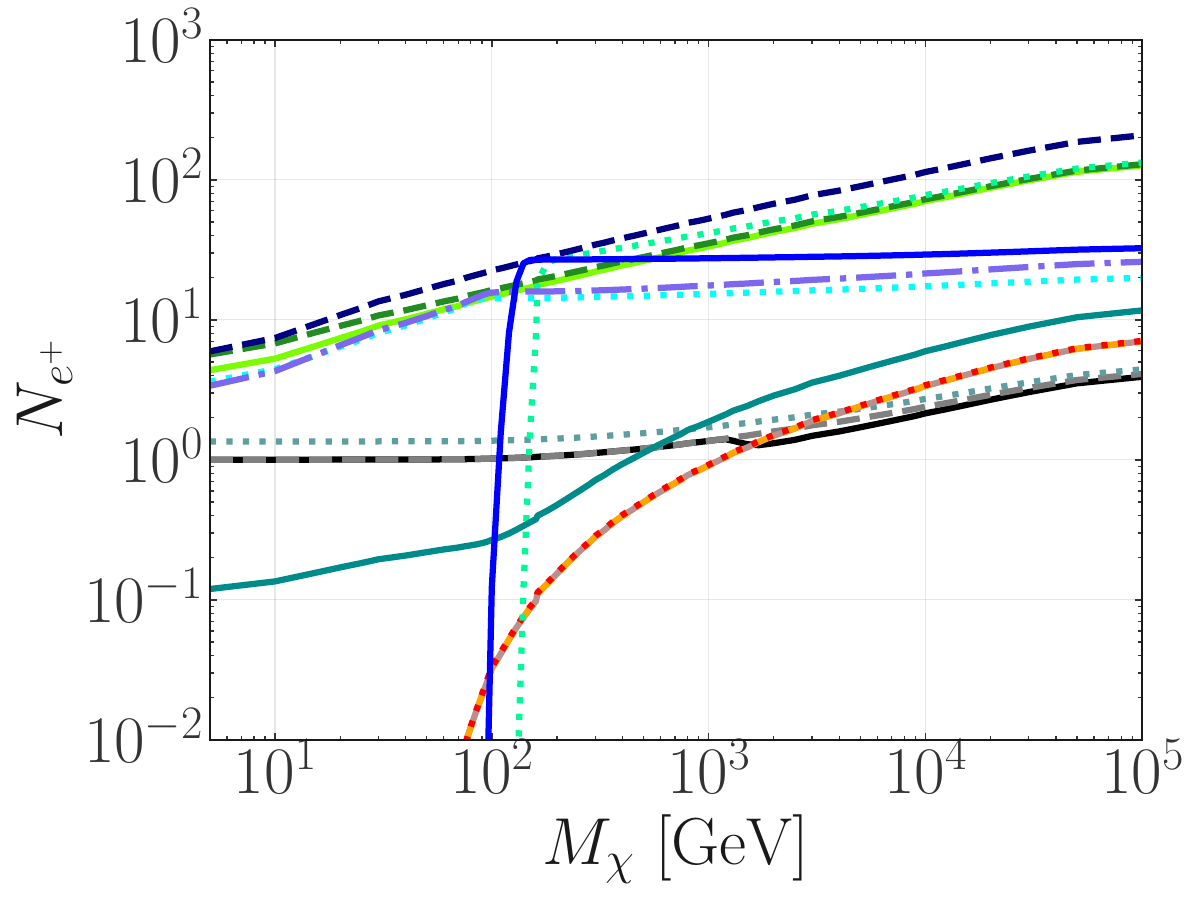}
\vfill
\includegraphics[width=0.32\linewidth]{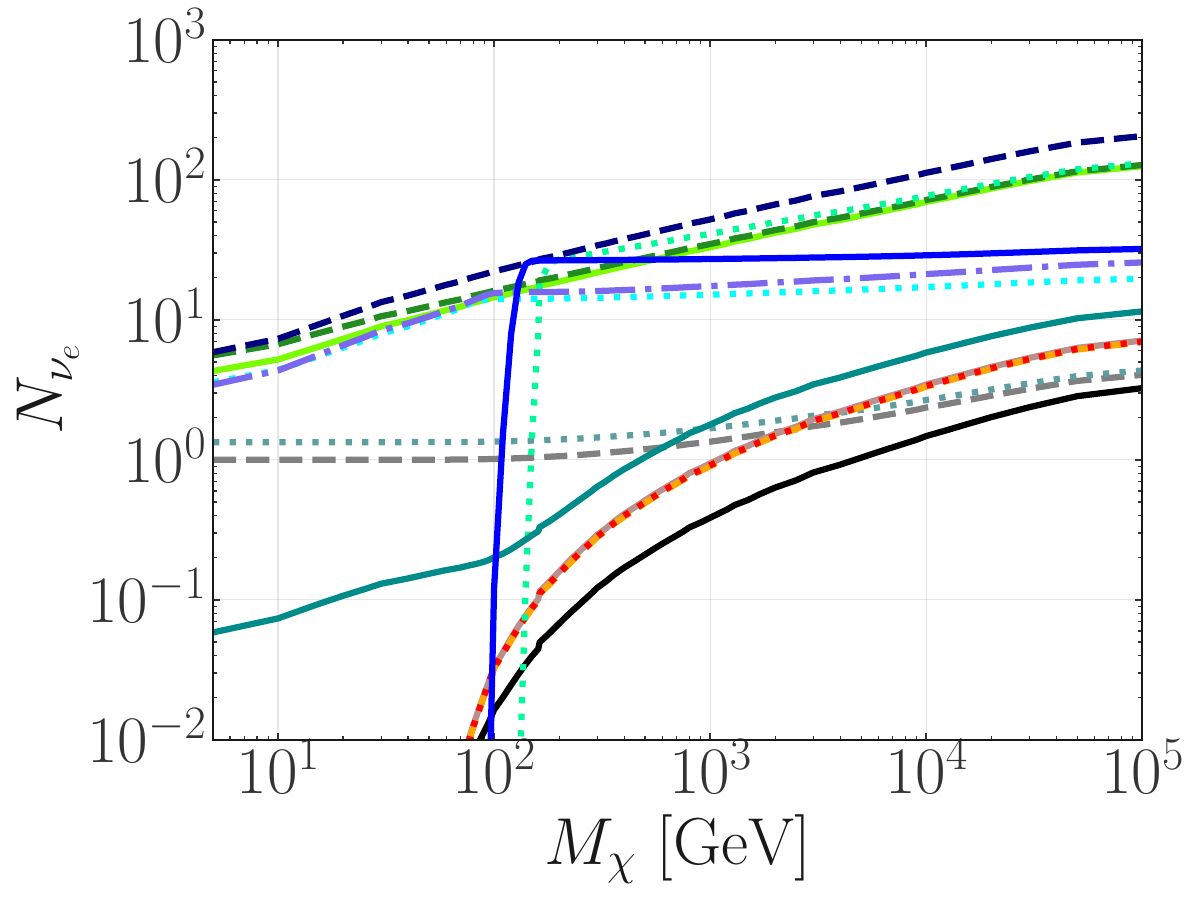}
\includegraphics[width=0.32\linewidth]{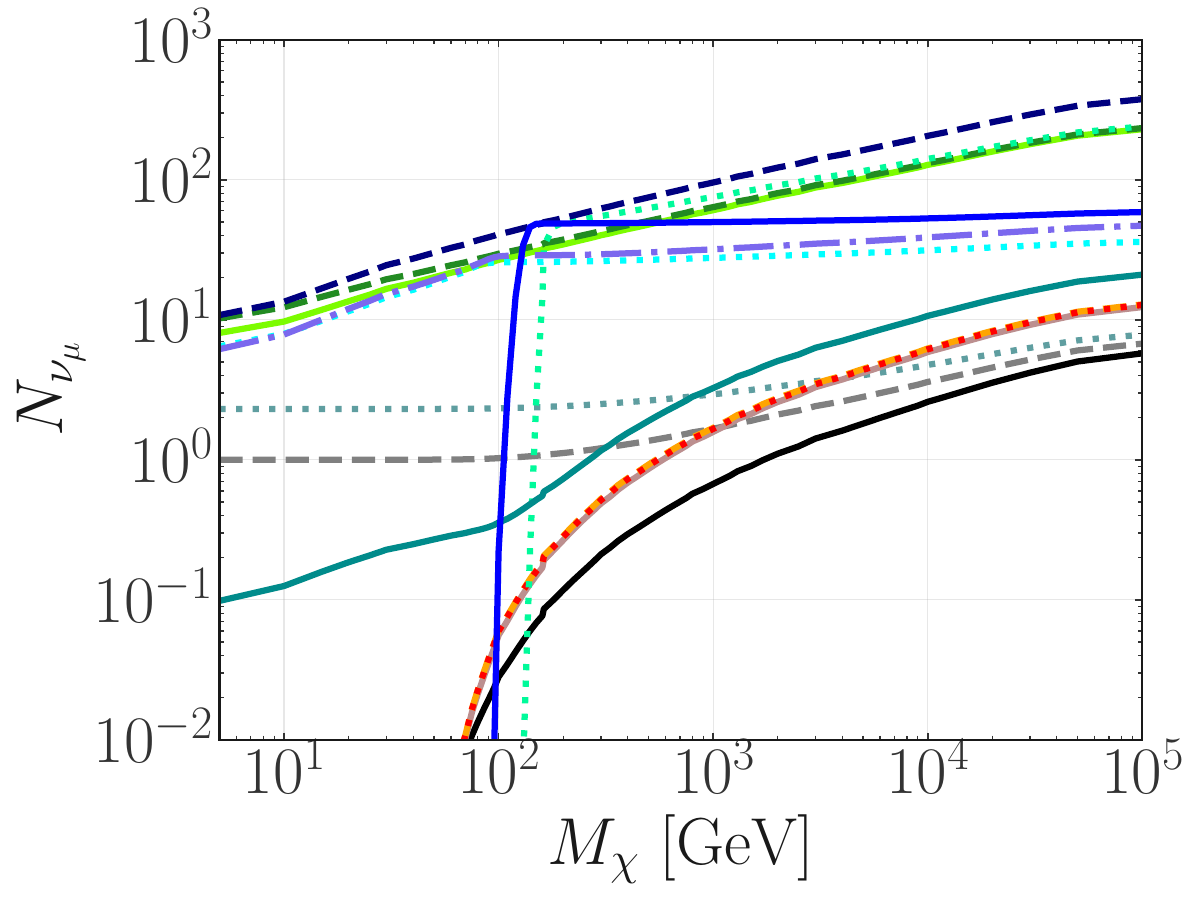}
\includegraphics[width=0.32\linewidth]{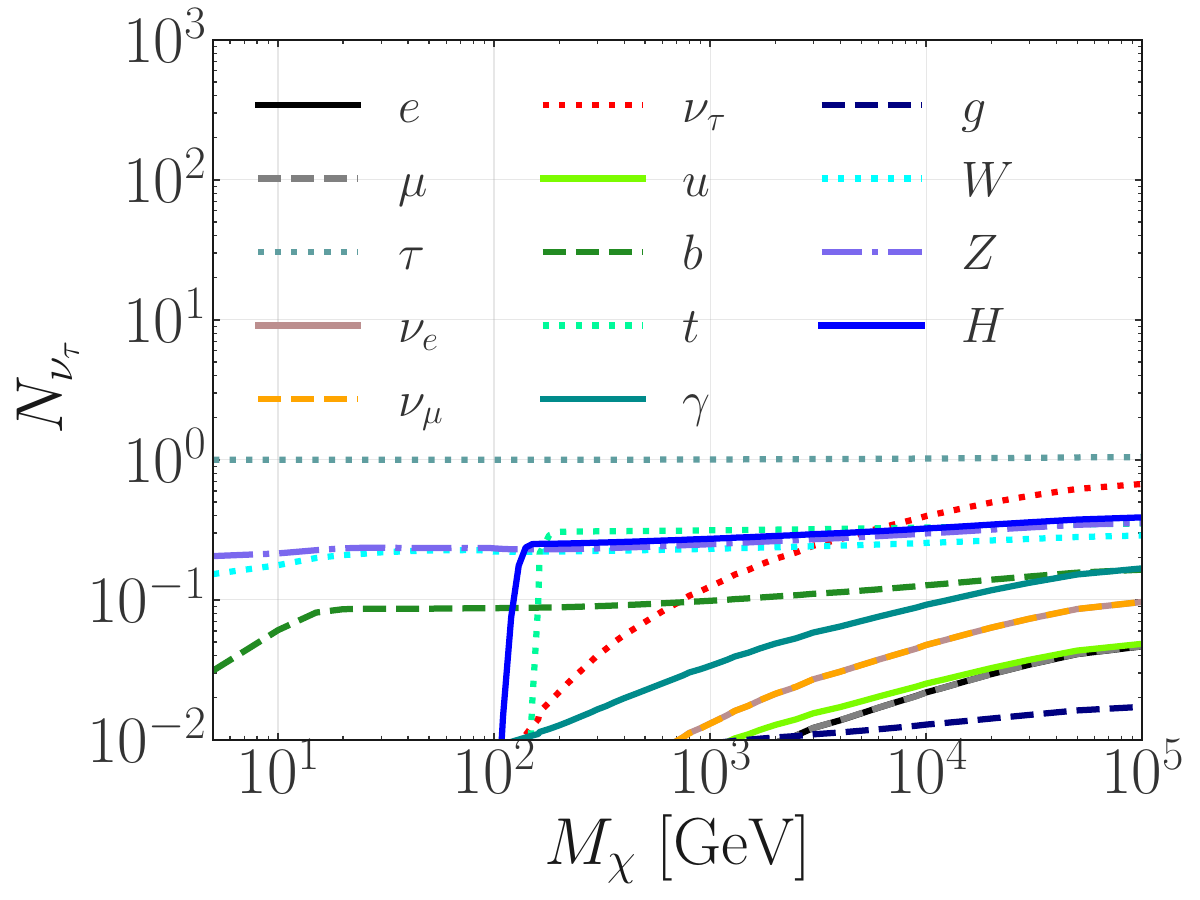}
    \caption{Mean multiplicity per annihilation for the production of cosmic messenger particles as a function of the DM mass for different annihilation channels. From top left to bottom right we show the mean yields for $\gamma$-rays, $\bar{p}$, $e^+$, $\nu_e$, $\nu_\mu$, and $\nu_\tau$.}
    \label{fig:multiplicity}
\end{figure}

In Fig.~\ref{fig:multiplicity}, we show the multiplicity for the production of the cosmic messenger particles from DM annihilation obtained for the different channels as a function of the DM mass. For the case of $\gamma$-rays, the channels involving quarks and gluons produce most of the photons together with the massive bosons. Instead, leptons and neutrinos produce much less $\gamma$ rays. The trend with the mass is similar for the production of positrons, neutrinos and antiprotons. The difference is about a normalization factor. In particular, the multiplicity in the production of $e^+$, $\nu_e$, and $\nu_{\mu}$ is around 0.5, 0.5, and 1.0 times the one for $\gamma$ rays, respectively,  whereas the multiplicity for $\bar{p}$ and $\nu_{\tau}$ yields a fraction of 0.3 and 0.003, respectively, of the multiplicity of $\gamma$ rays only.

\subsection{Comparison with \pppc and \hdms}

\begin{figure}[!t]
\centering
\includegraphics[width=0.49\linewidth]{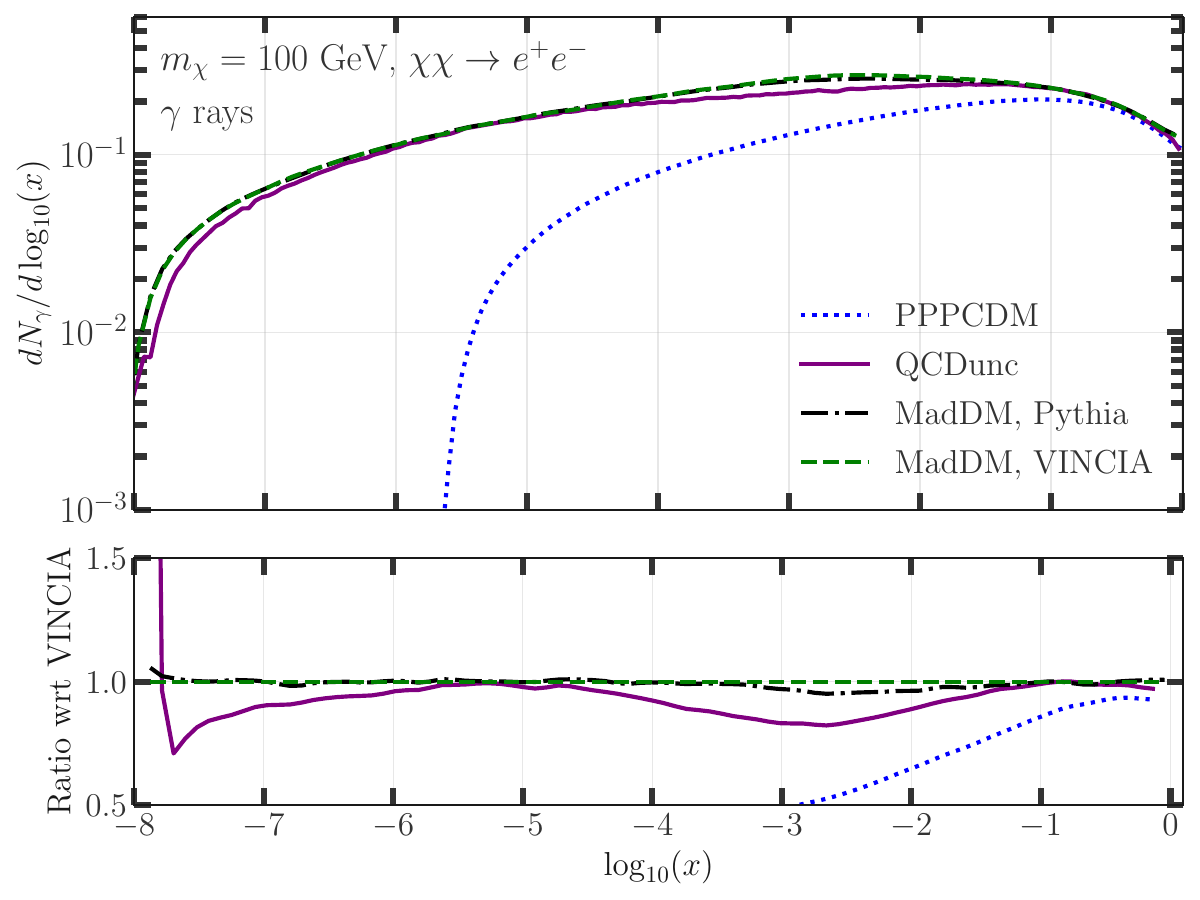}
\includegraphics[width=0.49\linewidth]{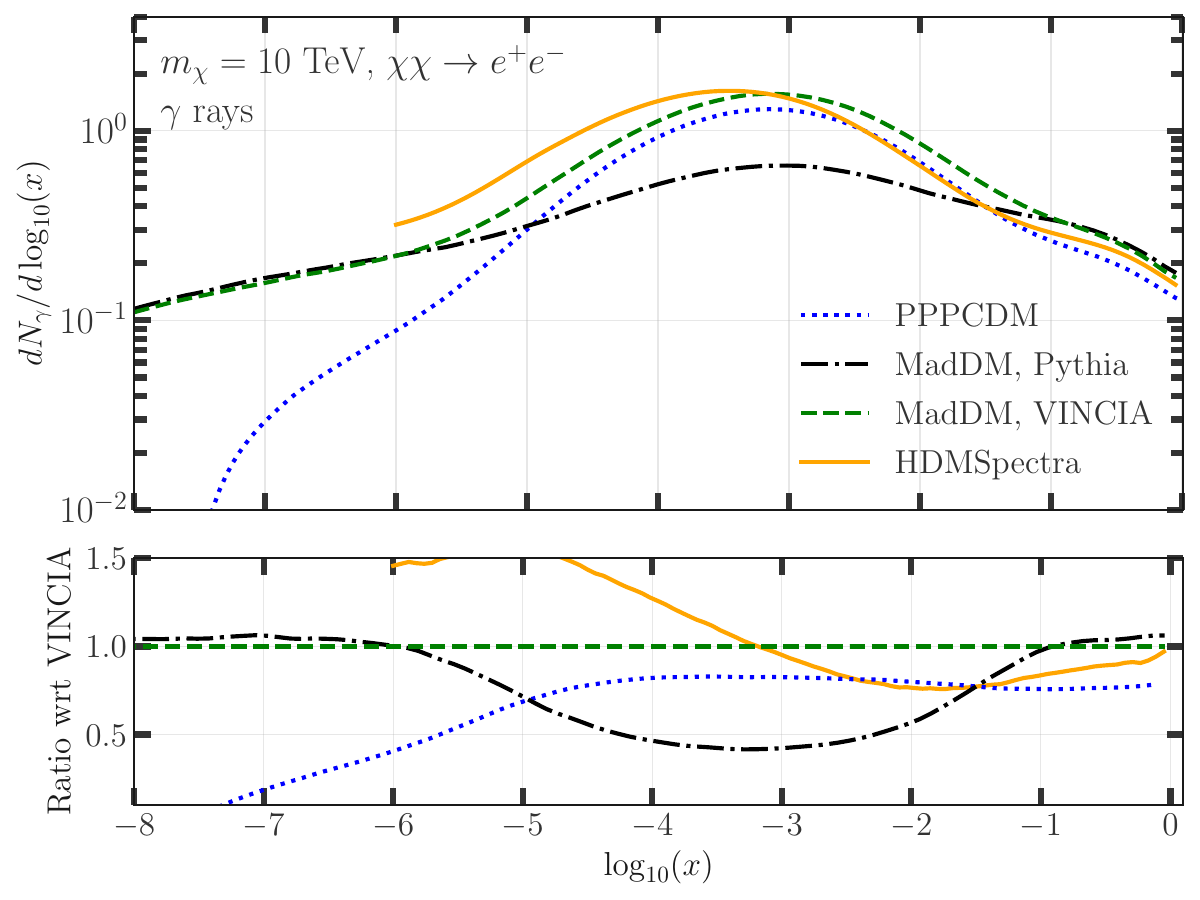}
\includegraphics[width=0.49\linewidth]{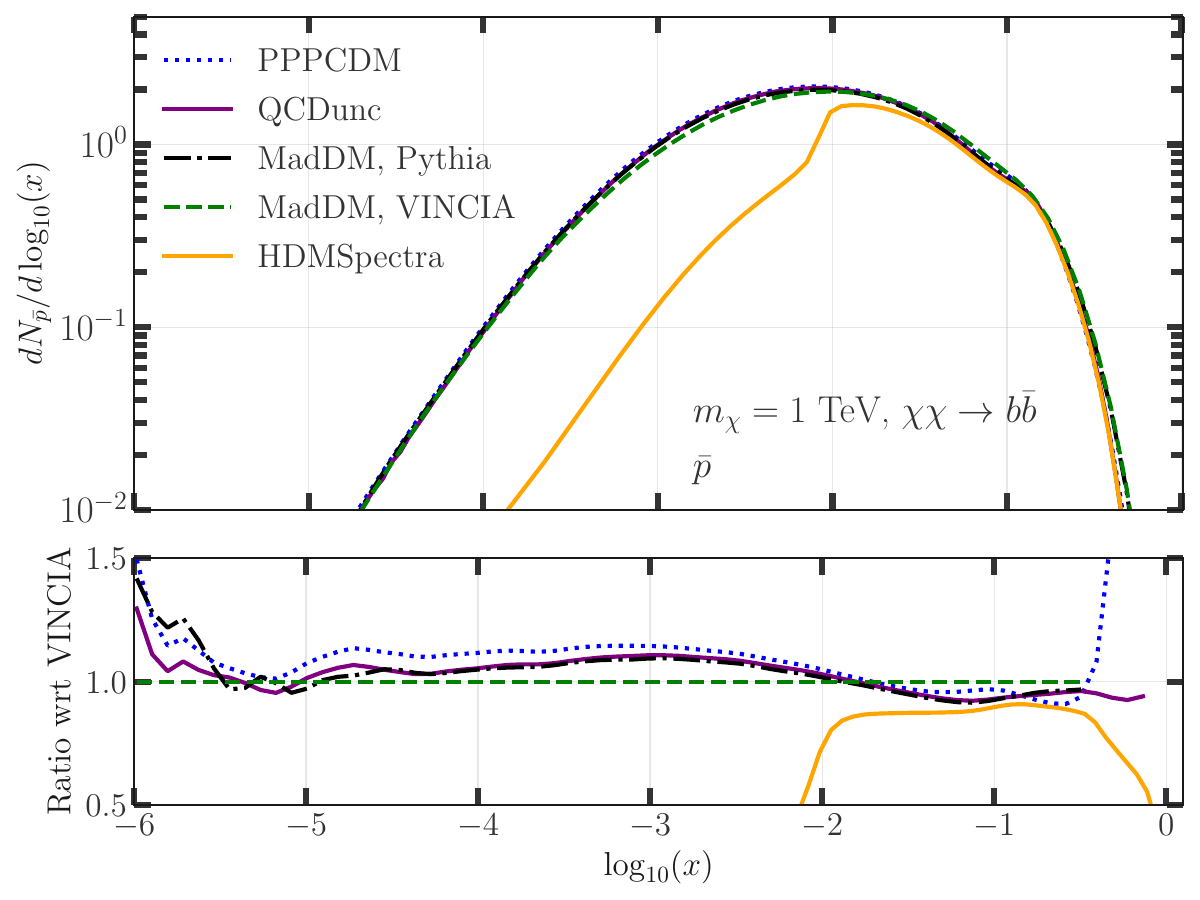}
\includegraphics[width=0.49\linewidth]{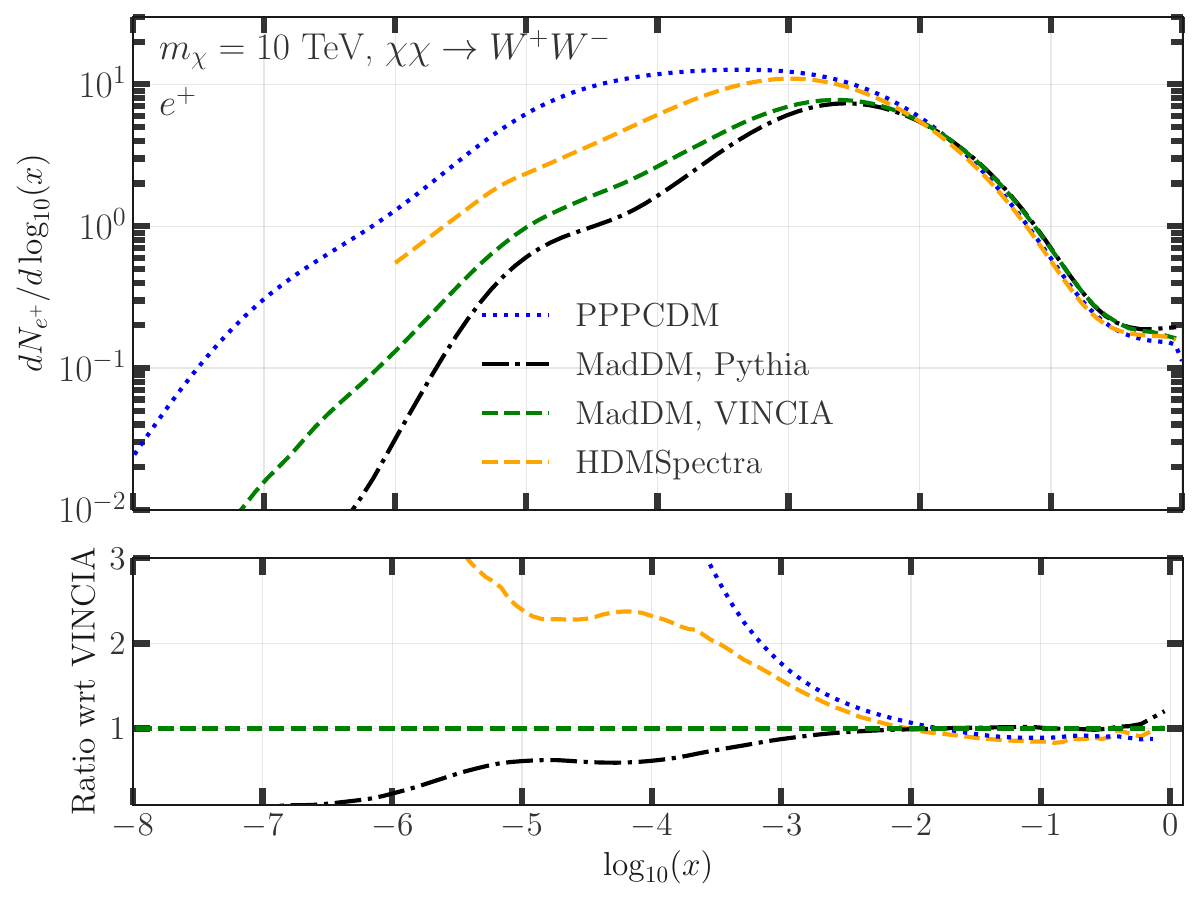}
\caption{Comparison among the spectra obtained with our analysis using the \Pythia and \vincia shower algorithms with the results of \pppc, \qcdunc and \hdms.}
\label{fig:comparison}
\end{figure}

In this section we compare the results of our spectra with the ones obtained in \pppc, \hdms and \qcdunc.

We start by discussing the case of the particle production for the $e^+e^-$ and $\mu^+\mu^-$ channels and DM masses below 1 TeV.
We show in the top left panel of Fig.~\ref{fig:comparison} the comparison of our results with respect to \pppc for the $e^+e^-$ and $m_{\chi}=100$ GeV. We do not show \hdms results because no spectra are provided from this reference for $m_{\chi}<500$ GeV.
We see that \pppc is systematically lower than our spectra for $\log_{10}(x)<-1$. The most important difference is that the \pppc results show a prominent cutoff for $\log_{10}(x)<-5$. The \pppc spectra are systematically smaller than ours at small $\log_{10}(x)$ also for the $\gamma$-ray, $e^+$ and neutrino production with the $\mu^+\mu^-$ channel.
At a DM mass of 100 GeV the main $\gamma$-ray production mechanism is FSR while the EWBR process produces a negligible contribution (see Sec.~\ref{sec:composition}). Therefore, the significant difference between our results and \pppc for the $e^+e^-$ and $\mu^+\mu^-$ channels should reside on the computation of the FSR. 
As an additional check, we generate the spectra using another BSM model, which is the DM simplified model with vector mediator and fermionic DM particle. We also use the {\tt main\_07.cc} provided in the examples of, {\tt \Pythia} which produces the spectra with a generic spin 0 resonance. Also, with these two additional models, the results are very similar to the ones we obtain. 
In order to investigate this discrepancy further, we run the simulation with the same \Pythia version used in \pppc, which is 8.135. We first use the default setup. The results we obtain with this \Pythia version are very similar to the one reported in \pppc (see top left panel of Fig.~\ref{fig:comparison}). Therefore, the deviation between \pppc and our results are due to the difference in the FSR calculation between version 8.1 and 8.3 \Pythia. 
In order to investigate this further, we change the value of the parameter {\tt TimeShower:pTminChgL} from the default value $10^{-4}$ to $10^{-6}$, which is the one used in the newer version of the code.
We therefore traced the discrepancy to reside in the lower threshold for this parameter adopted in the MC: by lowering its value, the discrepancy with our results vanishes, except for a deviation of the order of $20\%$ around $x\sim 10^{-3}$ which is likely due to the improved tuning.
The parameter {\tt TimeShower:pTminChgL} is very important for FSR because it represents a cutoff on the tranverse momentum ($p_T$) for the QED emissions of photons off charged leptons, as already mentioned in Refs.~\cite{Ambrogi:2018jqj,Bauer:2020jay}.
The results obtained by \qcdunc are overall similar to the ones we find except that at around $\log_{10}(x)\approx -3$ the \qcdunc spectrum has a deficit of about 20\%, which is due to the fact that Ref.~\cite{Jueid:2022qjg} does not include EWBR, which contributes exactly at those energies.

The disagreement in the low-energy part of the spectra for the $e^+e^-$ and $\mu^+\mu^-$ channel remains also for DM masses larger than 100 GeV\@. This is visible in the right top panel of Fig.~\ref{fig:comparison} where we show the case for 10 TeV\@. 
In particular, we can see that our results as well as the ones from \hdms, at small $x$ do not show a cutoff in the spectrum as in the case of \pppc.
This issue related to FSR in \pppc has a very relevant impact for the indirect DM searches in $\gamma$ rays with the data of {\it Fermi}-LAT or IACTs. In particular, for analysis of astrophysical objects for which the secondary emission from inverse Compton scattering is not important, and thus the prompt emission is the only one considered, \eg~with Milky Way dwarf spheroidal galaxies. This is particularly relevant for DM masses above 1 TeV for which the low energy tail is the important part of the spectrum for comparison between the theoretical predictions and experimental data for the flux.

In the top right panel of Fig.~\ref{fig:comparison} we show the $\gamma$-ray production for a DM mass of 10 TeV. 
At these masses the EWBR process starts to be important, and the spectra obtained with the three methods are different at the peak, which is located at $\log_{10}(x)$ between $-4$ and $-2$. This is visible comparing the result we find with default \Pythia and \vincia shower algorithm. The difference obtained between \vincia, \pppc and \hdms is at the level of $20$--$30\%$ and it is present for the spectra of all particles.
These discrepancies are due to the different way the EW shower of gauge bosons is implemented in the three analyses.
In particular, \pppc is missing subsequent radiation of gauge bosons and does not include the trilinear diagrams with three gauge bosons. More importantly, \pppc does not include the helicity of the particles when producing the shower. As we have seen in Sec.~\ref{sec:polar}, this can change the spectra at the level of $20$--$40\%$ depending on the energy scale of the process. 
Instead, \hdms includes all the EW corrections considered also in \vincia and takes into account the helicity of the particles. However, their matching procedure may lead to issues for the particle spectra in the low energy region.
There is a last ingredient that can make a difference, which is related to the fragmentation function parameters of \Pythia version used in  \pppc and \hdms which are different from the latest version of \Pythia with our tuning.

For the channel involving quarks, the difference between \pppc or \qcdunc and our model are at the level of $20$--$30\%$ in the most relevant energy ranges. This is true for the production of all particles and for all masses. These deviations are mainly due to the different versions of the \Pythia code. We remind the reader that our results are expected to be more robust because we use a newer version of \Pythia with a more refined shower algorithm that is tuned to data relevant to DM.\footnote{In our analysis, we employ data from resonant  $Z$-boson production only. As hadronization occurs at long distances compared to the hard-scattering process, this approach allows us to constrain the parameters of the hadronization model in a controlled manner. Similar challenges occur in fits of Parton Distribution Functions. Constraining the hadronization model using data at higher energies such as the LHC comes with additional complications and uncertainties.}
The comparison with the spectra of \hdms shows much bigger differences. In fact, the spectra of all the particles show a similar trend in the right-hand part of the peak while in the left-hand part there is a cutoff in the \hdms results. The reason for this discrepancy in the low energy range is originated from the way the mass effects are generated. In the \hdms, the mass effects are accounted for through matching at the EW scale. This leads to smaller yields for massive particle spectra that starts at scales of order the particle mass with respect to the EW scale, {\it i.e.} $10^{-2}$ for the antiprotons. This issue has been discussed in details in Appendix D.2 and shown in Figure 9 of Ref.~\cite{Bauer:2020jay}. 
This different trend could be due to differences between the treatment of the \hdms and \vincia’s EW showering. These differences include the treatment of spin interference, a different treatment of soft interference and the matching procedure at the EW scale, which is not required in \vincia as it performs all evolution in the broken description of the SM\@. For example, \vincia includes full soft coherence effects in QCD processes while \hdms consider it in an approximate way. Moreover, \hdms uses the massless parton approximation in which all the particles are massless. Therefore, mass effects in the analysis of \hdms needs to be taken into account. We refer the interested reader to Ref.~\cite{Bauer:2020jay} where this point is discussed in detail. Note that our model describes the photon data for $e^+ e^- \to q\bar{q}$ produced at the $Z$-pole very well. 
This data corresponds to the region for which the disagreement between our results and those of the \hdms are the most dramatic, \ie~photon energies between $1$ and $10$ GeV. 

There are other very important differences between our model, \pppc and \hdms also when considering the spectra for the channel involving $W^{\pm}$, $Z$ and $H$ bosons.
This is visible in the bottom right panel of Fig.~\ref{fig:comparison} for the spectrum of $e^+$ for DM mass of 10 TeV.
The high-energy part is the same for all the three modelings while below $\log_{10}(x)<-2$ they start to differ significantly. In particular, \pppc is the one that provides the largest yield. We think that the large differences at the low $\log_{10}(x)$ values are due to the treatment of resummations showers, EW showers which is absent in \pppc and to the matching between the analytical calculations and \Pythia used in the \hdms. We stress that more work is needed in order to understand the impact of matching and resummation on the spectra for very heavy DM.

\section{Conclusions}
\label{sec:conclusions}

We presented an improved prediction of DM annihilation spectra for cosmic messengers, specifically, $\gamma$ rays, 
positrons, antiprotons, and the three flavors of neutrinos, which are relevant for indirect detection of DM\@. We employed the \vincia shower algorithm based on the helicity-dependent Antenna functions implemented in the \Pythia code. \vincia includes processes that have not been included in \Pythia before, such as the trilinear boson interaction and the full soft-coherence in multiple FSR emissions. Furthermore, it takes into account the helicity of the particles during the entire showering process. For the most relevant parameters of the string hadronisation model, we  have performed an improved tuning  to the LEP data at the $Z$ boson resonance using measurements of pions, photons, hyperons and event shapes. 
The resulting spectra provide a new state of the art, as we demonstrate by a thorough comparison of our results to previous literature as well as a careful assessment of the underlying uncertainties.
The precision of our prediction reaches a 10\% level in the energy ranges most relevant for DM searches. 

We have generated the spectra from 5 GeV to 100 TeV for all annihilation channels into a pair of SM fermions and bosons including $\gamma Z$, $HZ$ and off-shell vector bosons previously not considered in publicly available spectra. To allow for applicability to a wide range of DM models, we have taken into account fixed helicity states for fermions and polarization for the EW gauge bosons.
However, this approach does not take into account effects such as ISR and IB, which requires a reevaluation of the hard process in the specific DM model under consideration. Upon simple rescaling, our results can also be applied to decaying DM covering the case of (pseudo)scalar and (axial)vector DM\@. 
Our results provide important input for the interpretation of upcoming searches for DM in the multi-TeV range performed with data from LHAASO, HAWC and the future Cherenkov Telescope Array Observatory. The tabulated spectra are publicly available on \href{https://github.com/ajueid/CosmiXs.git}{github}.

\acknowledgments

We thank Christian Bierlich, Peter Skands and all \Pythia authors team for the help provided with the code.
We also thank Marco Cirelli, Daniele Massaro, Olivier Mattelaer, Nicholas Rodd for insightful discussions and comments.
N.F.~and M.D.M.~acknowledge support from the Research grant {\sl TAsP (Theoretical Astroparticle Physics)} funded by Istituto Nazionale di Fisica Nucleare (INFN). The work of A.J. is supported by the Institute for Basic Science (IBS) under the project code, IBS-R018-D1. A.J. would like to thank the CERN Theory Department for its hospitality where this work has been finalized. C.A. acknowledges support by the F.R.S.-FNRS under the “Excellence of Science” EOS be.h project no. 30820817. J.H.~acknowledges support by the Alexander von Humboldt foundation via the Feodor Lynen Research Fellowship for Experienced Researchers and
Feodor Lynen Return Fellowship. R.R. acknowledges support from the Ministerio de Ciencia e Innovación (PID2020-113644GB-I00) and the GVA Research Project {\sl Sabor y Origen de la Materia (SOM)} (PROMETEO/2022/069).

\appendix

\section{Parton-shower algorithms: \Pythia vs \vincia}
\label{app:shower}

In \Pythia~8, the parton showers are based on the dipole type $p_\perp$-ordered evolution. This algorithm has been available since \Pythia~6.3 and used for both Initial State Radiation (ISR) and Final State Radiation (FSR) \cite{Sjostrand:2004ef,Sjostrand:2006za}. On the other hand, \Pythia~8 includes an implementation of the $\gamma \to f\bar{f}$ splittings and EW massive gauge boson emissions like $q \to q' W$ and $q \to q Z$ \cite{Christiansen:2014kba}. The EW showers are, however, switched off by default.

Since \Pythia~8.219, photon emissions from heavy resonances such as the $W$--bosons are included in the parton-shower machinery. The default treatment in \Pythia is based on a combination of DGLAP splitting kernels for QED+QCD radiation with dipole ($2\to 3$) kinematics~\cite{Sjostrand:2004ef}. Mass effects in the parton showers are available through matrix-element corrections (MECs) which are available for ISR \cite{Sjostrand:2004ef} and FSR \cite{Norrbin:2000uu}.
The strength of the QCD showers in \Pythia~8 is controlled by the effective value of the strong coupling constant $\alpha_S$. Note that in \Pythia~8 the value of $\alpha_S$ at the $Z$-pole is not equal to $\alpha_S(M_Z)^{\rm \overline{MS}} = 0.118$. There are two reasons for such a choice.  In the soft-gluon emission limit, the dominant terms for the splitting functions can be universally absorbed into the leading order splitting kernel by a translation to the Catani-Marchesini-Webber (CMW) scheme (also called the MC scheme) \cite{Catani:1990rr}.  This results in an increase of $\alpha_S(M_Z)$ by about $10\%$. Furthermore, in the recent tuning of \Pythia~8 an agreement with experimental data for the measurements of the $e^+ e^- \to 3~{\rm jets}$ is reached if $\alpha_S$ is increased by another $10\%$, see \cite{Skands:2010ak,Skands:2014pea}. Note that in \Pythia~8 there is the possibility to choose different values of $\alpha_S$ for ISR, FSR, MPI or the hard-scattering process. The default value of $\alpha_S(M_Z)$ in \Pythia~8 for FSR is $\alpha_S(M_Z) = 0.1365$ and the default choice of the RGE running is the one-loop order. We do not change these options in this study.

It is accepted that a scale proportional to the shower evolution scale ($p_\perp$) to be used for the evaluation of $\alpha_S$ at each branching (called thereafter the renormalisation scale). Uncertainty estimates can be performed by variation of the renormalisation scale by a factor of two in the positive and the negative directions: $\{\mu_+, \mu_-\} = \{1/2, 2\}~\mu_R$. However, this may destroy some of the universal corrections obtained in the CMW scheme. To solve this issue, a framework for the automated scale variations was recently developed in Ref.~\cite{Mrenna:2016sih} and was implemented since \Pythia~8.215. The formalism allows for compensation terms to reduce the effects of large variations while having an agreement with the CMW scheme. On the other hand, this formalism allows for variations of the non-singular terms of the splitting functions. The firstt application of this formalism to DM indirect detection has been done in Refs. \cite{Amoroso:2018qga, Jueid:2022qjg, Jueid:2023vrb} where uncertainties were found to be of order $10$--$20\%$ depending on the annihilation final state and the energy region. 

\vincia is a $p_\perp$-ordered parton-shower model for QED+QCD+EW emissions based on the Antenna formalism.  This formalism was first introduced in \textsc{Ariadne} event generator \cite{Lonnblad:1992tz} which is initially based on the colour dipole model \cite{Gustafson:1986db,Gustafson:1987rq} and notably used for LEP studies. The treatment of QCD showers in FSR is similar to the one used in \textsc{Ariadne}. In the case of ISR, \vincia extends the concept of backward evolution to the formalism \cite{Ritzmann:2012ca} through coherent Initial-Initial (II), Final-Final (FF) \cite{Brooks:2020upa} and Resonance-Final (RF) Antennae \cite{Brooks:2019xso}. Due to the fact that all these components are coherent and interleaved in a single sequence of decreasing $p_\perp$, \vincia possesses the unique property of soft coherence for all the physical situations. For the QED showers, the default Antenna functions include fully coherent multipole soft interference effects, which are added to the collinear DGLAP structure \cite{Kleiss:2017iir,Skands:2020lkd}. Such QED multipole treatment is fully interleaved with the QCD evolution as well. This feature is very unique to \vincia.

In addition, \vincia contains an implementation of EW showers which includes all the Higgs boson couplings and all the gauge-boson self-interactions \cite{Kleiss:2020rcg,Brooks:2021kji} allowing for all the $Z/W/H$ branchings. This EW shower module enables for inclusion of weak corrections in EW Sudakov form factors and resummation of multiple massive gauge-boson emissions and branchings. However, when activating this module only the collinear limits are implemented and not the full soft interference effects. To use the EW shower module, the helicity information on the produced partons needs to be provided, since this module is based on helicity-dependent shower \cite{Larkoski:2013yi,Fischer:2017htu}. This can be achieved by either providing the Les Houches Files (LHEF) as input to add parton showers which explicity contain the helicity information or internally via a \vincia option for hard-scattering matrix element calculations.

\vincia also includes interleaved resonance decays which means that short-lived heavy resonances such as the $W/Z/H$ bosons, the top quark or any beyond-the-SM (BSM) resonance that are produced either in the hard-scattering process or emitted in the EW evolution are treated to be stable particles until the evolution scale reaches a $p_\perp$ of order of the off-shellness scale. The system composed of a shower plus decay is then merged into the upstream system and the QED+EW+QCD evolution of the system continues starting from the offshellness scale. This picture is physically intuitive, as heavy resonances can not emit radiations at frequencies that are lower than the inverse of their lifetime. This unique feature of \vincia leads to dramatic impacts on the distributions of the reconstructed invariant mass, as compared to the case where resonance decays are not interleaved with the shower evolution (like in \Pythia simple shower). The mass effects are properly taken into account for FSR, and the corresponding Antenna functions have the appropriate limits in the quasi-collinear regions. On the other hand, \vincia supports the sector of Antenna showers where the phase space is divided into non-overlapping regions and in which case every sector receives only contributions from one Antenna branching function. This feature enables for straightforward inclusion of higher order corrections and also for multi-jet merging, which is also called sector merging (for more details, see Refs. \cite{Hoche:2021mkv,Campbell:2021svd}). The choice of $\alpha_S$ in \vincia for ISR and FSR is similar to that in the simple shower in \Pythia~8. On the other hand, \vincia allows for different choices of the scale factors for ISR and FSR emissions and splittings and different cut-off scales for the II, IF and FF emissions. \\

\section{Dark Matter energy spectra calculated with \maddm}
\label{app:spectrum}

In this section, we report the relevant commands that we use for generating the spectra with the \maddm code. The version of \maddm that we use is a custom one which includes the \vincia shower algorithm and the tuning we have derived in this paper. This version will be released in the comings months.

We remind the reader that we use three specific BSM models, which are:
\begin{itemize}
\item the Singlet scalar model with a Higgs portal (SHP) (see, \eg, \cite{DiMauro:2023tho}).
\item DM simplified model with a fermionic DM particle and a vector boson mediator with fermionic DM  (DMSimp1) (see, \eg, \cite{Arina:2018zcq}).
\item DMS   imp model with a CP-odd scalar mediator model with fermionic DM (DMSimp0) (see, \eg, \cite{Arina:2018zcq}).
\end{itemize}
In particular, we use the SHP model for all the channels except for the neutrinos, which are assumed to be massless and thus have coupling with the SM Higgs boson. Moreover, for the cases for which we want to get specific chirality or polarization states for fermions and bosons we use two cases we use the DMSimp model.
The syntax considered in \maddm to generate the spectra for the SHP model is the following: \\

\noindent
{\tt import model ScalarHiggsPortal\_NLO\_UFO} \\
{\tt define darkmatter n1} \\
{\tt generate indirect\_detection b b~} \\
{\tt output folder\_name} \\
{\tt launch folder\_name} \\
{\tt set indirect = flux\_source} \\
{\tt set vave\_indirect\_cont 1e-3} \\
{\tt set save\_output spectra} \\
{\tt set precise} \\
{\tt set sigmav\_method madevent}  \\
{\tt set nevents 5e6} \\
{\tt set msdm 1000} \\

In the first row the model is imported, then the DM particle is defined, and we request the calculation of the cross-section and spectra for the bottom channel. The mode is written in output in a folder and the analysis is launched in the same folder. With {\tt set indirect = flux\_source} we request the calculation of the flux at the source, and we set the DM velocity to $10^{-3}c$. We require that the spectra files are saved in the folder, and we use the precise method to calculation the cross-section using the {\tt madevent} model which fixes the DM relative velocity. Finally, we set the number of events and the DM mass value in units of GeV.

For the spectra with the neutrino channels, we use the DMSimp1 model by changing the couplings between the mediator and the neutrino to a value different from 0. 
We use for this scope the following commands: \\

\noindent
{\tt import model DMsimp\_s\_spin1\_MD} \\
{\tt define darkmatter ~xd} \\
{\tt generate indirect\_detection ve ve~} \\
{\tt output folder\_name} \\
{\tt launch folder\_name} \\
{\tt set indirect = flux\_source} \\
{\tt set vave\_indirect\_cont 1e-3} \\
{\tt set save\_output spectra} \\
{\tt set precise} \\
{\tt set sigmav\_method madevent}  \\
{\tt set gnu11 0.25}  \\
{\tt set gnu22 0.25}  \\
{\tt set gnu33 0.25}  \\
{\tt set nevents 5e6} \\
{\tt set msdm 1000} \\

In order to produce the spectra for the $ZH$ channel, we use the DMSimp0 model. The syntax used for this case is: \\

\noindent
{\tt import model DMsimp\_s\_spin0\_MD} \\
{\tt define darkmatter ~xd} \\
{\tt generate indirect\_detection h z} \\
....\\
where the part after the generate command is the same as in previous cases, shown above. \\

We use the same model also in the case we want to generate only left-handed electrons: \\

\noindent
{\tt import model DMsimp\_s\_spin1\_MD} \\
{\tt define darkmatter ~xd} \\
{\tt generate indirect\_detection e+ e-} \\
{\tt output folder\_name} \\
{\tt launch folder\_name} \\
{\tt set indirect = flux\_source} \\
{\tt set vave\_indirect\_cont 1e-3} \\
{\tt set save\_output spectra} \\
{\tt set precise} \\
{\tt set sigmav\_method madevent}  \\
{\tt set gVl11 0.25}  \\
{\tt set gVl22 0.25}  \\
{\tt set gVl33 0.25}  \\
{\tt set gAl11 -0.25}  \\
{\tt set gAl22 -0.25}  \\
{\tt set gAl33 -0.25}  \\
{\tt set nevents 5e6} \\
{\tt set msdm 1000} \\
where we have specified to take values of the $g_V$ and $g_A$ that are opposite. In case we desire to produce the opposite chirality, we have to reverse the sign of the axial couplings.

As discussed in the main text of the paper, for DM masses below the threshold of the massive boson, the off-shell contribution of the $W$, $Z$ and $H$ can be very relevant. In order to take into account this effect, we generate the diagrams with four final fermions as follows: \\
\linebreak
\noindent
{\tt import model ScalarHiggsPortal\_NLO\_UFO} \\
{\tt define darkmatter n1} \\
{\tt define ferm = u u~ d d~ c c~ s s~ t t~ b b~ e- e+ mu- mu+ ta- ta+ ve ve~ vm vm~ vt vt~} \\
{\tt generate indirect\_detection w+ w- > ferm ferm ferm ferm / g+ g- g0 ferm} \\
..... \\

In the last row, the part written as {\tt / g+ g- g0 ferm} is added to remove the goldstone bosons and the fermions from the internal legs in the diagrams.
We proceed with the same syntax also for the $HZ$ channel but with the model DMSimp0. \\

Finally, for the case of polarized EW bosons, {\it i.e.} $W_L W_L, W_T W_T, Z_L Z_L, Z_T Z_T$, the user needs to use the
following commands:\\

\noindent
{\tt import model ScalarHiggsPortal\_NLO\_UFO} \\
{\tt define darkmatter n1} \\
{\tt generate indirect\_detection w+\{0\} w-\{0\}} \\
{\tt ...} \\
and for transverse gauge bosons, {\tt w+\{0\} w-\{0\}} needs to be replaced by {\tt w+\{T\} w+\{T\}}. Similar commands can be used for the case of the $Z$-boson, {\it i.e.} {\tt generate indirect\_detection z\{0\} z\{0\}} for $Z_L Z_L$.

The output of \maddm is then passed to \Pythia to add parton showers and hadronization. Techincally, \maddm generated a {\tt LHEF} with the kinematics of the final particles in the diagram and \Pythia produce the showering and hadronization processes starting from these final particles. Below we list the main commands to activate the \vincia Antenna shower module with EW corrections. First, the following two commands need to be added \\

\noindent
{\tt PartonShowers:model = 2} \\
{\tt Vincia:ewMode       = 3} \\

The first command switch to the \vincia Antenna shower module, while the second command activates the fully-fledged EW corrections. At the run time, \Pythia will display all the changes with respect to the default configuration that is based on the \Pythia simple shower and the Monash tuning. To use the parameters of the hadronization model that are obtained in our tuning, the following needs to be added as well \\

{\tt StringZ:aLund = 0.337} \\
{\tt StringZ:bLund = 0.784} \\
{\tt StringPT:sigma = 0.296} \\
{\tt StringZ:aExtraDiquark = 1.246} \\

There are five particles that are considered to be unstable at astrophysical and cosmological timescales while they are stable at collider experiments, \ie~$\pi^\pm$, $\mu^\pm$, $K^\pm$, $K_L$ and $n$. To make these particles unstable, there are two possibilities: either increase the limit on the default value of the proper $c\tau$ for the particle to decaying: \\

\noindent
{\tt ParticleDecays:limitTau0 = on} \\
{\tt ParticleDecays:tau0Max = 10.} \\

The last command is the default option for which particles with $c\tau > 10~{\rm mm}$ are considered long-lived. For DM studies, we can change the value of $c\tau$ to very high values like $10^6$ or something like that. The other option is to explicitly ask \Pythia to decay these particles, \ie \\

\noindent
{\tt 13:mayDecay   = true      ! muon} \\
{\tt 211:mayDecay  = true      ! pi+-}  \\
{\tt 321:mayDecay  = true      ! K+-} \\
{\tt 130:mayDecay  = true      ! Klong} \\
{\tt 2112:mayDecay = true      ! neutron} \\

\bibliographystyle{JHEP}
\bibliography{paper.bib}

\end{document}